\documentclass[%
 reprint,
 amsmath,amssymb,
 aps,
]{revtex4-2}

\usepackage{graphicx}
\usepackage{subfig}
\usepackage{dcolumn}
\usepackage{bm}
\usepackage{color}
\usepackage{ulem}
\usepackage[dvipsnames]{xcolor} 

\begin{document}

\preprint{APS/123-QED}

\title{Optics Design of Vertical Excursion Fixed-Field Alternating Gradient Accelerators}

\author{S.~Machida}
\email{shinji.machida@stfc.ac.uk}
\affiliation{STFC Rutherford Appleton Laboratory, Harwell Campus, Didcot, OX11 0QX, United Kingdom}
\author{D.J.~Kelliher}
\affiliation{STFC Rutherford Appleton Laboratory, Harwell Campus, Didcot, OX11 0QX, United Kingdom}
\author{J-B.~Lagrange}
\affiliation{STFC Rutherford Appleton Laboratory, Harwell Campus, Didcot, OX11 0QX, United Kingdom}
\author{C.T.~Rogers}
\affiliation{STFC Rutherford Appleton Laboratory, Harwell Campus, Didcot, OX11 0QX, United Kingdom}


\date{\today}

\begin{abstract}
Vertical excursion fixed-field alternating gradient accelerators can be designed
with tunes that are invariant with respect to momentum and trajectories that are 
scaled images of each other displaced only in the vertical direction. 
This is possible using 
guiding fields that have a vertical exponential increase,
with a skew quadrupole 
component in the magnet body and a solenoid component at the magnet ends.
Because of the coupling this introduces, orbit and optics calculations and 
optimisation of parameters need to be performed numerically. 
In this paper, idealised magnetic fields are calculated 
from first principles, taking into account end fields. 
The parameter dependence of the optics and the dynamic aperture of the ring are 
calculated for the example of a ring with an approximately 25\,m circumference 
that accelerates proton beams from 3\,MeV to 12\,MeV.
The paper reports for the first time the design of such
an accelerator lattice using tools specifically devised to analyse transverse 
coupled optics without the need for approximations.

\end{abstract}

\maketitle


\section{Introduction}
The idea of a vertical excursion Fixed-Field Alternating Gradient accelerator 
(vFFA)~\cite{Ohkawa1,Teichmann}
can be traced back to the discovery of the original FFA concept~\cite{Ohkawa2, Symon, Kolomensky}, 
although relatively little emphasis was placed on the vFFA option at the time. 
In a vFFA, the beams move vertically when accelerated in contrast to the original FFA where the orbits stay in a horizontal plane.
The idea has since been revived independently by Brooks, who has pointed out several potential applications 
of such a machine~\cite{Brooks}.

To avoid any confusion, we use the notation hFFA and vFFA in this paper to distinguish between
the horizontal and vertical arrangements.

When used to accelerate electrons in the ultra-relativistic regime, a
vFFA can deliver continuous electron beams because the path length is 
independent of momentum, in contrast with ordinary cyclotrons.
Such a machine has been called an electron cyclotron~\cite{Ohkawa1}.
Recently a continuous proton beam accelerator has been proposed by Mori~\cite{Mori} for 
the production of high power muon beams based on the vFFA concept.
 
The feature of a finite dispersion function only in the vertical direction is unique among particle accelerators.
The momentum compaction factor is zero as a result of the zero horizontal 
dispersion function.
The path length is independent of the beam momentum so that the beam dynamics 
behaviour is closer to that of a linear accelerator.
This is advantageous for muon acceleration at ultra-relativistic energies in a muon collider or a neutrino factory.
There is no need to ramp the magnetic field strength or to change the RF 
acceleration frequency according to the beam momentum.
If adopted as a synchrotron radiation source,
light rays of different radiation spectra are separated vertically  
because the beam momentum is a function of vertical position.
The zero momentum compaction factor including higher order terms of momentum deviation also opens up novel lattice design strategies towards short bunch operation \cite{martin}.
For crystalline beam formation in a ring accelerator,
zero dispersion in the horizontal direction could eliminate the use of taper 
cooling~\cite{Okamoto}.
These are just a few examples of possible vFFA applications.
A practical feature is that the footprint of the main magnets in a vFFA is reduced compared to a 
cyclotron or an hFFA, although the magnets are taller.


Unlike an hFFA, a vFFA separates the shape of the ring (as seen from above) from the so-called scaling conditions.
The scaling conditions are, first, the shape of orbits for different momenta in hFFAs are photographic enlargement of each other, 
whereas in vFFAs the orbits are identical apart from a vertical translation.
Secondly, effective focusing strength normalized by the particle's momentum is constant for all the momenta. Those two conditions make the transverse tune constant over the whole range of acceleration.
The magnetic field variation in an hFFA, which applies to each field component, is 
\begin{equation}
\label{eq:eq1}
B/B_{0}=\left( r/r_0 \right)^k
\end{equation} 
where $r_0$ is the reference radius measured from the machine centre and 
$B_{0}$ is the magnetic field at the reference radius.
$k$ is the geometrical field index defined as 
$k=\left( r/B \right) \left( \partial B/\partial r\right)$,
which determines the field strength as a function of the distance from the 
machine centre. 
In a vFFA the magnetic field satisfies the scaling conditions
\begin{equation}
\label{eq:eq2}
B/B_{0}=\exp\big(m(y-y_0)\big)
\end{equation} 
where $y_0$ is the reference position in the vertical coordinate and $B_{0}$ is the magnetic field at the reference position.
$m$ is the normalised field gradient defined as $m=\left( 1/B \right) \left( \partial B/\partial y\right)$.
This does not involve the geometry of the ring.
As long as stable optics exist, the ring shape can be arbitrary~\cite{IPAC2019}.

Note that there is a family of FFAs which do not obey the scaling conditions, referred to as `non-scaling'.
However, here we consider only a vFFA design where the scaling conditions hold true.

vFFAs rely on a magnetic field where all three components increase 
exponentially in the vertical direction. As shown in section~\ref{sec:magnets},
it is possible to create such magnetic fields with a semi-analytical approach consistent with Maxwell's equations.
With a proper combination of these magnets closed orbits and stable optics can 
be found corresponding to a ring accelerator.

As Brooks pointed out~\cite{Brooks},
the beams are predominantly confined by a skew quadrupole component in the 
middle of the lattice magnets.
The particle motion in the horizontal and vertical planes is not independent.
However the source of the coupling is not only from the skew quadrupole 
component in the magnet body.
As described in section \ref{sec:magnets}, there 
is a potentially strong solenoid field at both ends of the magnet which 
introduces additional mixing between the horizontal and vertical motions.
In fact, the maximum magnetic field is driven by this longitudinal end field in 
our design.
When the magnet is long compared with the fringe field fall-off and the bending 
angle per cell is small, 
a 45~degree rotation of the physical horizontal and vertical coordinates is 
sufficient to decouple the particle motion, as in the case of~\cite{Brooks}.
This is not the case, however, when the magnet is short and the bending angle per cell is of the order of 10~degrees,
which is the case for the majority of accelerator designs up to a few GeV and is discussed below.
Edge focusing, weak focusing and higher order multipoles also contribute.
The beams experience a mixture of normal and skew components as they pass through each cell. 

The complication due to this mixing makes it challenging to derive
essential accelerator parameters  analytically such as transverse tune and beta functions.
In this paper the stability of the optics and the beam 
dynamical behaviour in a vFFA are determined by numerical modelling.
The $4{\times}4$ transfer matrix is obtained from particle tracking.
The eigenvalues and eigenvectors of the matrix elements are used to analyse the 
coupled motion.
Further particle tracking is used to explore the stable phase space volume.

Our future plan at the ISIS Neutron and Muon Source Facility aims for a proton 
driver of 1.2\,GeV kinetic energy to provide 1.25\,MW beam power with a ring 
of about 150\,m circumference~\cite{Thomason}.
With vFFA optics, a reasonable choice for the ring would have a total of 25 cells, each of 6\,m length.
To support the design work, construction of  a 25\,m circumference vFFA with 10 
focusing cells is being considered as a scaled-down, prototype ring. 
It is planned that this ring will use 
the ISIS Front End Test Stand (FETS) as an injector~\cite{Letchford} and has therefore been provisionally accorded the name FETS-FFA.

The purpose of this paper is to discuss the procedure used in the design of the prototype vFFA.
However the methodology can be applied to any similar vFFA, including the applications discussed above. 

\section{Magnetic Field Model and Lattice Geometry}
 \label{sec:magnets}

\subsection{Magnet Field Model}
In a vFFA, the scaling condition requires magnetic fields that increase 
exponentially in the vertical direction.
In the following, the $x$-axis is horizontal, the $y$-axis is vertical and the $z$-axis is in the longitudinal direction.
We set the (arbitrary) reference position $y_0=0$ in Eq.~(\ref{eq:eq2}).
The mid-plane of a vFFA rectangular magnet is defined where the horizontal coordinate is zero.
The horizontal field component becomes zero when the current source of the magnetic field has mirror symmetry about the mid-plane.
All three components of the magnetic field are described by a polynomial expansion to order $N$ in $x$
and the vertical field variation on the mid-plane is described by a function of $g(z)$:
\begin{equation}
\label{eq:eq3}
\begin{aligned}
B_{x}\left(x,y,z \right)&=B_0\exp(my)\sum_{i=0}^Nb_{xi}\left(z\right)x^i,\\
B_{y}\left(x,y,z \right)&=B_0\exp(my)\sum_{i=0}^Nb_{yi}\left(z\right)x^i,\\
B_{z}\left(x,y,z \right)&=B_0\exp(my)\sum_{i=0}^Nb_{zi}\left(z\right)x^i.
\end{aligned}
\end{equation}
Maxwell's laws can be used to derive the recursive relations between the coefficients
\begin{align*}
b_{x0}\left(z \right)&=0, \\
b_{y0}\left(z \right)&=g(z), \\
b_{z0}\left(z \right)&=\frac{1}{m}\frac{d g}{d z},
\end{align*}
and
\begin{align*}
b_{x,i+1}\left(z \right)&=-\frac{1}{i+1}\left( m b_{yi}+\frac{d b_{zi}}{d z} \right), \\
b_{y,i+2}\left(z \right)&=\frac{m}{i+2}b_{x,i+1}, \\
b_{z,i+2}\left(z \right)&=\frac{1}{i+2}\frac{d b_{x,i+1}}{d z} .
\end{align*}
 
When a magnet is sufficiently long, 
in the middle of a magnet in the longitudinal direction where $g(z)$ can be taken as constant and all the derivatives of $g(z)$ taken as zero, 
Eq.~(\ref{eq:eq3}) reduces to the standard multipole series expansion of 
$B_x = -B_0 \exp \left( my \right) \sin \left(mx\right)$, 
$B_y = B_0 \exp \left(my\right) \cos \left(mx\right)$, 
$B_z = 0$. 
It is worth noting that, when it is expanded again as a series, normal multipoles and skew multipoles alternate in turn, 
e.g. the first term is a normal dipole, the second term is a skew quadrupole, the third term is a normal sextupole, the fourth term is a skew octupole as Eq.~(\ref{eq:eq4}) shows. 
In practice, it suggests that the magnetic fields of Eq. (\ref{eq:eq3}) may be fabricated with a combination of normal and skew multipoles up to some orders.
\begin{eqnarray}
\label{eq:eq4}
-\frac{B_{x}}{B_0} &=&  \exp\left( my \right)\sin\left( mx \right) \nonumber \\
&=& 0+mx+\frac{m^2}{2!}\left(2xy\right)+\frac{m^3}{3!}\left(-x^3+3xy^2\right) \nonumber \\
&+& \frac{m^4}{4!}\left( -4x^3y + 4xy^3 \right) + ..., \nonumber \\
\frac{B_{y}}{B_0} &=& \exp\left( my \right)\cos\left( mx \right) \\
&=& 1+my+\frac{m^2}{2!}\left(-x^2+y^2\right)+\frac{m^3}{3!}\left(-3x^2y+y^3\right)  \nonumber \\
&+& \frac{m^4}{4!}\left( x^4-6x^2y^2+y^4 \right) + ... \nonumber.
\end{eqnarray}

\begin{figure}[h]
\centering
\subfloat[$x=0$\,m]{
\includegraphics[width=.45\columnwidth]
{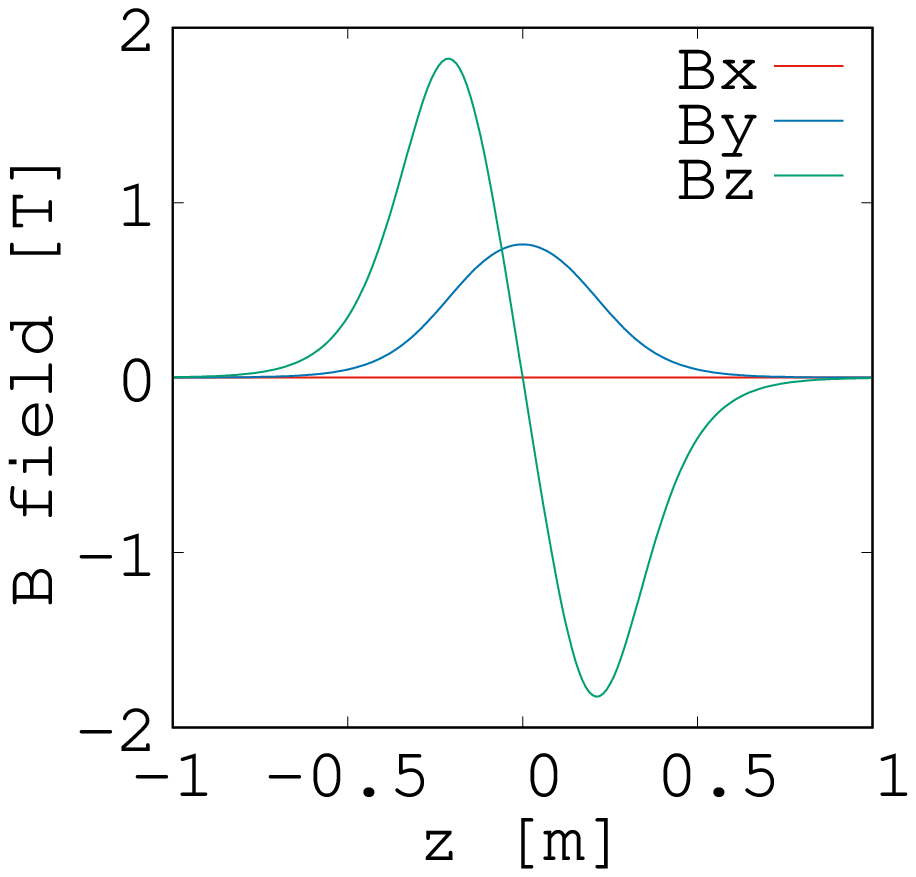}
}
\subfloat[$x=0.05$\,m]{
\includegraphics[width=.45\columnwidth]
{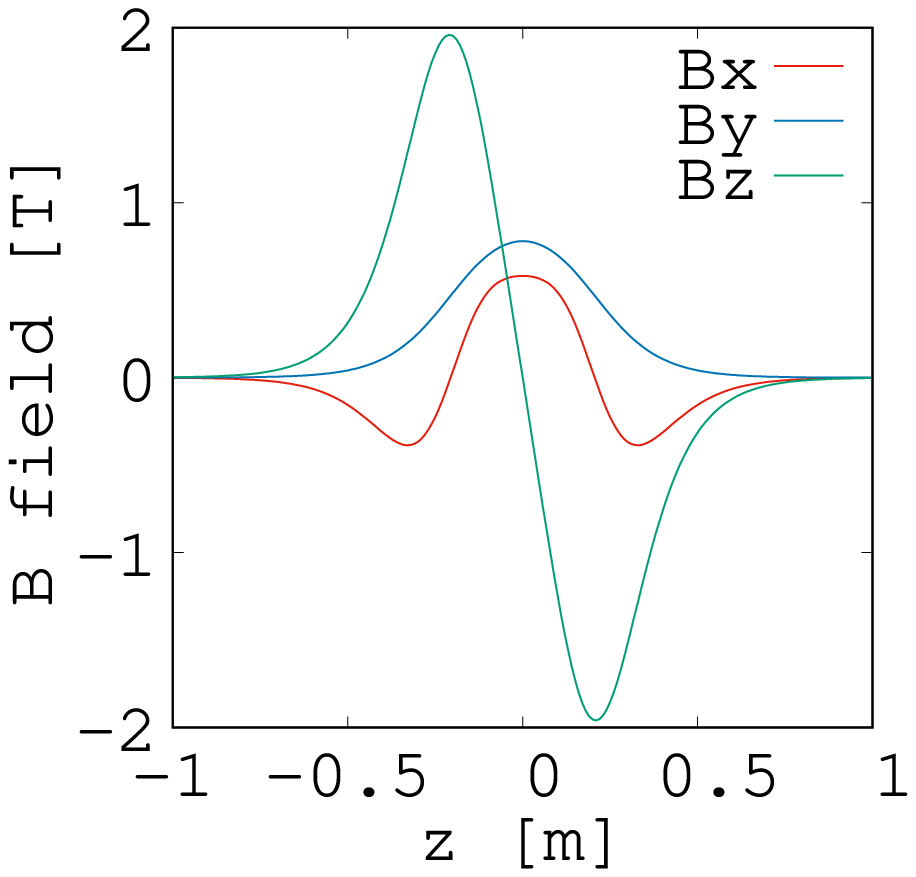}
}
\hspace{0mm}
\subfloat[$z=0$\,m]{
\includegraphics[width=.45\columnwidth]
{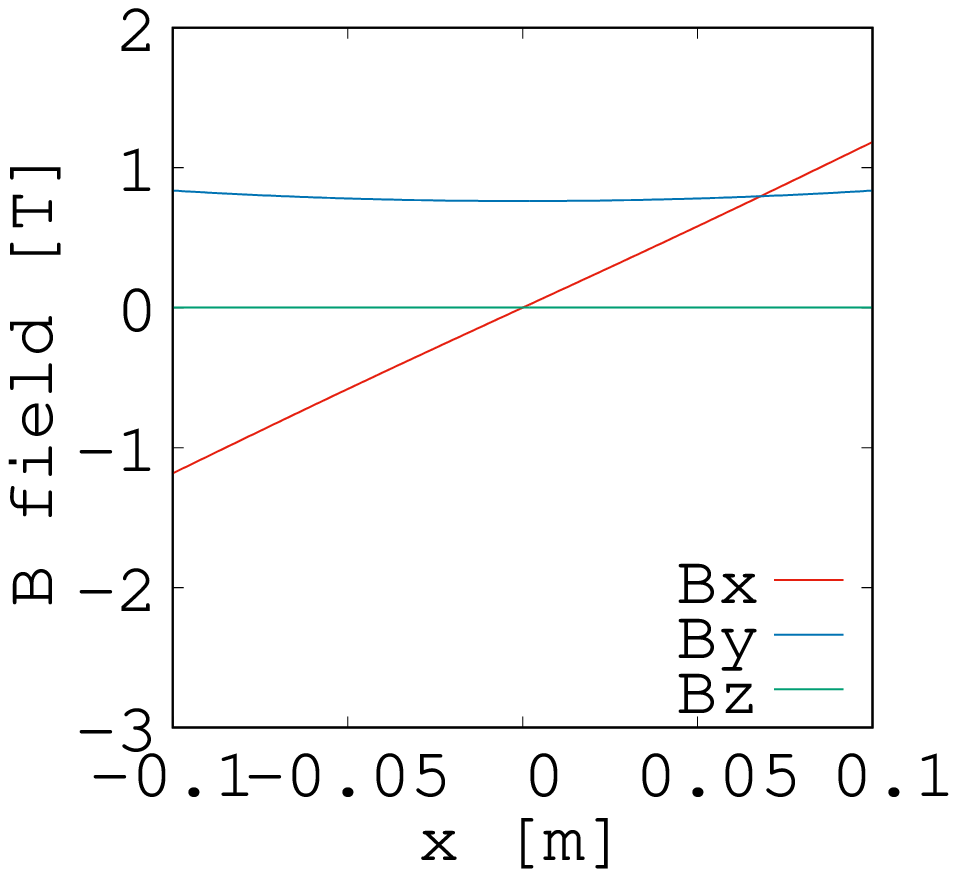}
}
\subfloat[$z=0.20$\,m]{
\includegraphics[width=.45\columnwidth]
{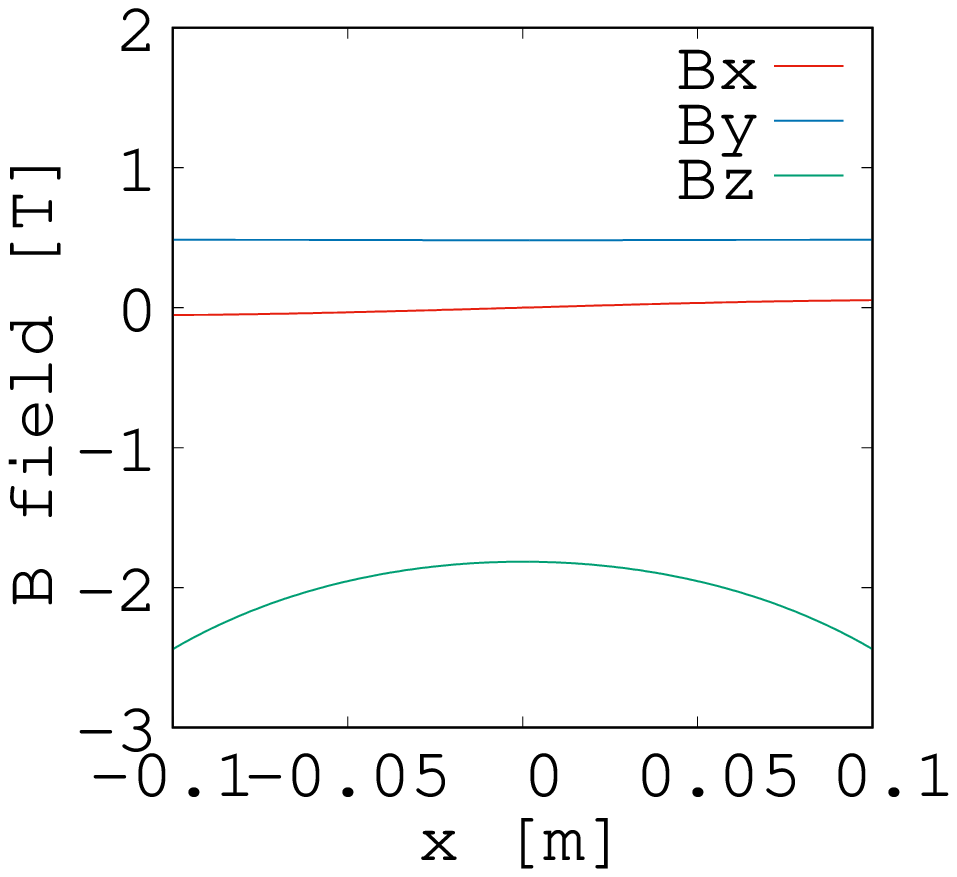}
}
\caption{\label{fig:fig1} $B_x$, $B_y$ and $B_z$ fields along the longitudinal axis 
((a) and (b)) and the horizontal axis ((c) and (d)). The centre of the 
magnet is at $x=0,z=0$. In all cases, $y=0$. The field profiles in (b) are shown off-midplane 
to demonstrate a non-zero horizontal field $B_x$. The field profile in (d) is shown in the magnet 
end region to demonstrate a non-zero longitudinal field $B_z$.
}
\end{figure}

Fringe fields are an important feature of FFA optics, and in this study we chose to model the magnetic field with a hyperbolic tangent function,
\[g(z) = \frac{1}{2}\left[ \tanh\left( \frac{z+M/2}{L} \right)-\tanh\left( \frac{z-M/2}{L} \right) \right]\]
where $M$ is the magnet length and $L$ is a parameter related to the fringe field extent. The field is symmetric with $g(z)=g(-z)$. More 
complicated field profiles will be considered as the hardware design matures.
To illustrate the field profiles, the 3D magnetic fields are shown in 
Fig.~\ref{fig:fig1} where the parameters of Table~\ref{tab:tab1} are used.
The order of polynomial $N$ is determined numerically by 
examining the convergence properties of the field and, where tracking is performed, the tune.
\begin{table}[h]
\caption{Dimension and strength of the magnet model shown in Fig.~\ref{fig:fig1}.}
\label{tab:tab1}
\begin{ruledtabular}
\begin{tabular}{lcdr}
Parameter & Value\\
\colrule
Magnet length ($M$) & 0.4\,m\\
Fringe field extent ($L$) & 0.2\,m\\
$B_0$ & 1\,T\\
Normalised field gradient ($m$) & 1.28\,m$^{-1}$\\
Order of polynomials $N$ in Eq.~(\ref{eq:eq3}) & 10\\
\end{tabular}
\end{ruledtabular}
\end{table}

In practice the vFFA magnet can be a superposition of many field maps described 
by Eq.~(\ref{eq:eq3}).
As long as the field strength changes according to $\exp(my)$ in all 
directions, the combined fields satisfy the scaling conditions.
For example in section~\ref{sec:optimisation} two such fields are 
superimposed with coordinates shifted in the horizontal direction by $+x_c$ and 
$-x_c$ respectively while maintaining the scaling conditions.

\subsection{Number of Cells}
Unlike ordinary accelerators where nonlinear magnetic fields exist either through 
fabrication errors or for correction of chromaticity around the central momentum,
FFAs have strong nonlinearity in the magnetic fields in order to correct
the chromaticity for the whole momentum range.
Systematic nonlinear resonances should be avoided as much as possible.
In addition, one of the eventual goals of our study is to accelerate high 
intensity beams in the vFFA where nonlinearity of the space charge potential is 
a concern, especially strong space charge driven 4$^{th}$ order resonances~\cite{sc4th}.

To maximise the resonance-free space in the transverse tune diagram,
the total number of cells is chosen as a multiple of 5.
With this number of cells,
systematic 5$^{th}$ order resonances always coincide with integer resonances.
It may be possible to find resonance-free regions of tune space 0.25 in cell tune away from 
5$^{th}$ order magnetic and 4$^{th}$ order space charge resonances.

A superperiod structure can also be devised in a FFA ring by introducing several families of magnets.
However, in our design, the ring comprises a simple repetition of a unit cell.
We design the prototype 12 \,MeV ring with 10~cells.

\subsection{Geometric Layout}
FODO and FDF triplet lattices are considered in this paper. 
The lattice parameters are 
shown in Table~\ref{tab:tab2}. The geometry of the rings is
described by a regular polygon with each side corresponding to a cell in the 
FDF triplet case or half-cell in the FODO case. Figures 2(a) and 2(c) may be 
used for guidance.

\begin{table}[h]
\caption{\label{tab:tab2}%
Geometrical parameters of FODO and FDF triplet focusing lattices. Magnets were
placed symmetrically about the centre of each side of the polygon.}
\begin{ruledtabular}
\begin{tabular}{lccdr}
Parameter & FODO & FDF Triplet \\
\colrule
Bend angle per cell & 36\,deg & 36\,deg\\
Cell length & 1.25\,m (half) & 2.50\,m\\
Bd magnet length & 0.50\,m & 0.24\,m\\
Bf magnet length & 0.50\,m & 0.40\,m\\
Space between Bd and Bf & 0.75\,m & 0.08\,m\\
Bd horizontal wedge angle & -20\,deg & 0\,deg\\
Bf horizontal wedge angle & +46\,deg & 0\,deg\\
Double coil shift ($x_c$) & $\pm$0.02\,m & $\pm$0.02\,m\\
Fringe field extent ($L$) & 0.15\,m & 0.20\,m\\
Relative displacement ($x_s$) & $\pm$0.17\,m & 0\,m\\
Tilt angle ($t_f$) & 0\,deg & 0\,deg\\
\end{tabular}
\end{ruledtabular}
\end{table}

In the FODO lattice, a combination of a magnet with 
normal bending angle (Bf) and a magnet with reverse 
bend (Bd) is placed centrally on each side 
of the polygon. 
The number of polygon sides is twice the number of cells.
Bd is displaced from the polygon edge by a distance $x_s$ radially towards the
ring centre while Bf is displaced by $x_s$ radially away from the ring centre 
to follow the scalloped orbit of the beam.

The net bending angle per cell in the FODO is 36\,degrees.
As Bd bends the beams away from the ring centre,
Bf must bend the beam by more than 36\,degrees.
To keep the beam round the gap centre of each magnet,
a sector type magnet with curvature following the beam orbit is preferable.
The sector-type magnet is modelled by splitting the Bf and Bd magnets into seven
segments. 
The cell tune variation is less than $10^{-3}$ when seven or more segments are used, which was considered an acceptable precision for the model.
Each segment is a rectangular type magnet but with finite angle 
between neighbouring segments. The fields of all the rectangular magnets are 
superimposed in the modelling.

In a FDF triplet lattice,
Bf, Bd and Bf magnets are positioned symmetrically about the centre of each 
side of the polygon.
All three magnets have rectangular shape.
A sector magnet has not been considered to keep the magnet fabrication simple.
As the bending angle per magnet is less than in the FODO lattice, the beam
does not deviate so strongly from the magnet midplane.
Lattice optimisation with the displacement parameter $\pm x_s$, the tilt 
angle $t_f$ of Bf with respect to Bd, magnet length and fringe field extent is 
discussed in section~\ref{sec:optimisation} and the Appendix.

\section{Closed Orbit and Stable Optics}

\subsection{Finding a Closed Orbit}
Given the lattice geometry, the normalised gradient $m$ and field strengths 
$B_{0d}$ and $B_{0f}$ from Eq.~(\ref{eq:eq3}) of the Bd and Bf magnets 
respectively, the periodic orbit of a single cell (closed orbit) is numerically 
found by an algorithm minimising
the difference between 4D transverse phase space coordinates at the entrance 
and the exit of the cell.
The entrance and the exit are chosen at the centres of the long straight 
sections, where the field leakage from the magnets is at a minimum but not 
necessarily zero.
The magnetic fields from the magnet in the cell and from magnets in the 
upstream  and downstream cells are superimposed to account for this field leakage.

\begin{figure}[h]
\centering
\subfloat[FODO: orbit]{
\includegraphics[width=.45\columnwidth]
{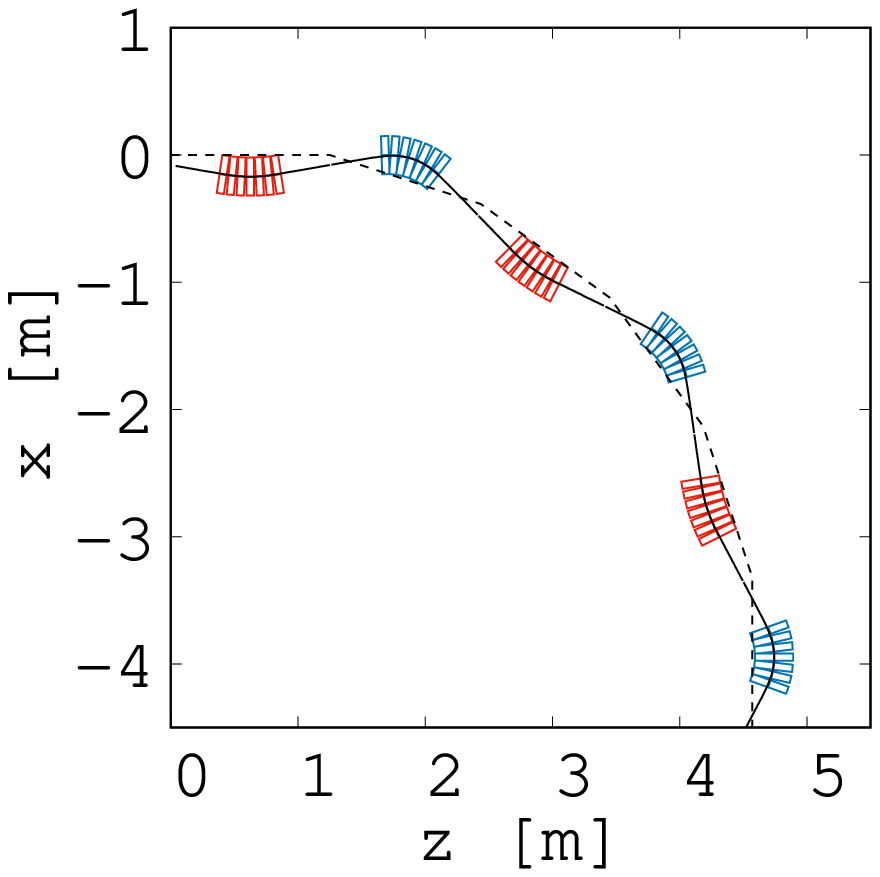}
}
\subfloat[FODO: fields]{
\includegraphics[width=.45\columnwidth]
{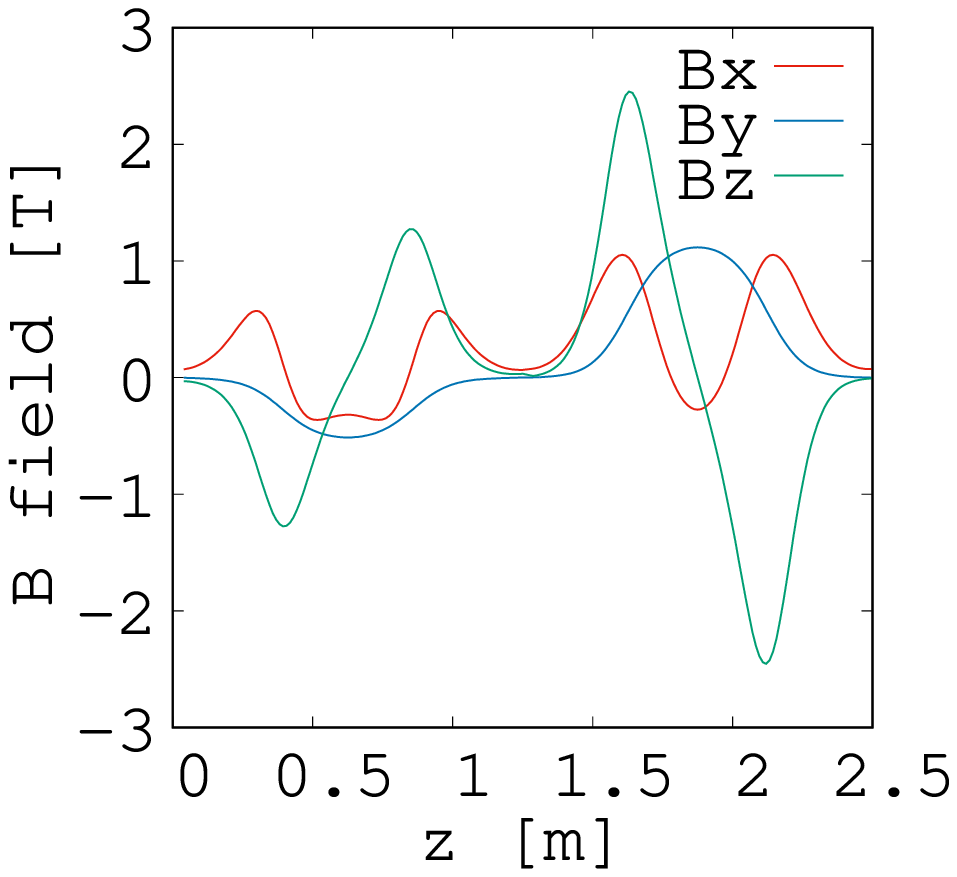}
}
\hspace{0mm}
\subfloat[FDF triplet: orbit]{
\includegraphics[width=.45\columnwidth]
{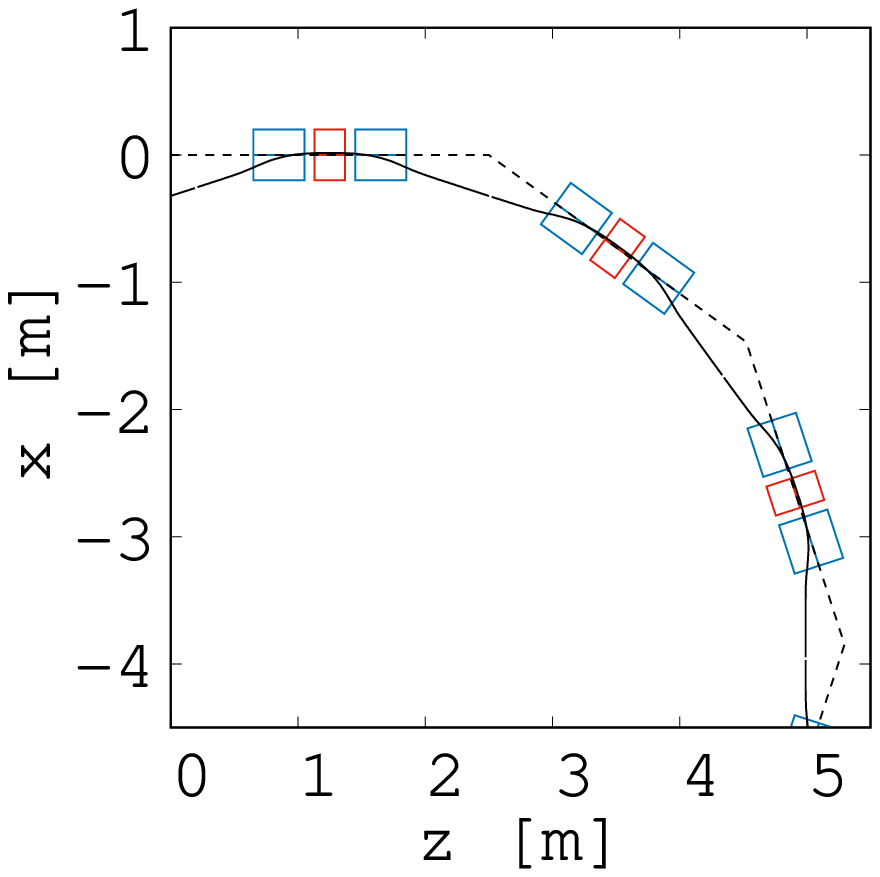}
}
\subfloat[FDF triplet: fields]{
\includegraphics[width=.45\columnwidth]
{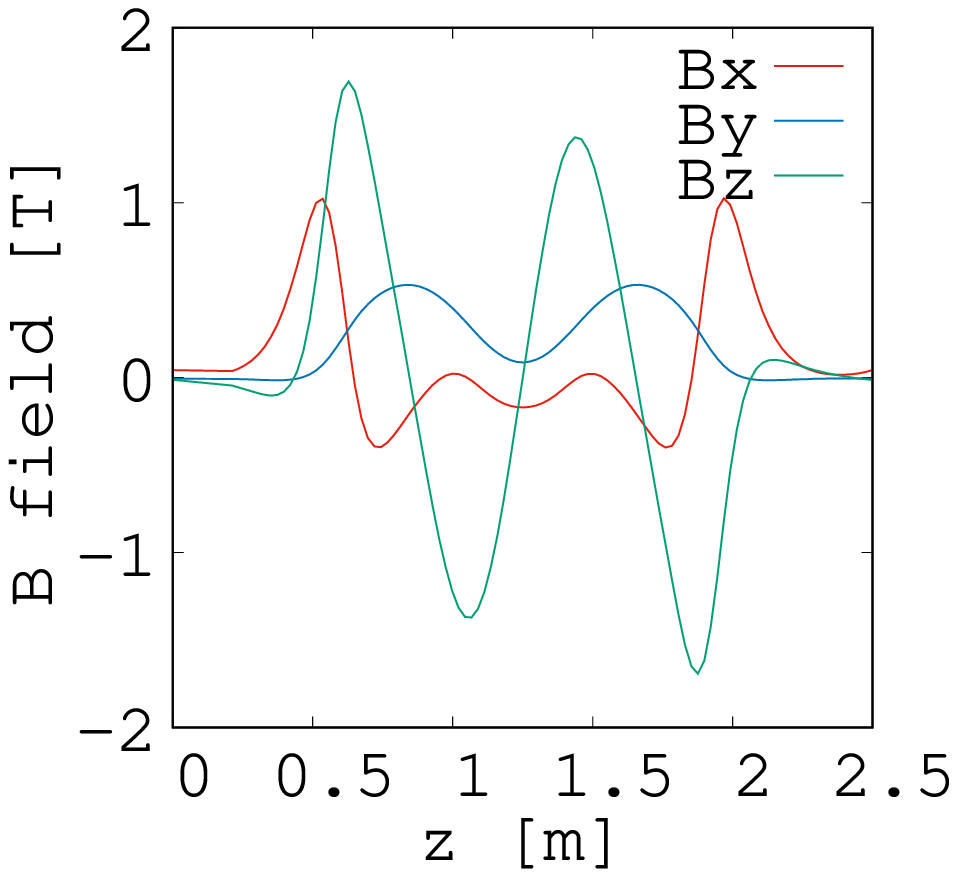}
}
\caption{\label{fig:fig2} Closed orbits for FODO (a) and FDF triplet lattices 
(c). 
The dashed line indicates the polygon with 20 sides for the FODO lattice and 
with 10 sides for the FDF triplet. 
Bd and Bf magnets are marked with by red and blue boxes respectively. 
In the FODO lattice, seven rectangular magnets mimic a sector magnet.
Bd is shifted inward with respect to one side of polygon by $x_s$ and Bf is 
shifted outward by $x_s$.
The magnetic fields along the orbit in a cell for 12\,MeV beam energy are shown 
for FODO (b) and FDF triplet (d).}
\end{figure}

The closed orbits for the FODO lattice and FDF triplet are shown in Fig.~\ref{fig:fig2}.
Magnetic fields along the orbit are also shown where $B_x$, $B_y$ and $B_z$ are defined with respect to the coordinate parallel to a side of the polygon.

\subsection{Orbits with Different Momenta}
\begin{figure}[h]
\centering
\includegraphics[width=.8\columnwidth]
{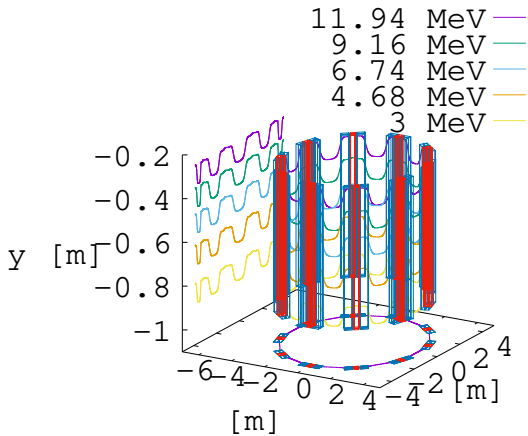}
\caption{\label{fig:topandside} Closed orbits for different momenta in the FDF 
triplet lattice. The locations of the Bd and Bf magnets are indicated by red 
and blue boxes respectively.  The ratio of magnet strength 
$B_{0d}/B_{0f}$=-0.20 and the value of the normalised field gradient $m$ is 
fixed at 1.28 m$^{-1}$. Path length is constant for the entire momentum range, in other 
words, the momentum compaction factor is zero.}
\end{figure}

Because of the finite horizontal field in a vFFA, particles travelling at a fixed energy do not stay in a horizontal plane but experience an oscillation in the vertical direction as the beam goes round the ring.
Orbits for different energies are displaced vertically, for example as the beam is accelerated.
Projections of the closed orbits for different particle momenta, shown in Fig.~\ref{fig:topandside}, reveal interesting features.
Since the magnetic field increases exponentially, the vertical projection shows that
the orbit separation with a constant increase of momentum shrinks logarithmically.
Horizontal projections of the orbits are actually identical, despite the vertical shift, and this confirms the earlier observation that the momentum 
compaction factor is zero over the entire momentum range.

\subsection{Finding Stable Optics}
Once the closed orbit is found, the next step is to determine whether the 
motion around it is stable by calculating the transverse eigentunes.
Particles are tracked through the cell with an offset in each transverse 
coordinate in order to determine the 4$\times$4 transverse transfer matrix $T$.
This transfer matrix clearly shows non-zero  
off-diagonal matrix elements (e.g. $T_{13}$ and $T_{14}$) because of the skew 
quadrupole component from the body field and solenoid fields at the magnet ends.
The former comes from the exponentially increasing field in 
Eq.~(\ref{eq:eq3}).

\begin{figure}[h]
\centering
\includegraphics[width=.7\columnwidth]
{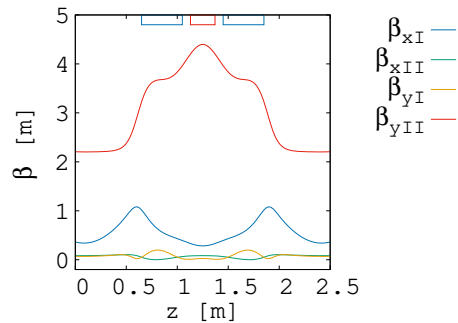}
\caption{\label{fig:fig4} Beta functions $\beta_{xI}$, $\beta_{xII}$, 
$\beta_{yI}$, $\beta_{yII}$ according to the procedure by 
Willeke-Ripken~\cite{ripken}.  
The magnet positioning is shown at the top, the red box corresponding to the Bd magnet and the blue box to the Bf magnet. }
\end{figure}

Two transverse tunes, denoted by $q_u$ and $q_v$, are obtained as arguments of 
the conjugate pairs of complex eigenvalues of the 4$\times$4 transfer matrix 
$T$,
The amplitude functions (the $\beta$-functions) of the FDF triplet lattice are 
calculated from the eigenvectors according to the Willeke-Ripken 
formalism~\cite{ripken} and are shown in Fig.~\ref{fig:fig4}.
In the following discussion, we will focus mainly on the FDF triplet lattice design.

\section{Parameter search}
\label{sec:optimisation} 

\subsection{Orbit Control}
In conventional accelerators, local orbit correction is provided by dipole 
magnets that are excited individually depending on the distortion along the 
beam trajectory.
The orbit is corrected in the horizontal direction by a dipole magnet with a 
vertical field and in the vertical direction by a dipole with a horizontal field.
In a vFFA, a vertical orbit corrector could be designed in the same way, but the horizontal orbit corrector may require a dipole magnet with a large gap, probably more than half a metre,
spanning the orbit excursion in the vertical direction, and this would be challenging to implement. 

\begin{figure}[h]
\centering
\subfloat[orbit shift in FDF triplet]{%
\includegraphics[width=.45\columnwidth]
{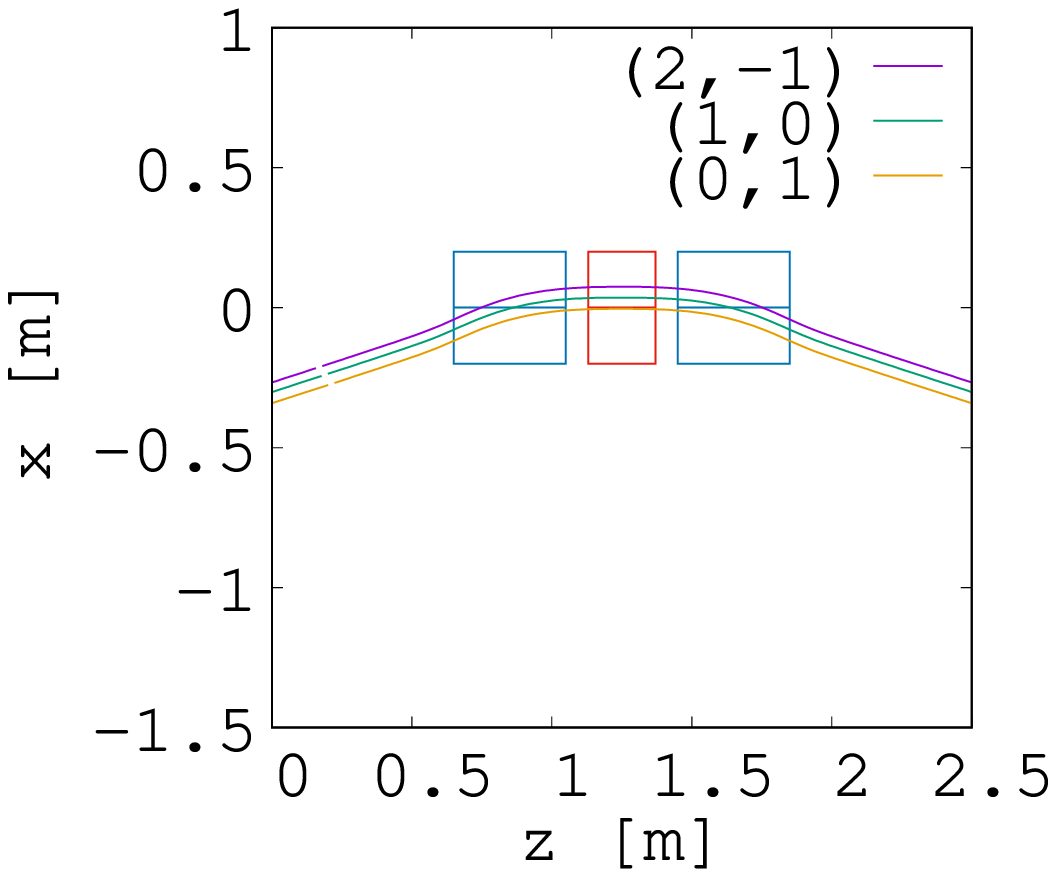}}
\subfloat[tune shift in FDF triplet]{%
\includegraphics[width=.45\columnwidth]
{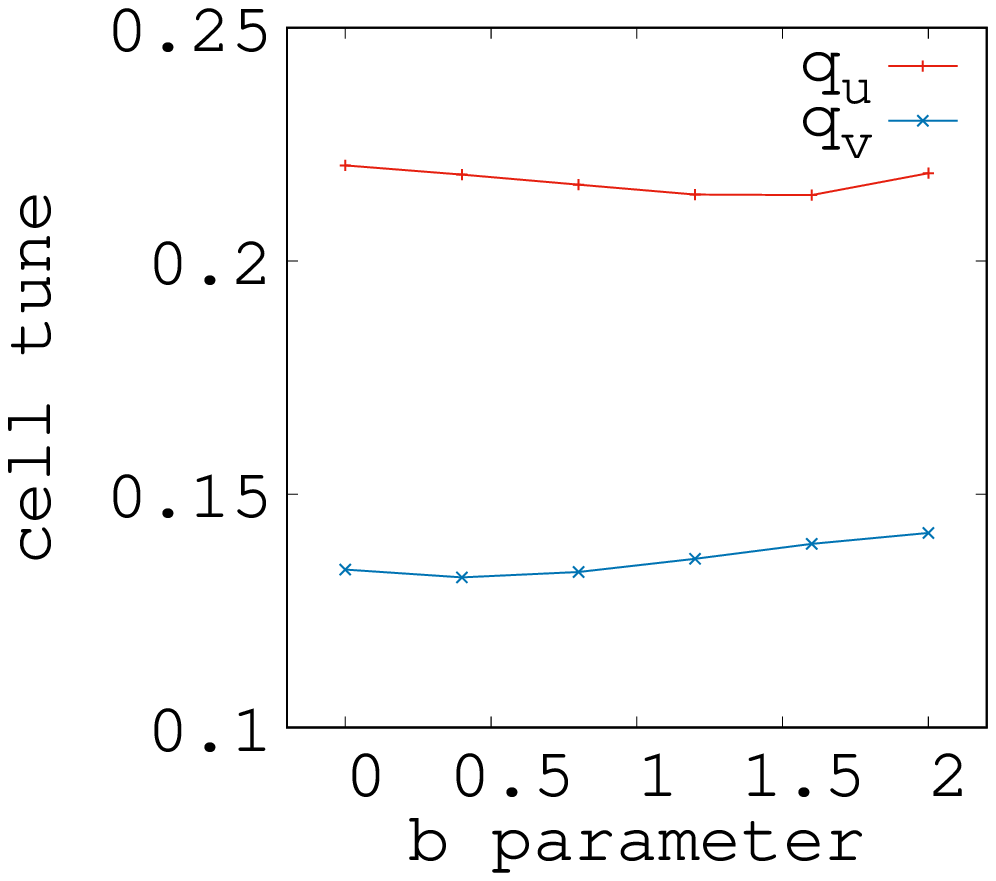}}
\hspace{0mm}
\subfloat[orbit shift at Bd in FODO]{%
\includegraphics[width=.45\columnwidth]
{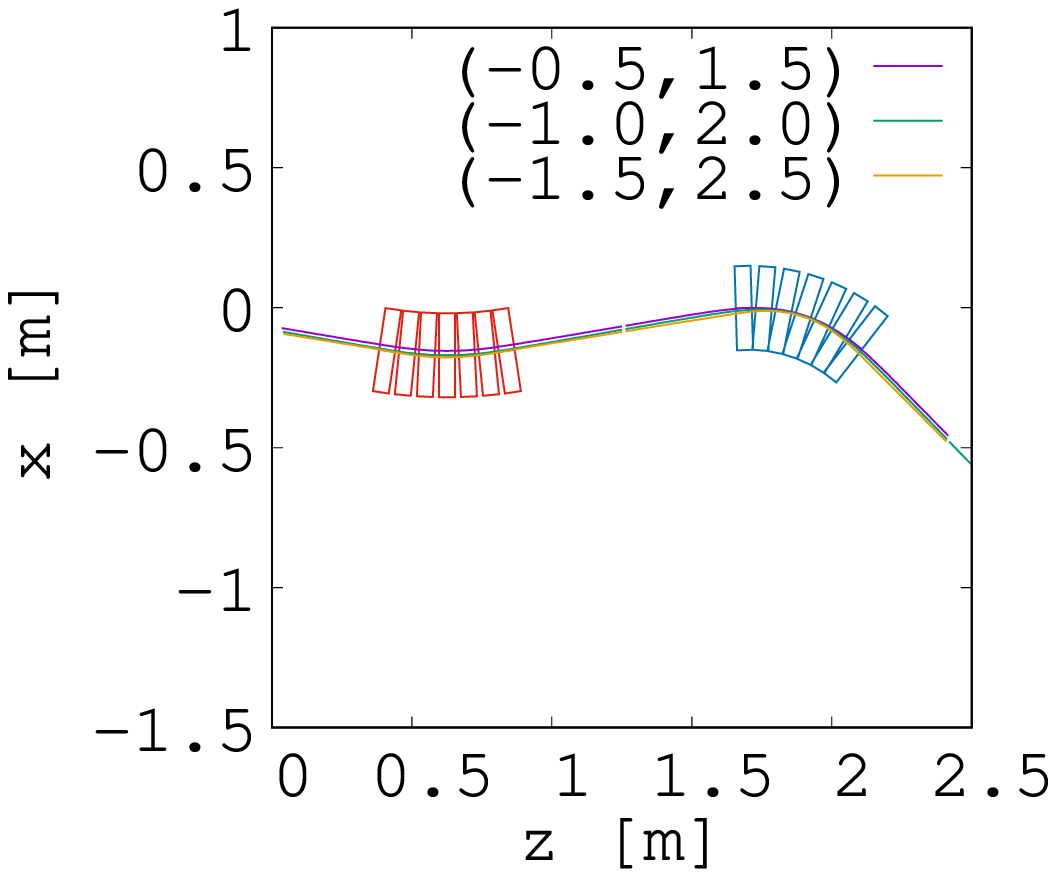}}
\subfloat[orbit shift at Bf in FODO]{%
\includegraphics[width=.45\columnwidth]
{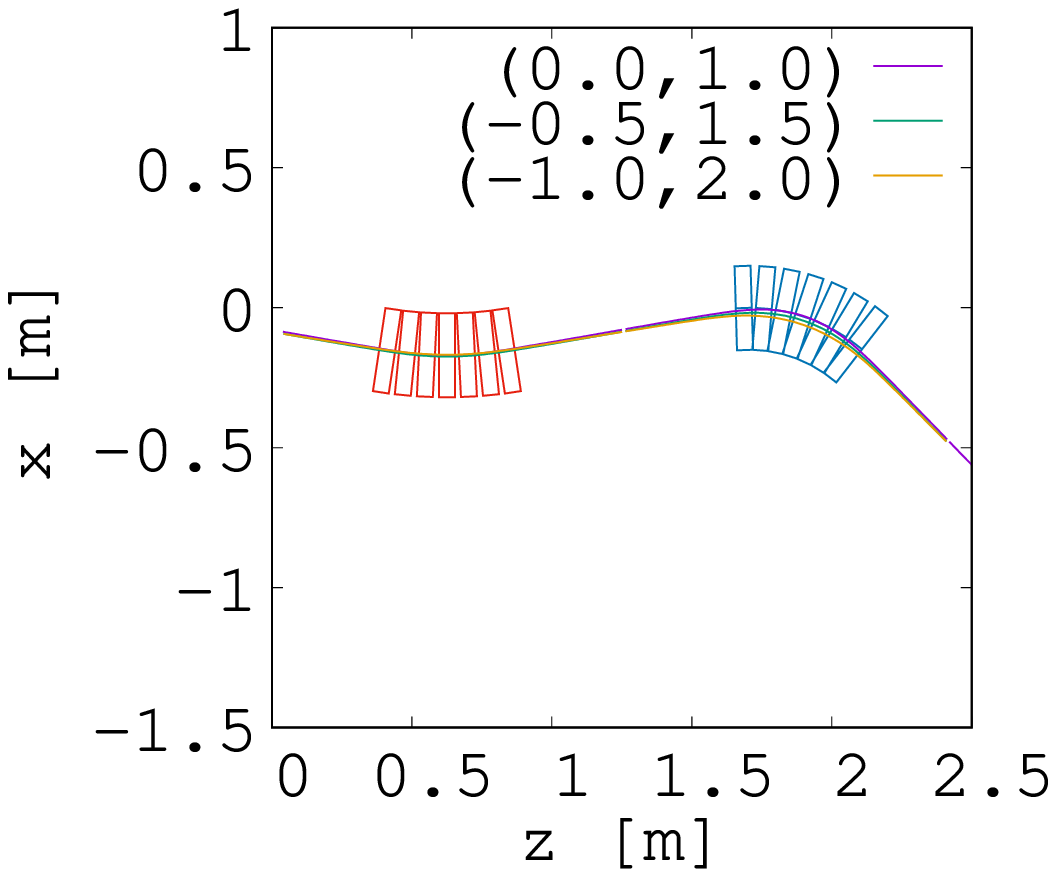}}
\caption{\label{fig:fig5} Orbit shift controlled by a double coil design. 
A bracket $(b,1-b)$ indicates the relative excitation of outer coil $b$ to 
inner coil $1-b$. The total excitation is kept at unity.
By having different excitation patterns of outer and inner coils, the orbit position can be controlled with little change of tune.}
\end{figure}

The double coil design mentioned above enables horizontal orbit correction.
Two pairs of coils, shifted horizontally with respect to their 
midplanes, can be excited with different currents. Tuning the currents can
enable tuning of the profile of the vertical field along the horizontal 
direction while maintaining the scaling condition.
The effect on cell tune is small as we expect from the usual method of orbit 
correction using dipole magnets. 
In Fig.~\ref{fig:fig5}, horizontal orbits are shown with
different coil excitations together with the corresponding variation in tune. 
The orbit moves by several centimetres while the variation in cell tune is less than 0.01.

\subsection{Optics Optimisation}
In an hFFA, two tuning parameters are used to change the orbit and optics: the 
field index $k$, defined in Eq.~(\ref{eq:eq1}), and the ratio of the field 
strengths of the normal (Bf) and reverse (Bd) bending magnets. The field index 
$k$ defines the focusing and defocusing strength of the magnet body. The field 
ratio controls the ratio of normal and reverse bending fields, which
enables manipulation of the orbit scalloping and hence focusing and defocusing 
action at the edge of the magnets.

\begin{figure}[htb]
\centering
\subfloat[$m$ and $B_{0d}/B_{0f}$ space.]{
\includegraphics[width=.9\columnwidth]
{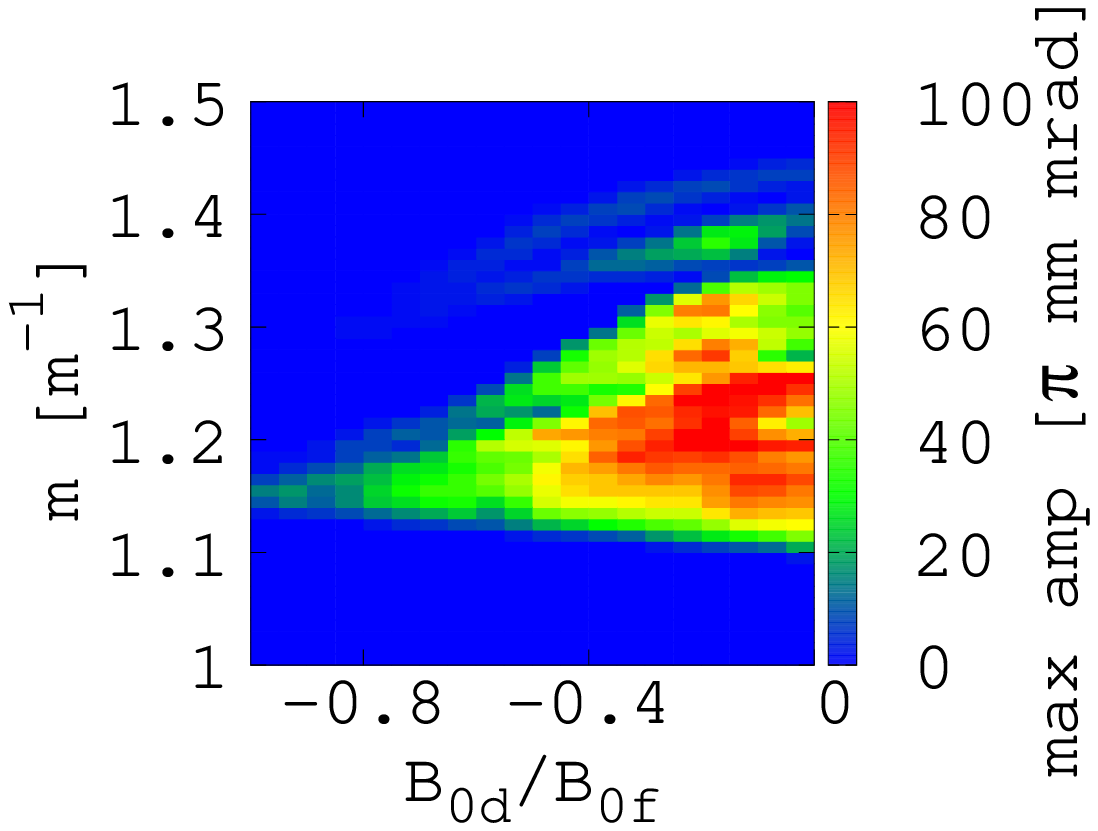}
}\\[1ex]
\subfloat[$q_u$ and $q_v$ space.]{
\includegraphics[width=.9\columnwidth]
{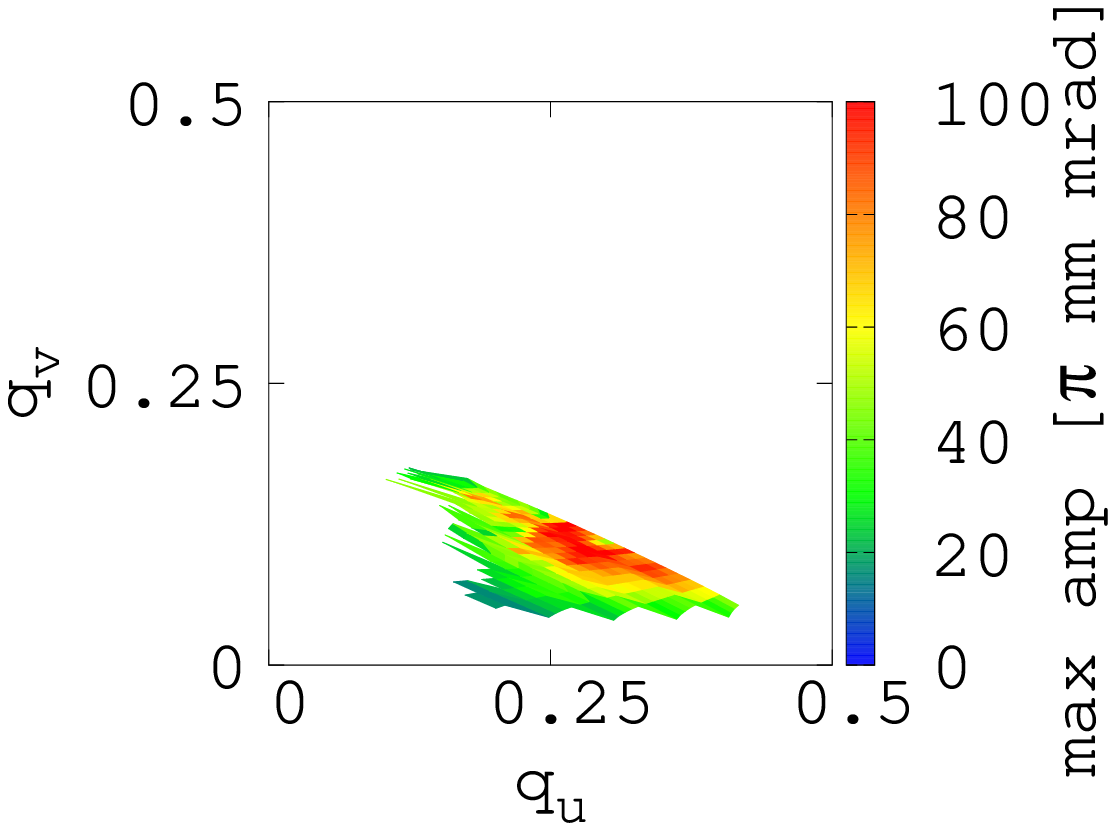}
}
\caption{\label{fig:fig6} Stable area in $m$ and $B_{0d}/B_{0f}$ space and $q_u$ and $q_v$ space. 
The colour code indicates the maximum amplitude of betatron oscillation 
$A=(i_\mathrm{max}+j_\mathrm{max})A_0$ where $A_0=5\pi$ mm\,mrad in normalised 
space, $i_\mathrm{max}$ and $j_\mathrm{max}$ are the indices 
corresponding to the largest value of $i$ for $u$ space or $j$ for $v$ space 
for which the particle survives 100 turns.}
\end{figure}

Similarly in a vFFA, the normalised field gradient $m$ and the ratio of the 
field strengths of Bf and Bd are the main parameters used to adjust optical features. 
Figure~\ref{fig:fig6} shows the area in $(m,B_{0d}/B_{0f})$ space that gives 
stable optics and significant dynamic aperture for the FDF triplet lattice.
The dynamic aperture on the corresponding $(q_u,q_v)$ space is also shown,
where $q_u$ and $q_v$ are the tunes in the decoupled space 
$(u,p_u,v,p_v)$~\cite{parzen}.

Dynamic aperture, the region of phase space in which the beam can be 
transported around the ring, is estimated by particle tracking. 
Particles are tracked having initial
values $(u,p_u,v,p_v)=(\sqrt{i u_0},0,0,0)$ and $(u,p_u,v,p_v)=
(0,0,\sqrt{jv_0},0)$.
$u_0$ and $v_0$ are chosen to correspond to a single particle amplitude of 
$A_0=5\,\pi$\,mm\,mrad in normalised space and $i$ and $j$ are positive integers. 
The dynamic aperture is 
estimated as 
$A =(i_\mathrm{max}+j_\mathrm{max})A_0$ where $i_\mathrm{max}$ and 
$j_\mathrm{max}$ correspond to the largest values of $i$ and $j$ 
for which the particle survives 100~turns. A more detailed study of dynamic aperture is 
described in section~\ref{sec:aperture}.

Similar figures for slightly different geometries are shown in Appendix~A. 
The baseline parameters of the FDF triplet lattice, $B_{0d}/B_{0f}=-0.20$ and $m=1.28$ m$^{-1}$,
were chosen for further study, owing to the excellent dynamic aperture.

\subsection{Edge Focusing}

During the process of identifying the triplet configuration and physical 
dimension of the magnets, the edge focusing can be adjusted in simulation by 
moving the magnets.
We have looked at the effect on the area of stable optics coming from two 
variations: one through the tilt angle $t_f$ of Bf magnets with respect to Bd magnet, and
the other from the relative horizontal displacement, $x_s$, between the Bd and 
Bf magnets.

\begin{figure}[h]
\centering
\subfloat[$t_f$=-2\,deg]{%
\includegraphics[width=.45\columnwidth]
{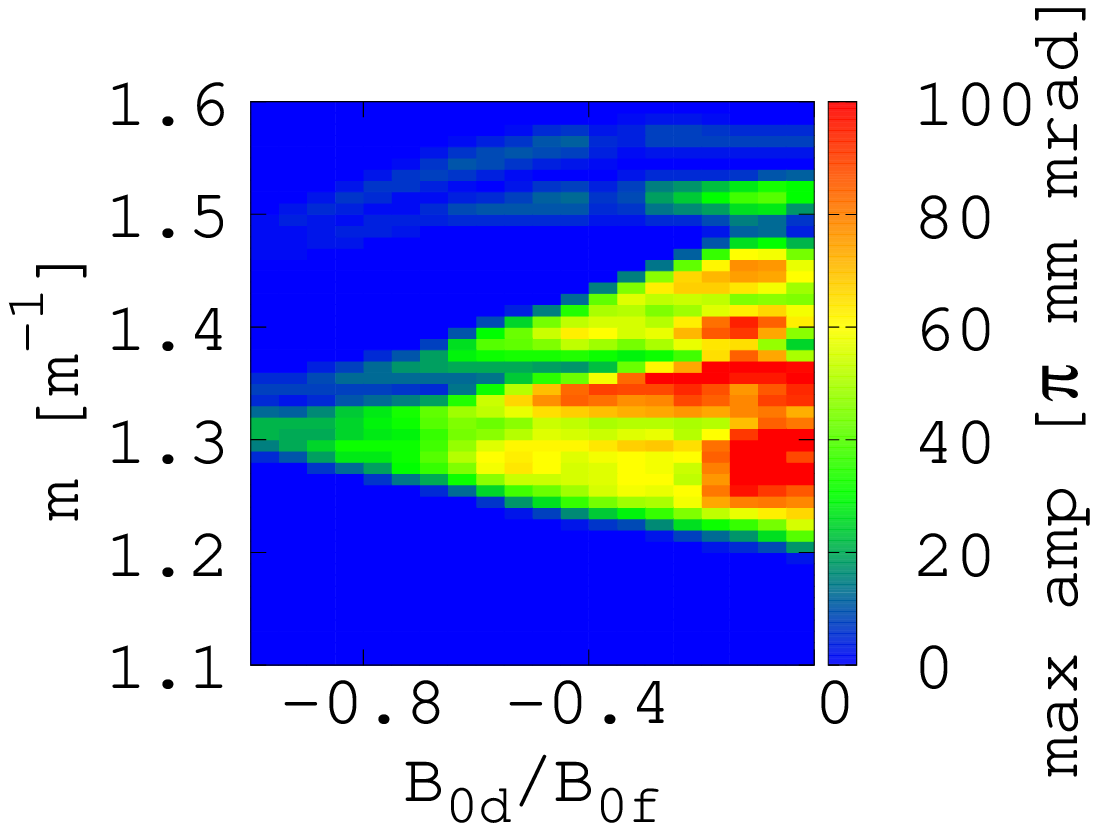}
}
\subfloat[$t_f$=-2\,deg]{%
\includegraphics[width=.45\columnwidth]
{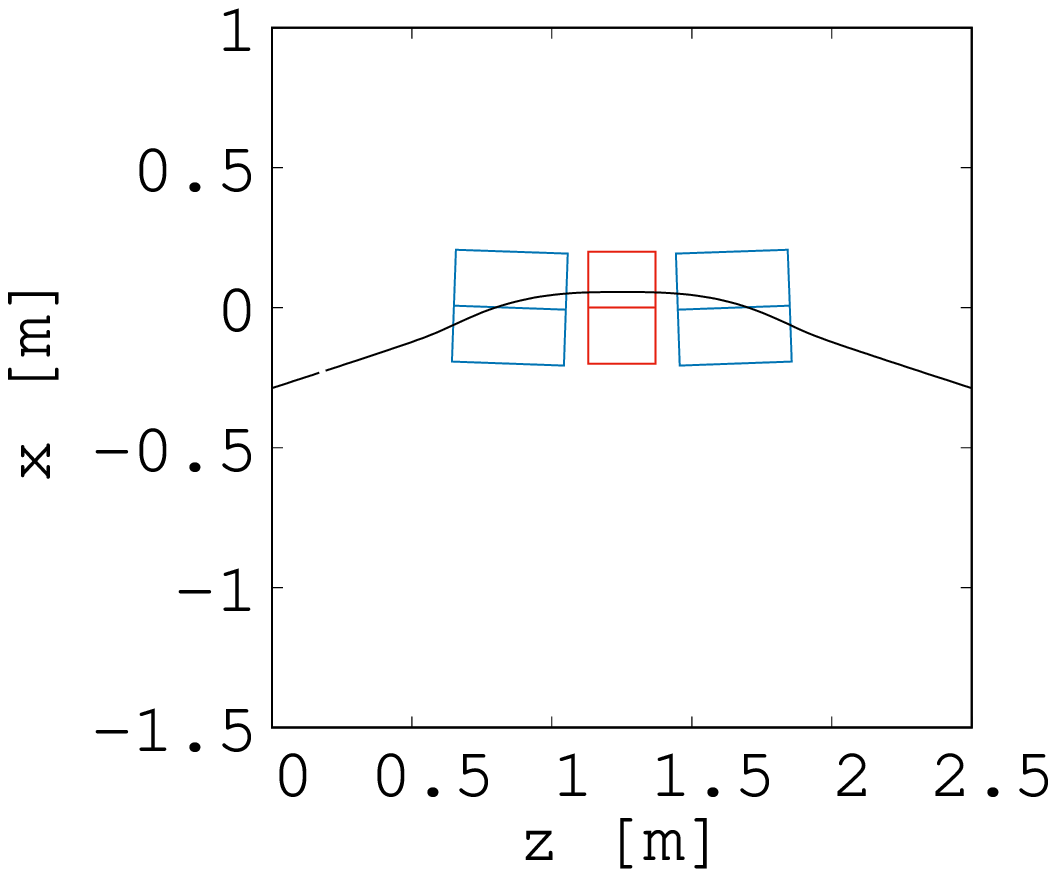}
}
\hspace{0mm}
\subfloat[$t_f$=0\,deg]{%
\includegraphics[width=.45\columnwidth]
{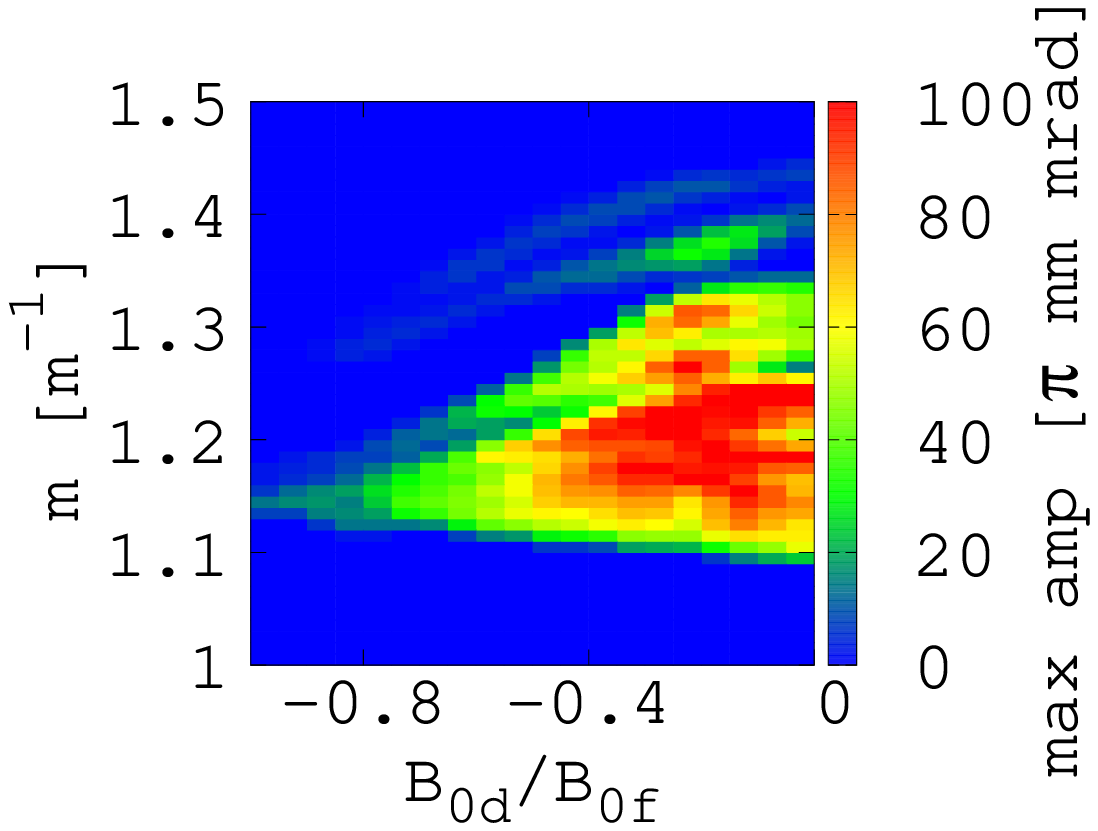}
}
\subfloat[$t_f$=0\,deg]{%
\includegraphics[width=.45\columnwidth]
{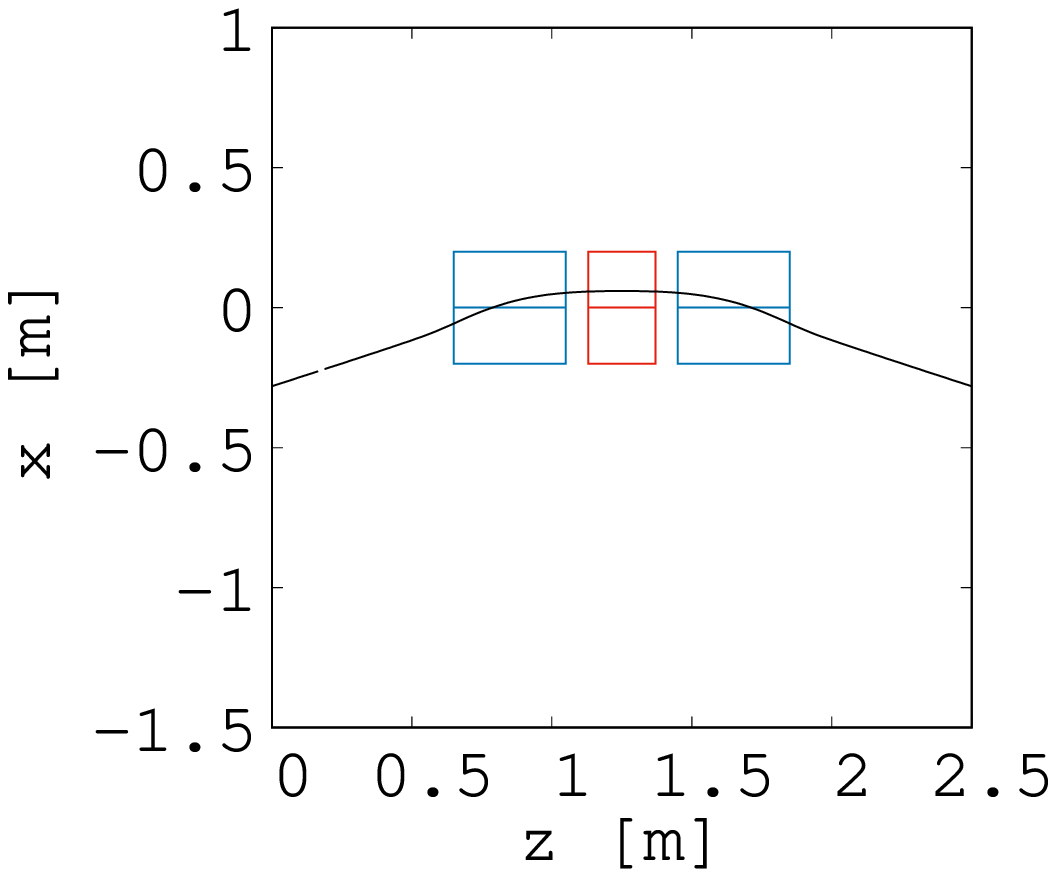}
}
\hspace{0mm}
\subfloat[$t_f$=2\,deg]{%
\includegraphics[width=.45\columnwidth]
{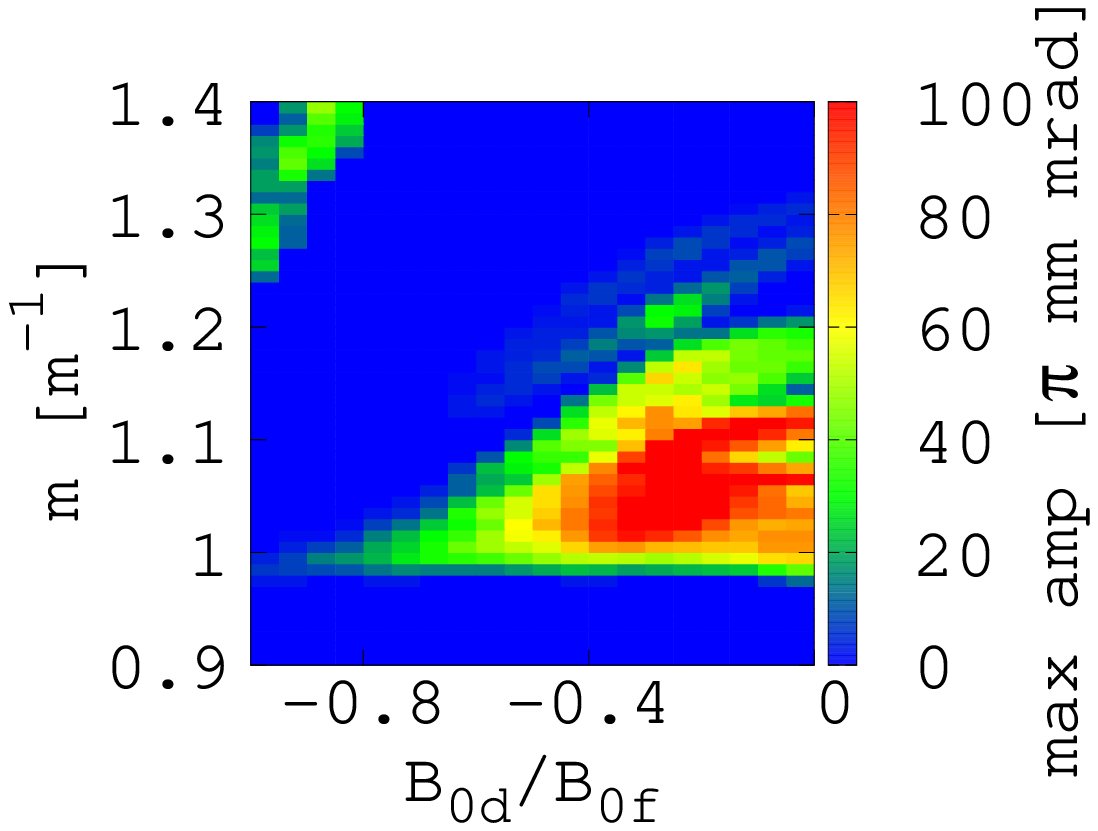}
}
\subfloat[$t_f$=2\,deg]{%
\includegraphics[width=.45\columnwidth]
{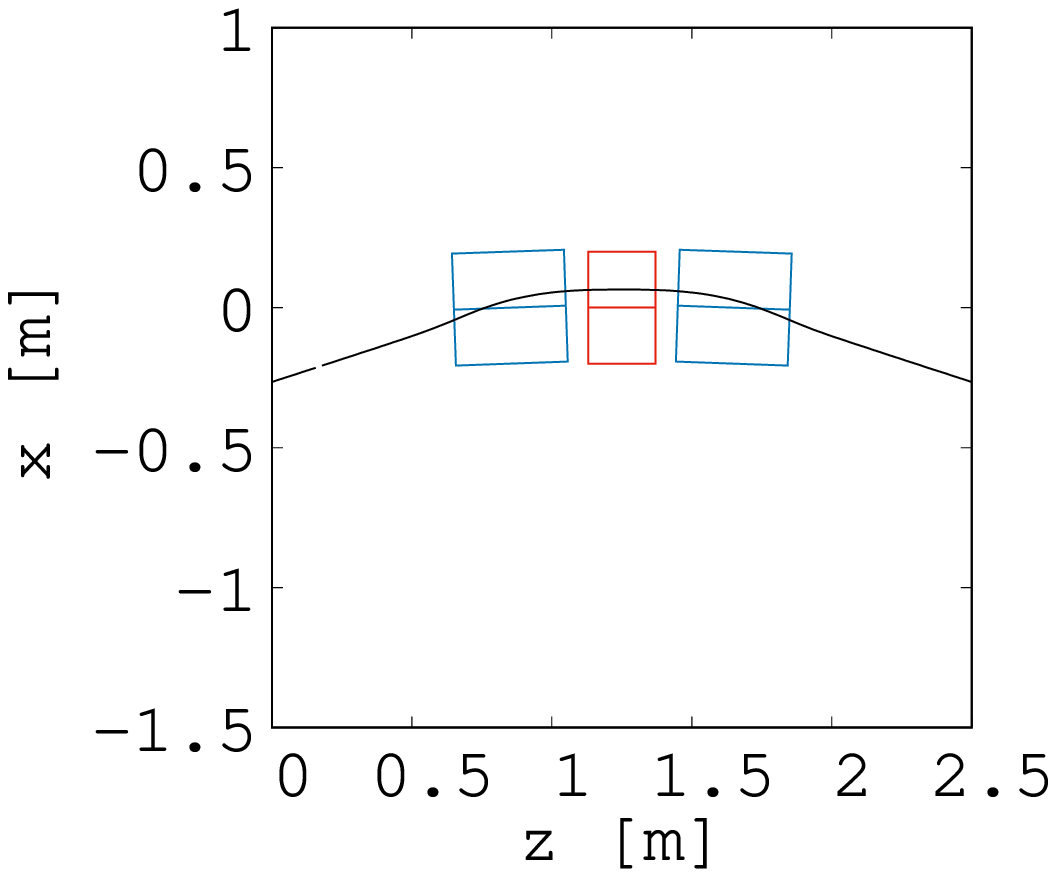}
}
\hspace{0mm}
\subfloat[$t_f$=4\,deg]{%
\includegraphics[width=.45\columnwidth]
{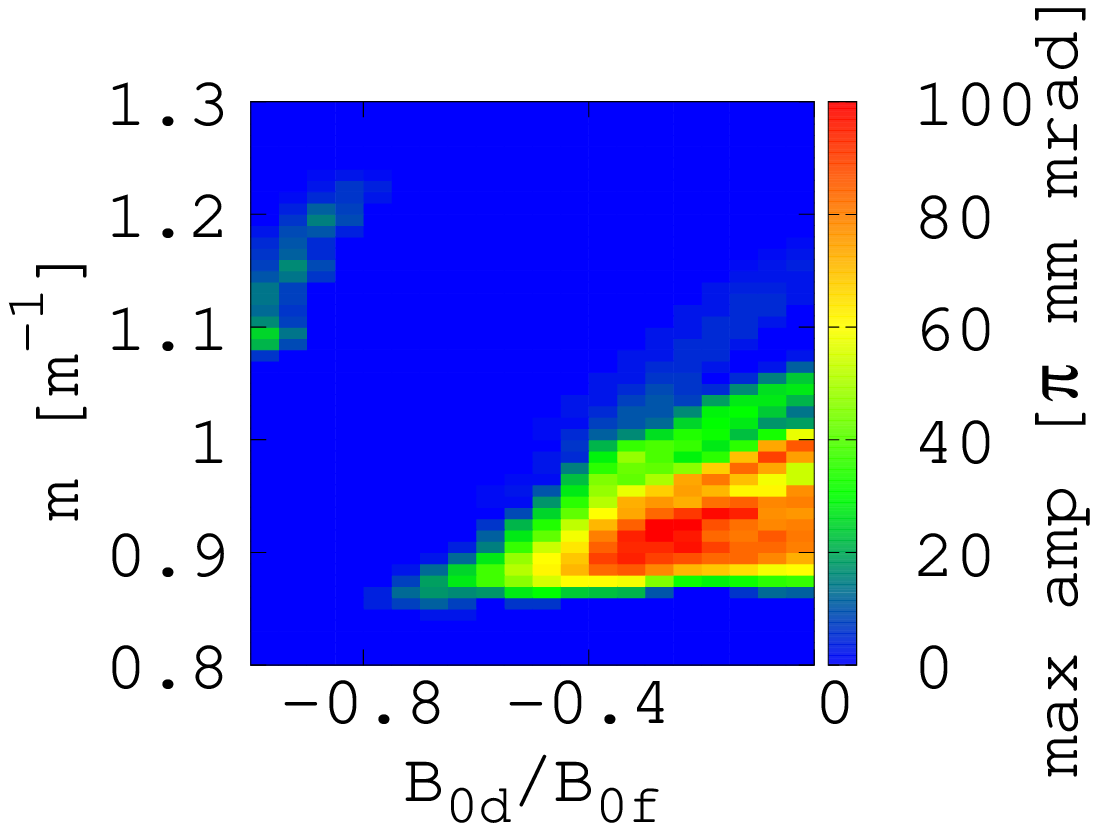}
}
\subfloat[$t_f$=4\,deg]{%
\includegraphics[width=.45\columnwidth]
{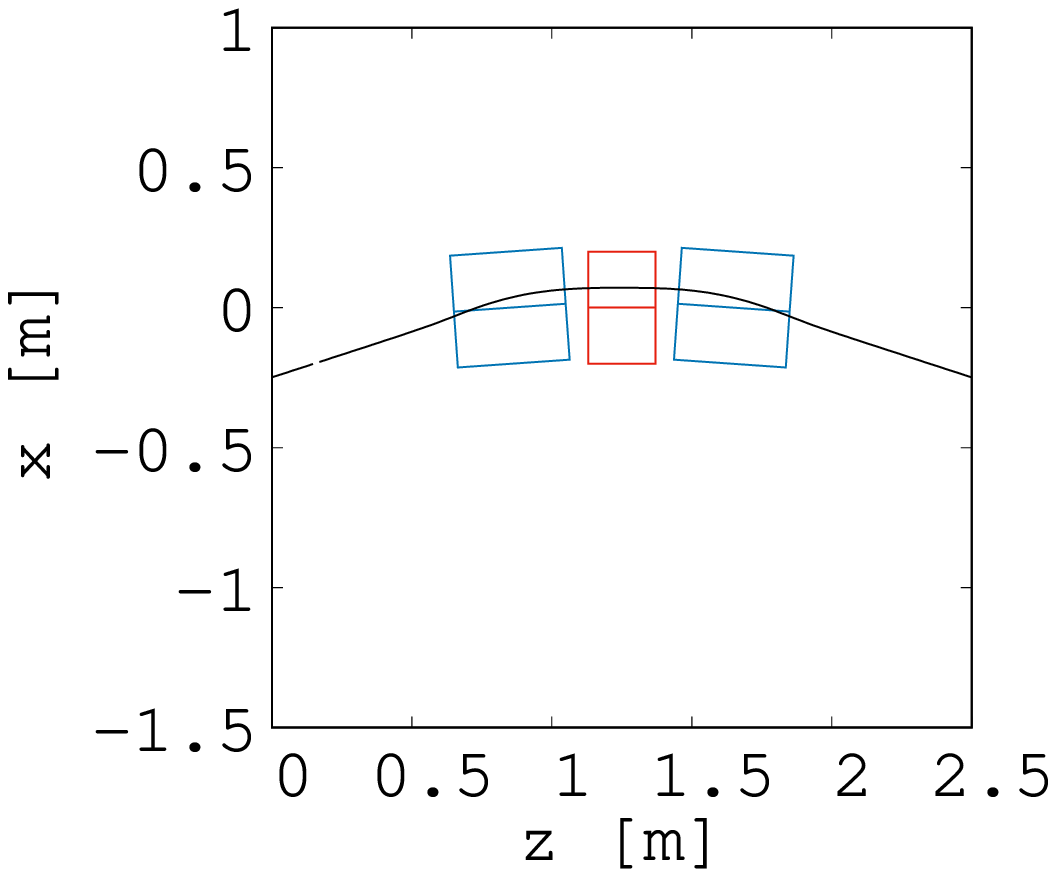}
}
\hspace{0mm}
\subfloat[$t_f$=6\,deg]{%
\includegraphics[width=.45\columnwidth]
{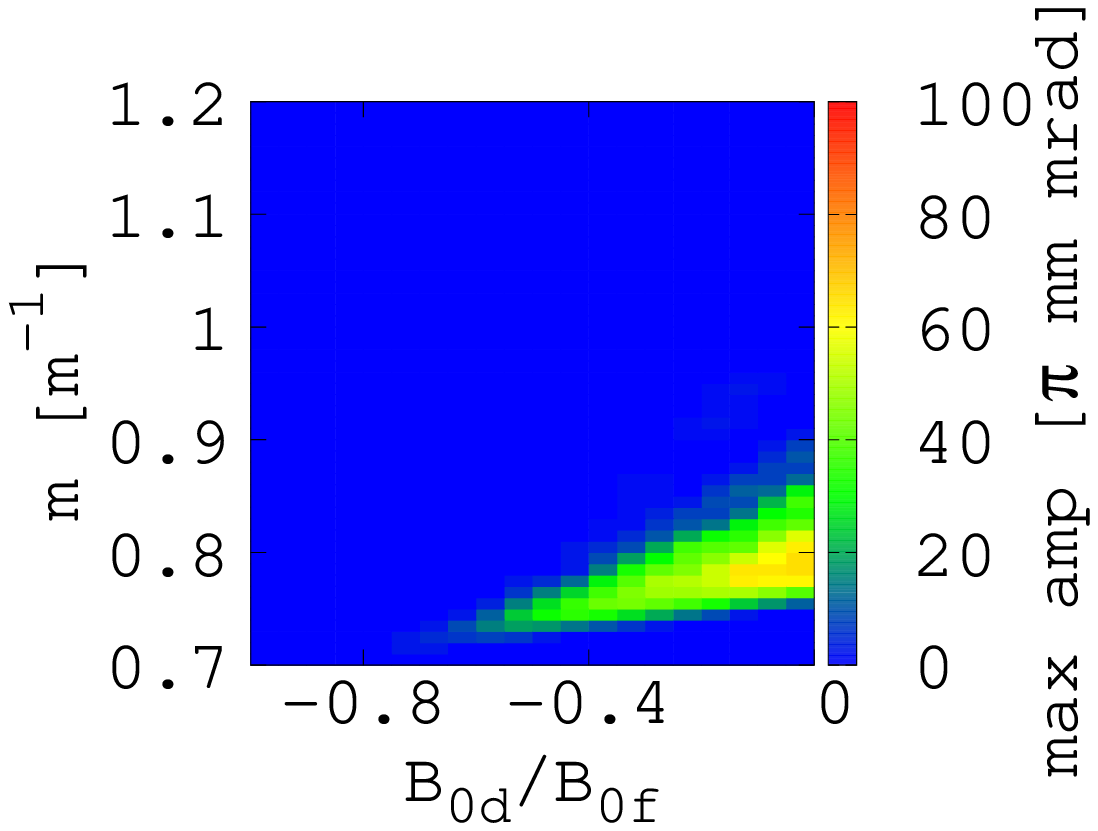}
}
\subfloat[$t_f$=6\,deg]{%
\includegraphics[width=.45\columnwidth]
{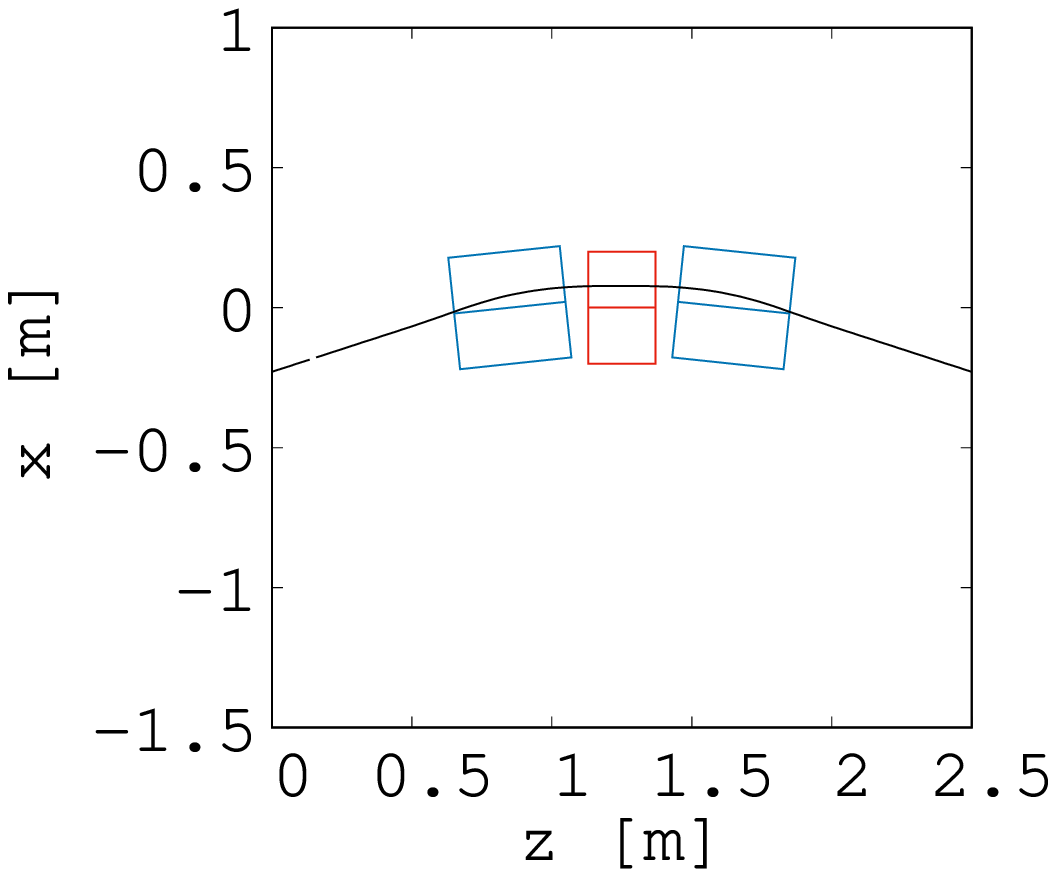}
}
\caption{\label{fig:fig7} Stable area in $m$ and $B_{0d}/B_{0f}$ space depending on the tilt angle $t_f$ of Bf with respect to Bd and representative orbit for each case. 
Parameters of the representative orbits are listed in Table~\ref{tab:tab3}.}
\end{figure}

\begin{figure}[h]
\centering
\subfloat[$x_s$=-0.04\,m]{%
\includegraphics[width=.45\columnwidth]
{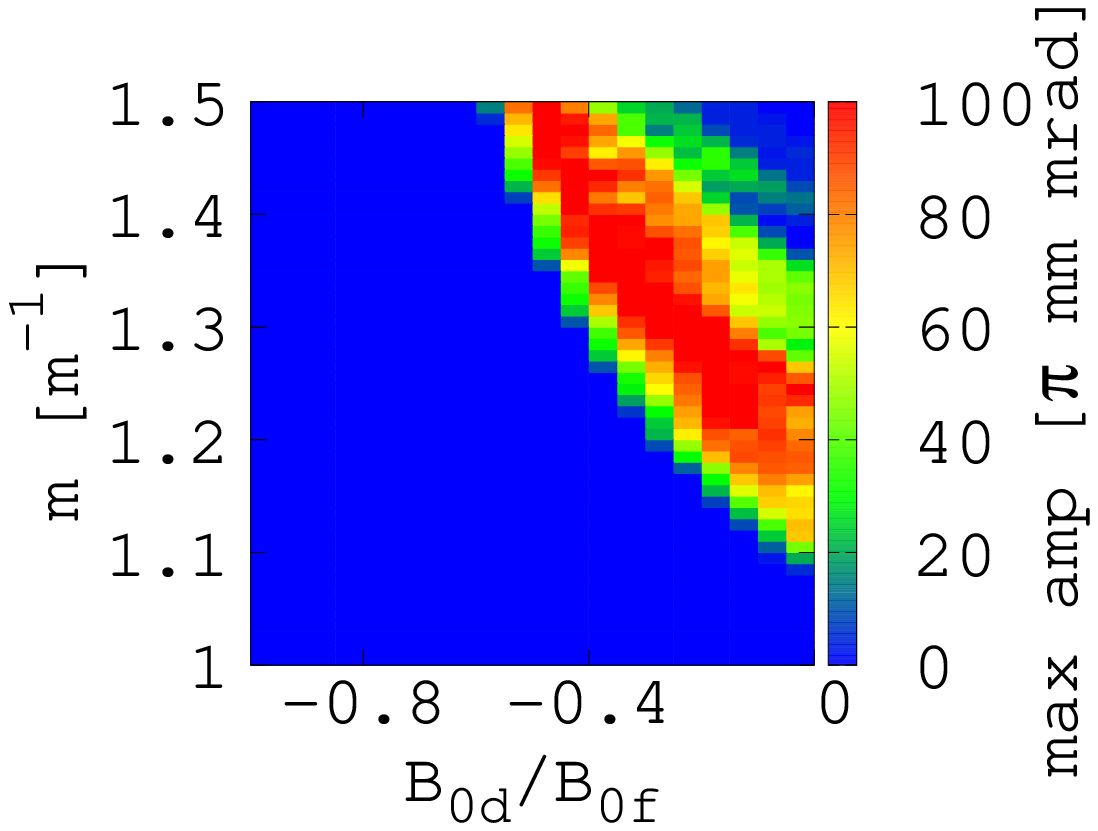}
}
\subfloat[$x_s$=-0.04\,m]{%
\includegraphics[width=.45\columnwidth]
{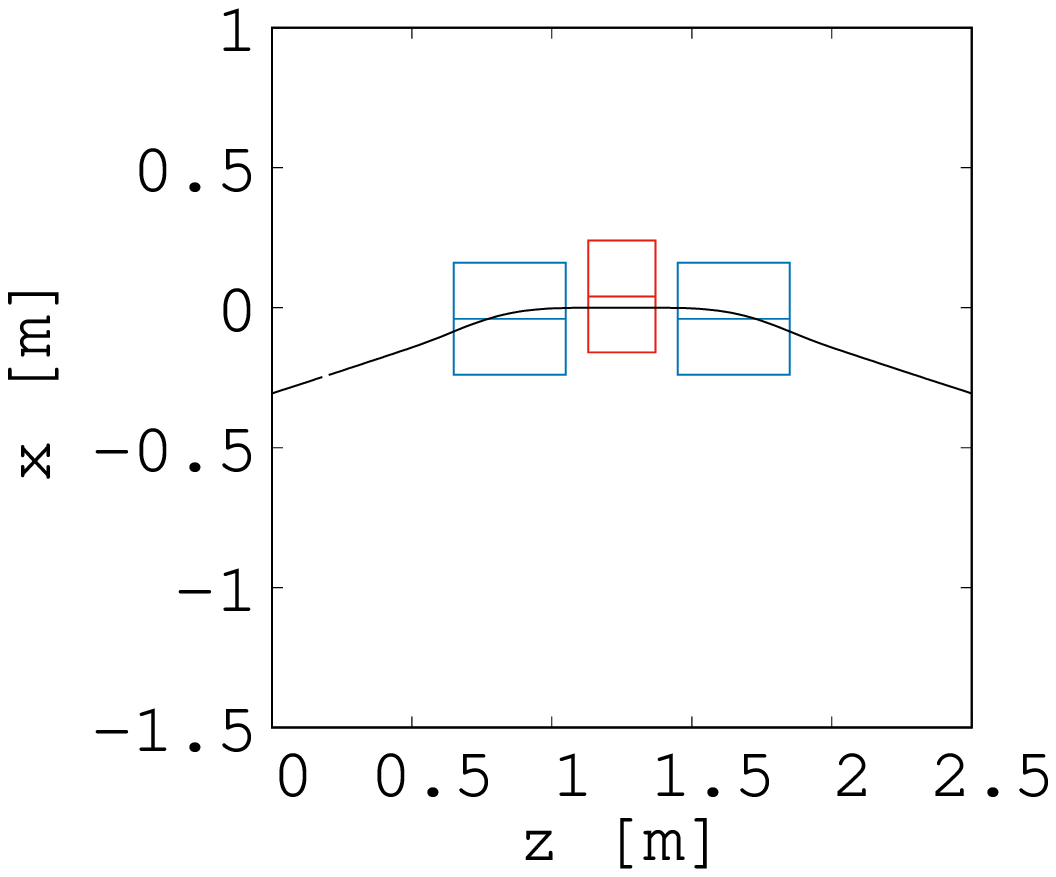}
}
\hspace{0mm}
\subfloat[$x_s$=-0.02\,m]{%
\includegraphics[width=.45\columnwidth]
{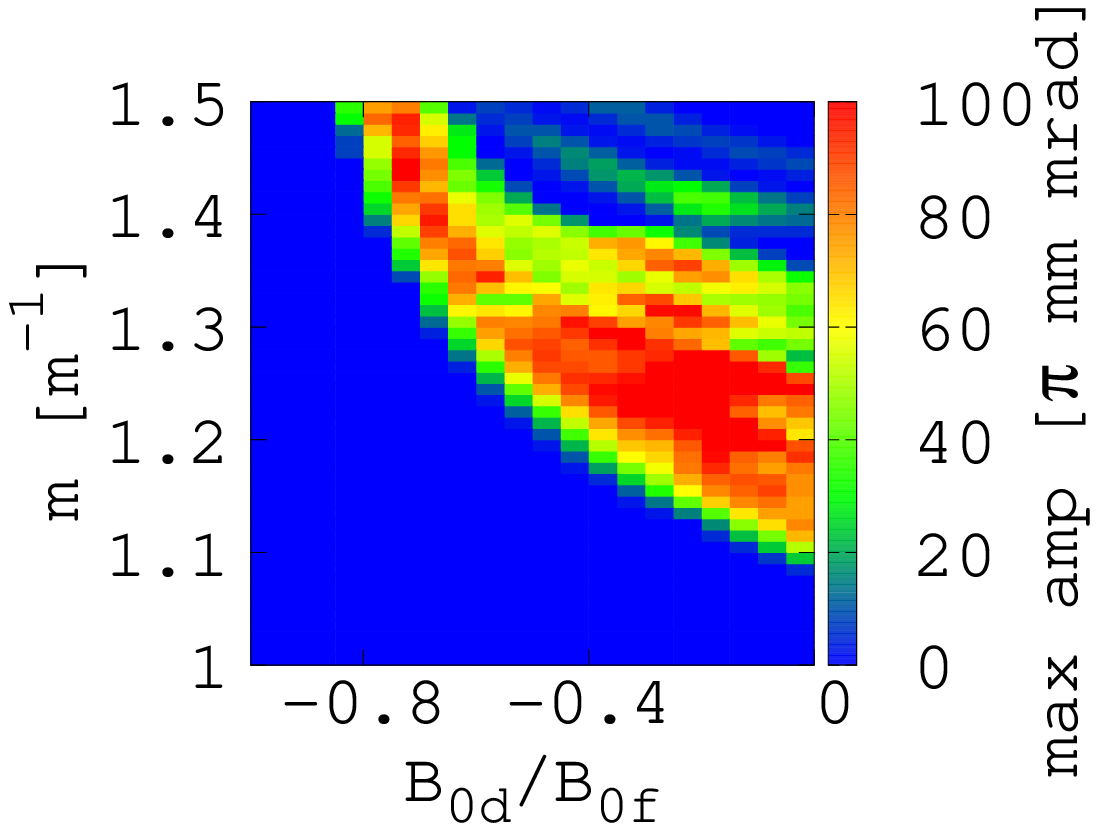}
}
\subfloat[$x_s$=-0.02\,m]{%
\includegraphics[width=.45\columnwidth]
{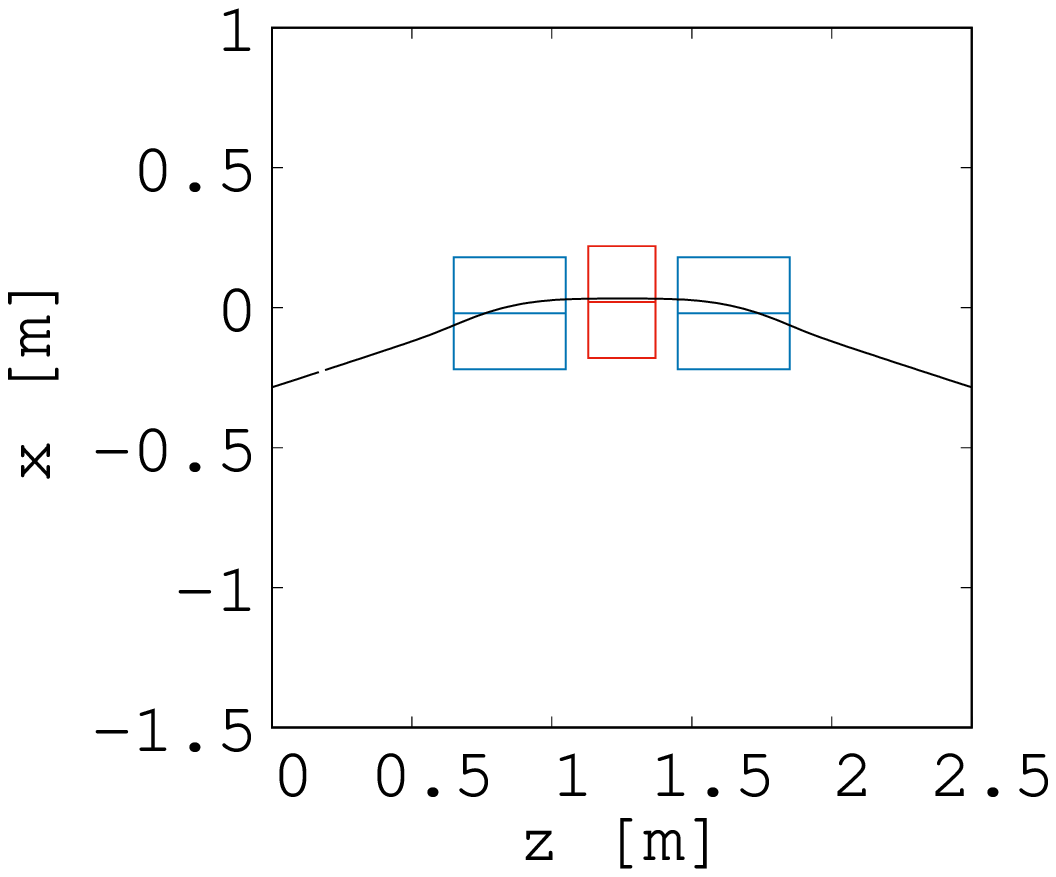}
}
\hspace{0mm}
\subfloat[$x_s$=0.0\,m]{%
\includegraphics[width=.45\columnwidth]
{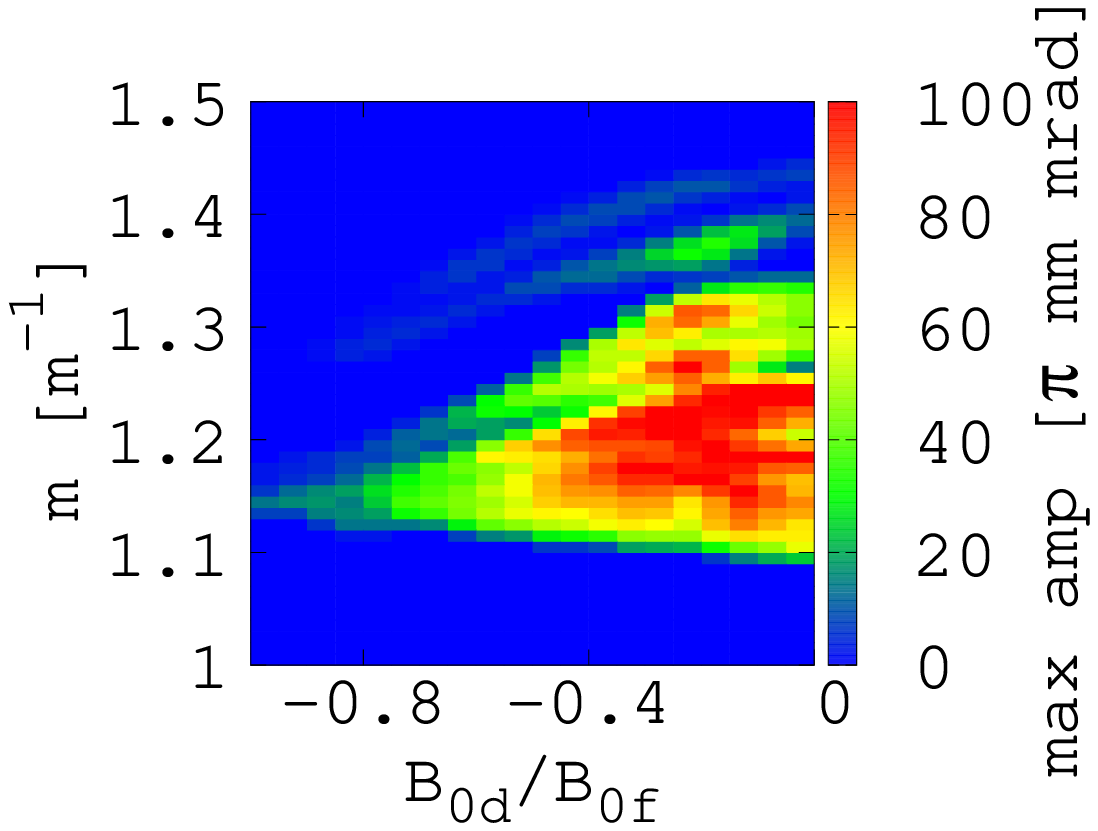}
}
\subfloat[$x_s$=0.0\,m]{%
\includegraphics[width=.45\columnwidth]
{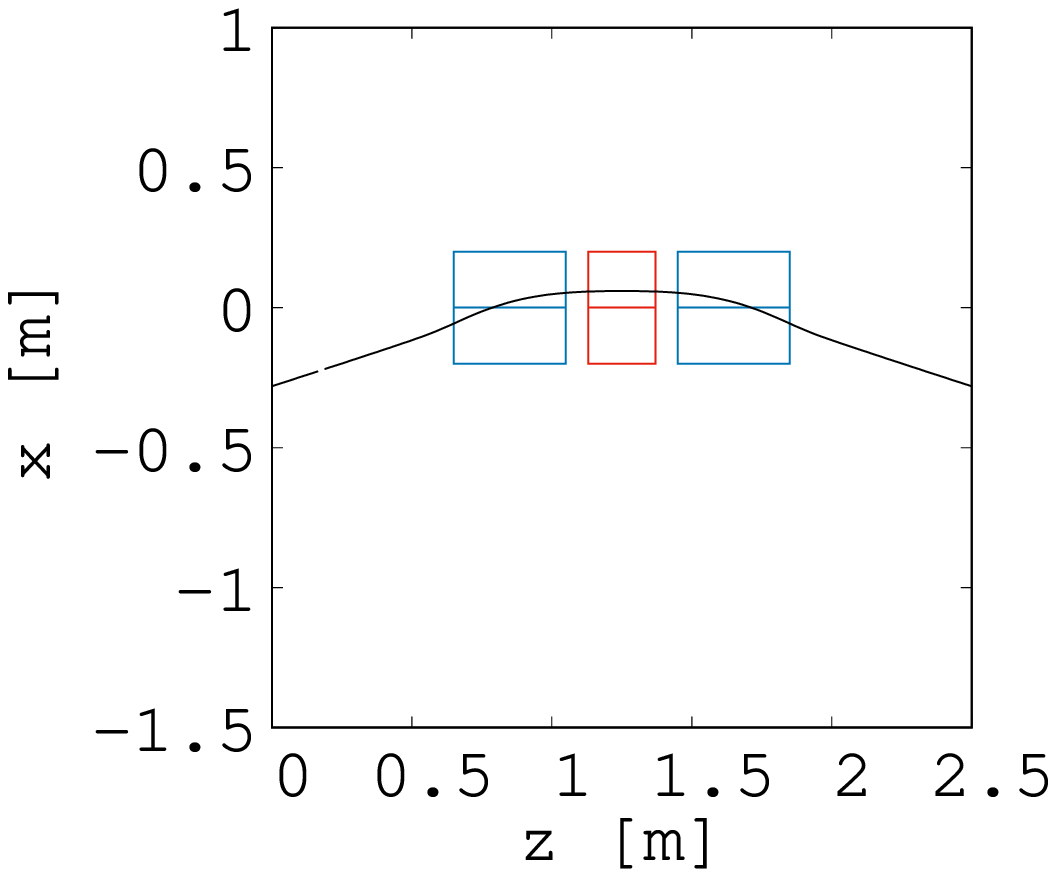}
}
\hspace{0mm}
\subfloat[$x_s$=0.02\,m]{%
\includegraphics[width=.45\columnwidth]
{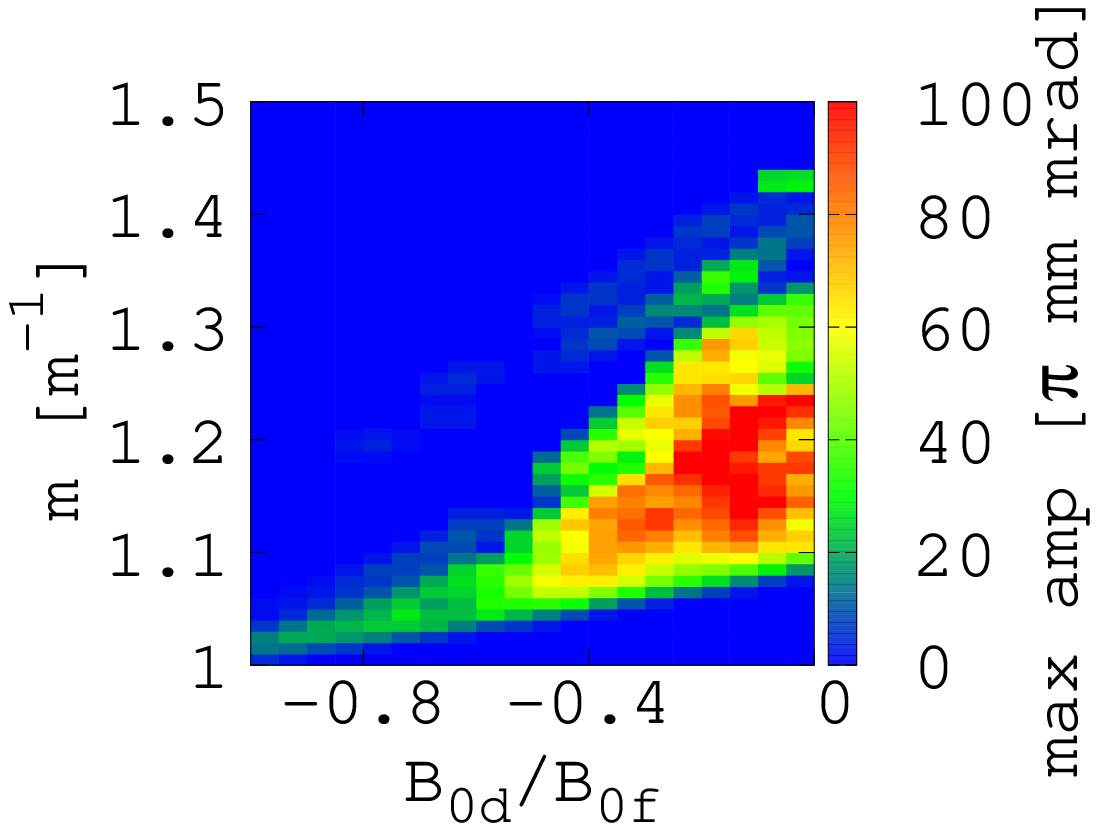}
}
\subfloat[$x_s$=0.02\,m]{%
\includegraphics[width=.45\columnwidth]
{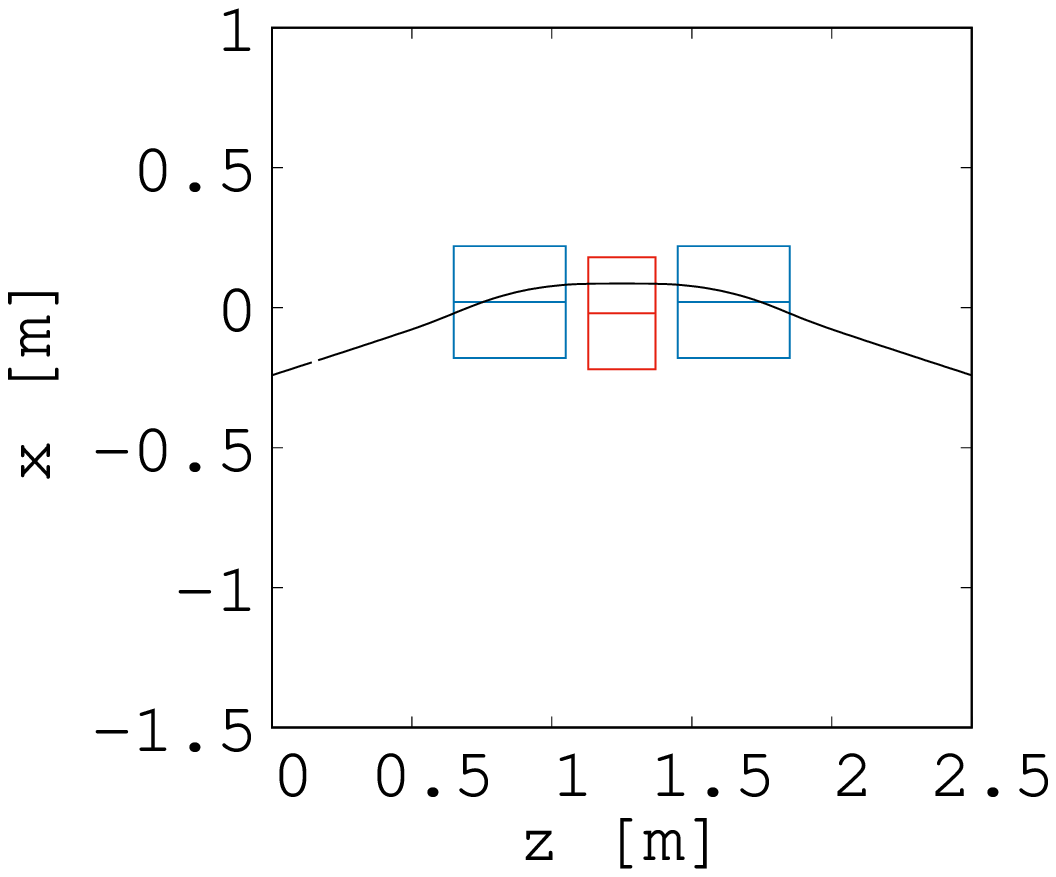}
}
\hspace{0mm}
\subfloat[$x_s$=0.04\,m]{%
\includegraphics[width=.45\columnwidth]
{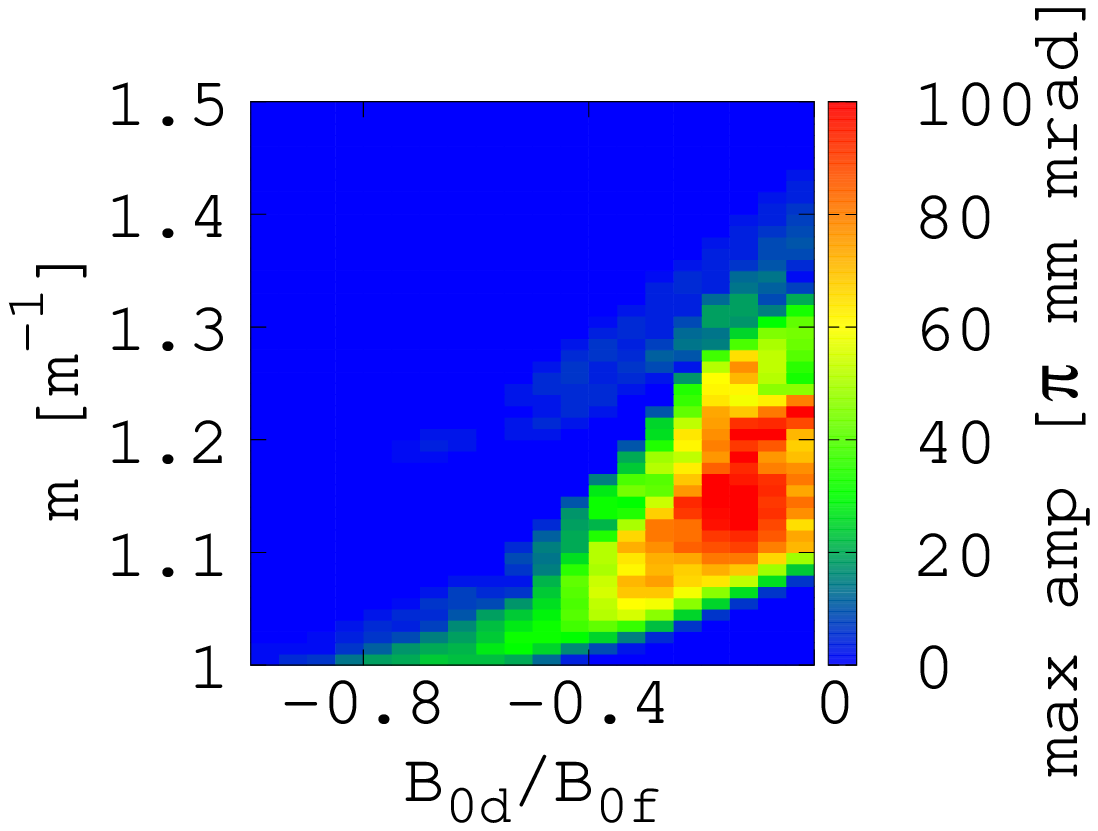}
}
\subfloat[$x_s$=0.04\,m]{%
\includegraphics[width=.45\columnwidth]
{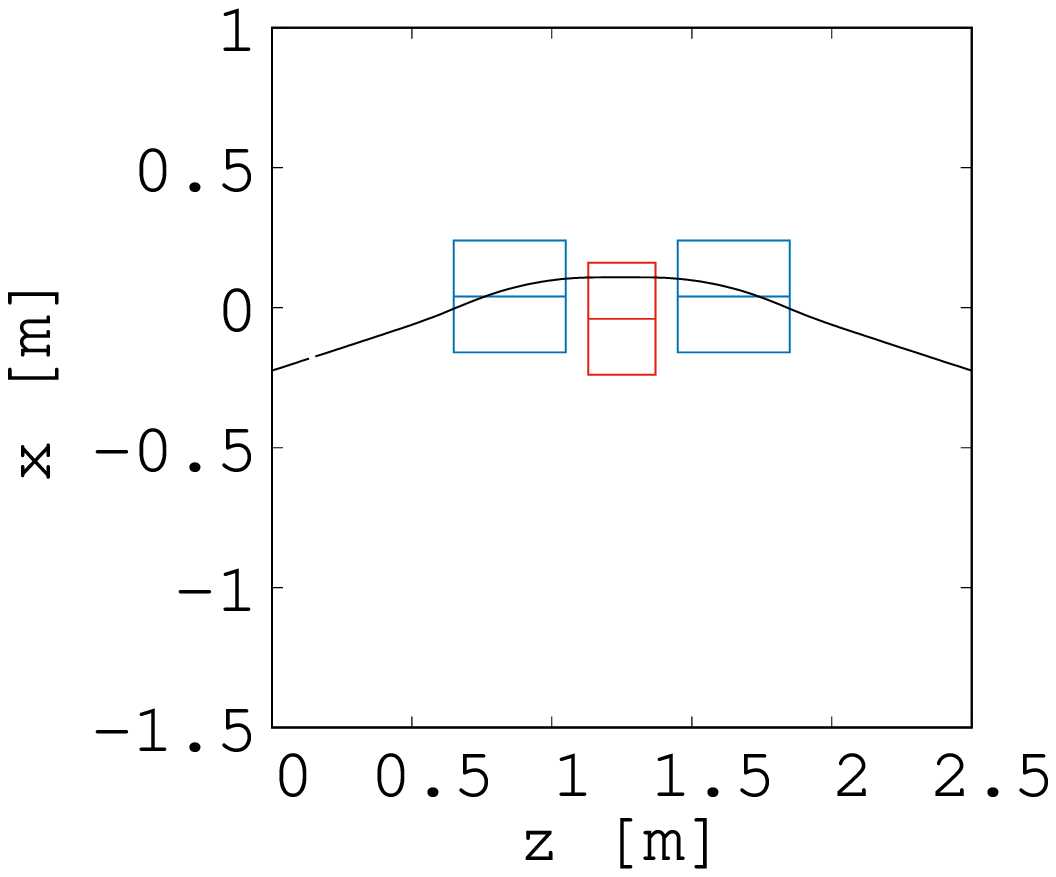}
}
\caption{\label{fig:fig8} Stable area in $m$ and $B_{0d}/B_{0f}$ space depending on radial displacement $x_s$ of Bf with respect to Bd and representative orbit for each case. Parameters of the representative orbits are listed in Table~\ref{tab:tab4}. }
\end{figure}

Figure~\ref{fig:fig7} shows the stable area when $t_f$ changes from -2 to 
+6~degrees. A positive angle indicates Bf is relatively more aligned with the 
orbit.
Figure~\ref{fig:fig8} shows the stable area when $x_s$ changes from -0.04 to 
+0.04\,m.
A positive displacement indicates Bf moves away from the ring centre and Bd 
moves towards the ring centre.
With a tilt imposed on Bf, the range of $m$ giving stable optics is reduced.
With a displacement $x_s$, a stronger correlation between $m$ and 
$B_{0d}/B_{0f}$ appears.

\begin{table}[h]
\caption{\label{tab:tab3}%
Parameters of the representative orbits in Fig.~\ref{fig:fig7}.
}
\begin{ruledtabular}
\begin{tabular}{lccccdr}
$t_f$ [deg] & $B_{0d}/B_{0f}$ & $m$ [m$^{-1}$] & $q_u$ & $q_v$ \\
\colrule
-2 & -0.20 & 1.42 & 0,2285 & 0.1577\\
0 & -0.20 & 1.28 & 0.2141 & 0.1394\\
 2 & -0.20 & 1.10 & 0.2264 & 0.1048\\
 4 & -0.20 & 0.93 & 0.2224 & 0.0736\\
 6 & -0.20 & 0.76 & 0.2164 & 0.0386\\
\end{tabular}
\end{ruledtabular}
\end{table}

\begin{table}[h]
\caption{\label{tab:tab4}%
Parameters for the representative orbits in Fig.~\ref{fig:fig8}.
}

\begin{ruledtabular}
\begin{tabular}{lccccdr}
$x_s$ [m] & $B_{0d}/B_{0f}$ & $m$ [m$^{-1}$] & $q_u$ & $q_v$ \\
\colrule
-0.04 & -0.40 & 1.44 & 0.2266 & 0.1131\\
-0.02 & -0.40 & 1.31 & 0.2245 & 0.1133\\
 0.00 & -0.40 & 1.22 & 0.2220 & 0.1147\\
+0.02 & -0.40 & 1.14 & 0.2342 & 0.1089\\
+0.04 & -0.40 & 1.10 & 0.2259 & 0.1155\\
\end{tabular}
\end{ruledtabular}
\end{table}

\section{Aperture}
\label{sec:aperture}

\subsection{Dynamic Aperture}
\begin{figure}[h]
\centering
\subfloat[$x$-$p_x$]{%
\includegraphics[width=.45\columnwidth]
{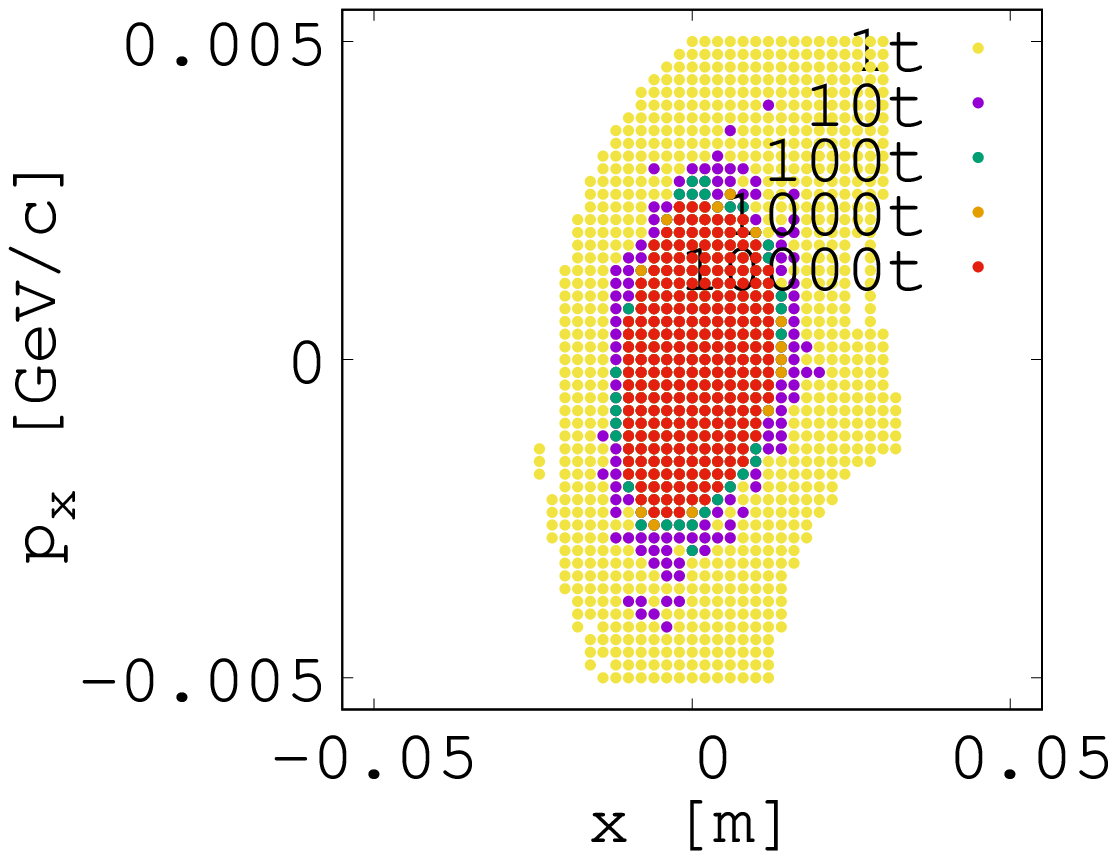}
}
\subfloat[$y$-$p_y$]{%
\includegraphics[width=.45\columnwidth]
{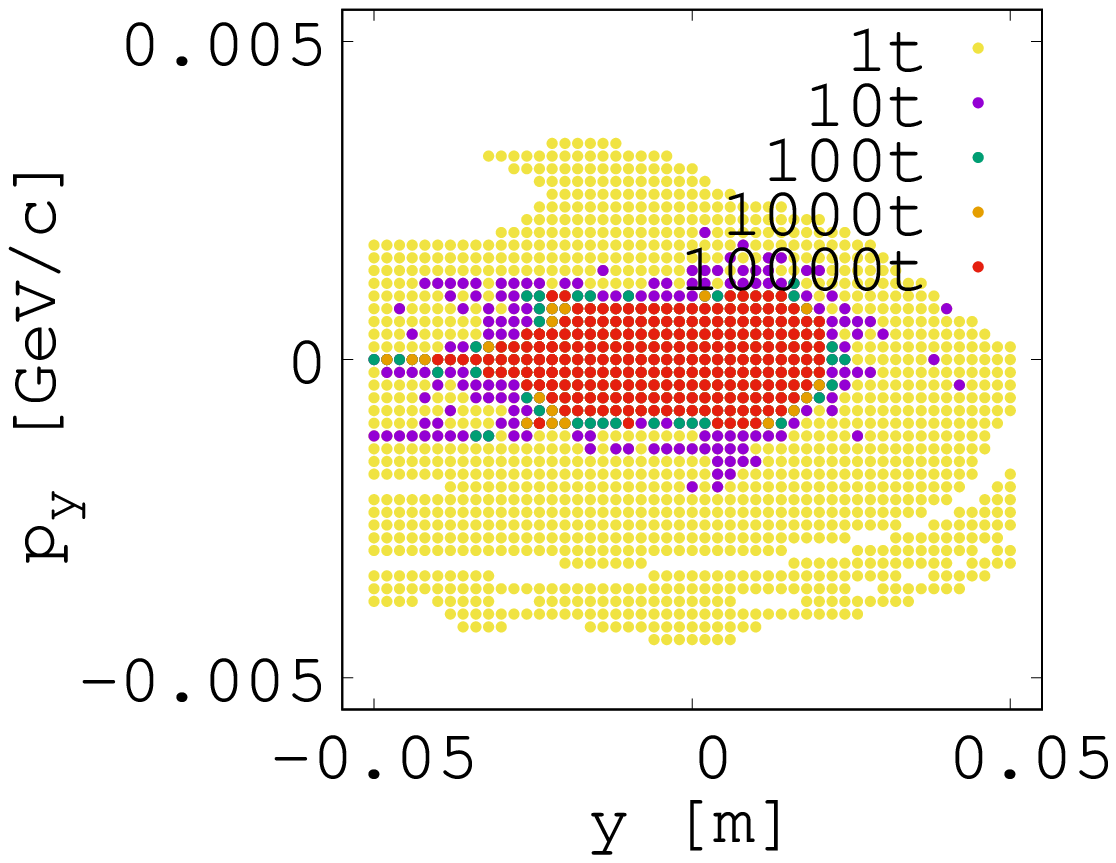}
}
\hspace{0mm}
\caption{\label{fig:fig9} Dynamic aperture in physical $x$-$p_x$ and $y$-$p_y$ planes. Results 
from tracking through a different number of turns: 1, 10, 100, 1000, 10000 turns are superimposed.
The dynamic aperture defined by tracking for 100 turns and 1000 turns has only a small difference and converges when the number of turns is more than 1000.}
\end{figure}

Dynamic aperture is estimated by particle tracking, but it is not straightforward to give a precise interpretation of the concept when the optics between the two transverse planes is coupled.
Two steps were considered to provide an estimate that 
would be practically useful for accelerator design. As in the 
algorithm described above, particles were tracked having initial
coordinate values $(x,p_x,y,p_y)=(i x_0,0,0,0)$ and $(x,p_x,y,p_y)=(0,0,jy_0,0)$
with respect to the closed orbit for a number of integer values of $i$ and $j$.
This gives an estimate of the horizontal and vertical phase space
where beams can be injected into the ring.
Because of the coupling between the two planes,
particles starting in one physical plane explore the 4D phase space,
and do not remain in the 2D sub-space, but may not exhaustively explore the full 4D region.
Figure~\ref{fig:fig9} shows the surviving particles.
To investigate the dependence of this dynamic aperture estimate on the number 
of turns, results with a different number of tracking turns are superimposed in 
the plots. 
The figure shows that there is only a small difference between the dynamic 
aperture defined by tracking for 100 turns and 1000 turns and almost no 
difference if the number of turns is more than 1000. 

\begin{figure}[h]
\centering
\subfloat[$u$-$p_u$]{%
\includegraphics[width=.45\columnwidth]
{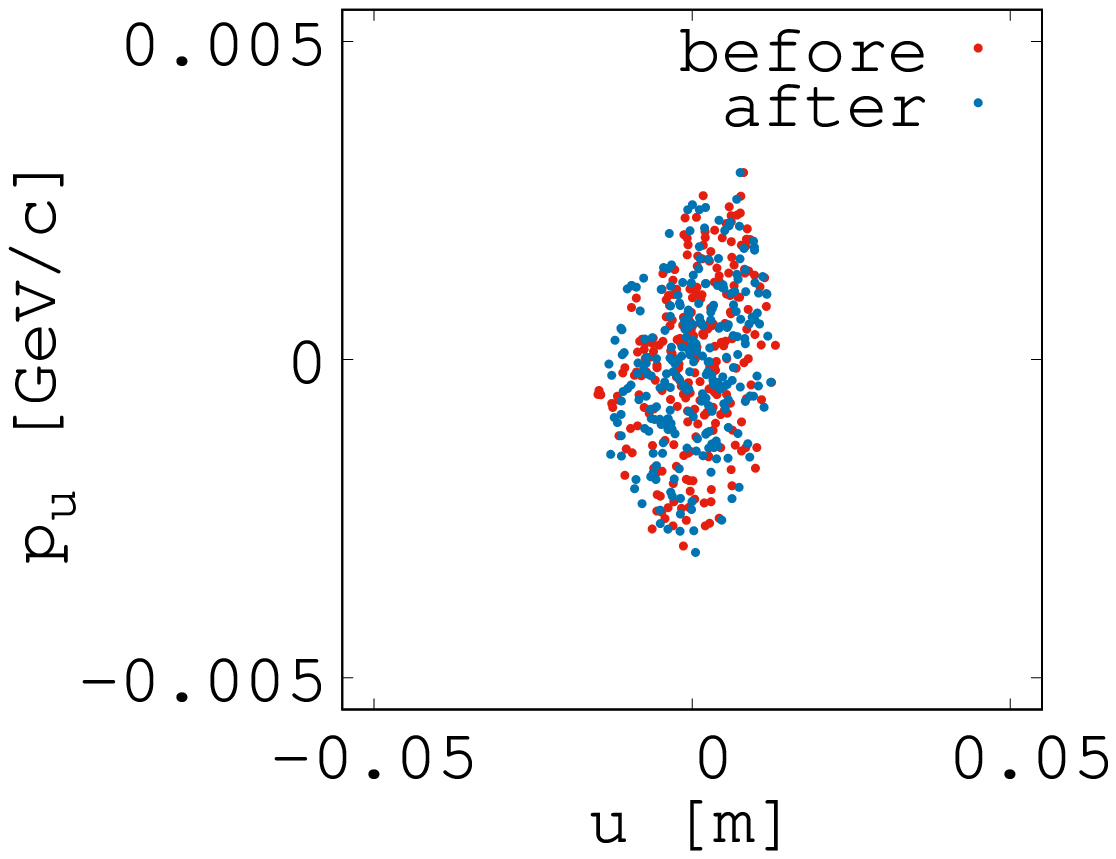}
}
\subfloat[$v$-$p_v$]{%
\includegraphics[width=.45\columnwidth]
{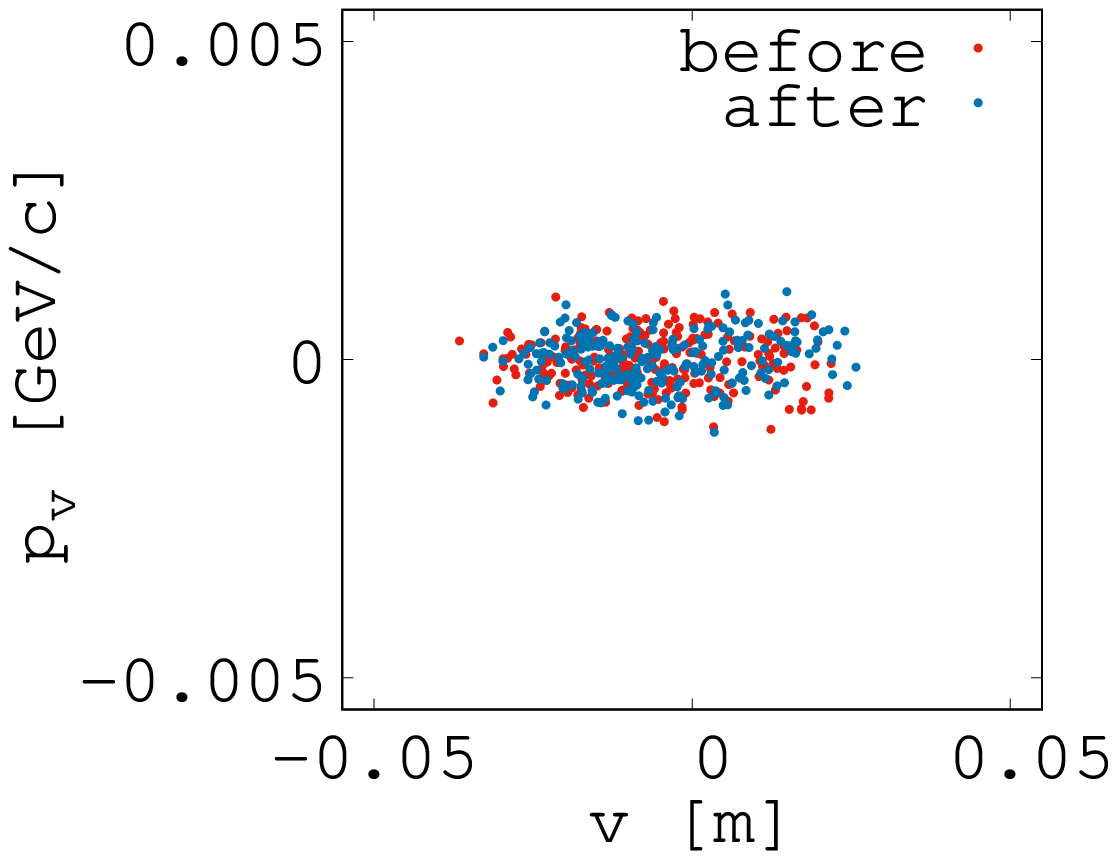}
}
\hspace{0mm}
\subfloat[$u$-$v$]{%
\includegraphics[width=.45\columnwidth]
{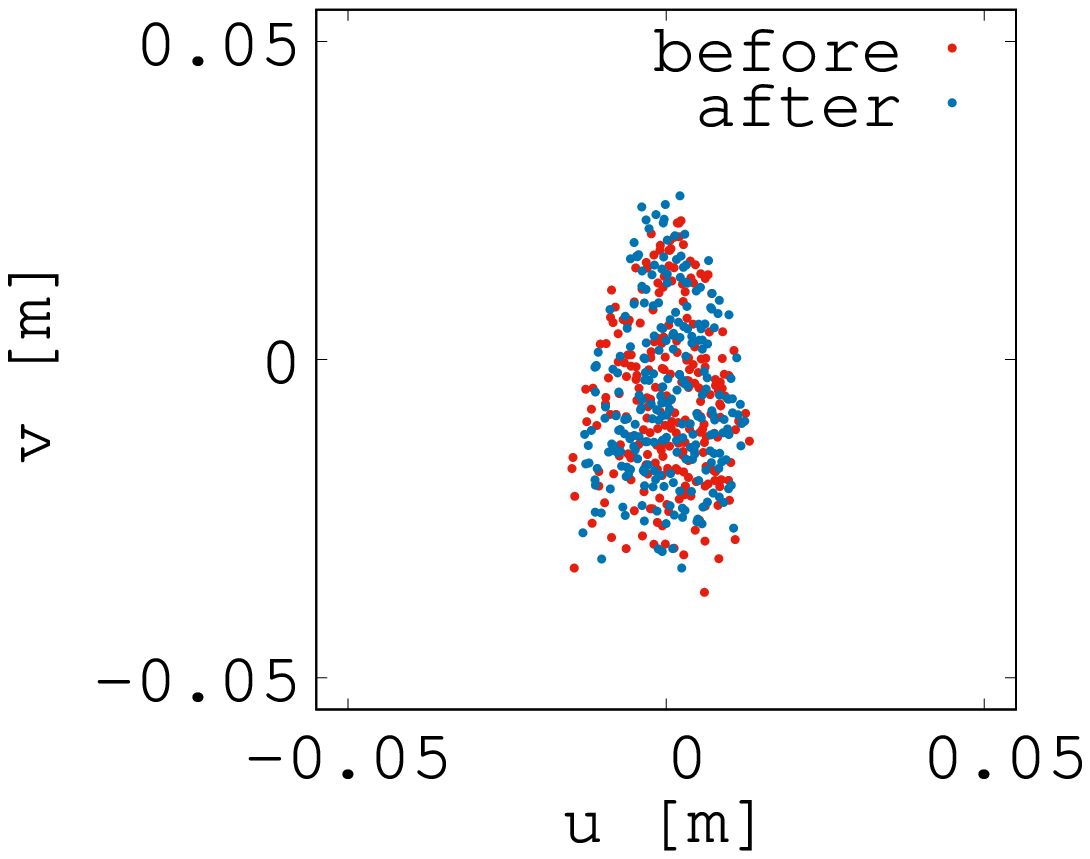}
}
\subfloat[$p_u$-$p_v$]{%
\includegraphics[width=.45\columnwidth]
{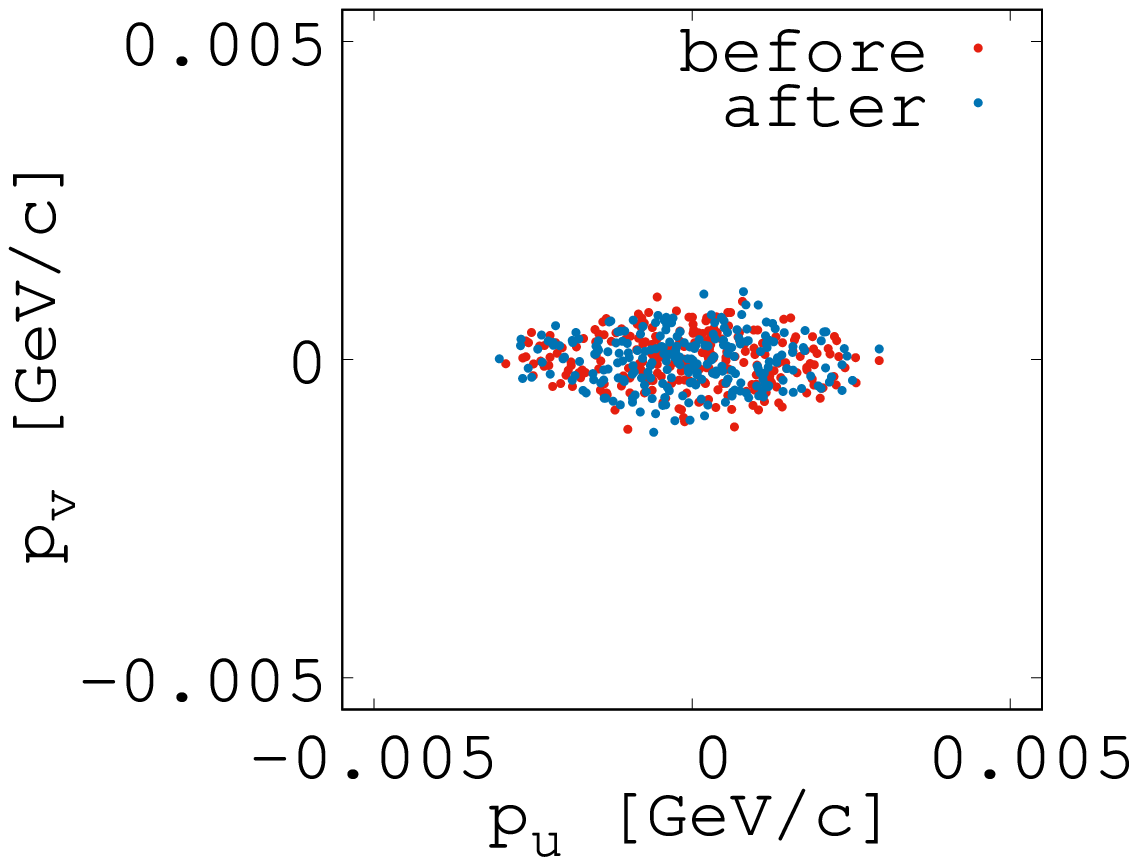}
}\hspace{0mm}
\hspace{0mm}
\subfloat[$u$-$p_v$]{%
\includegraphics[width=.45\columnwidth]
{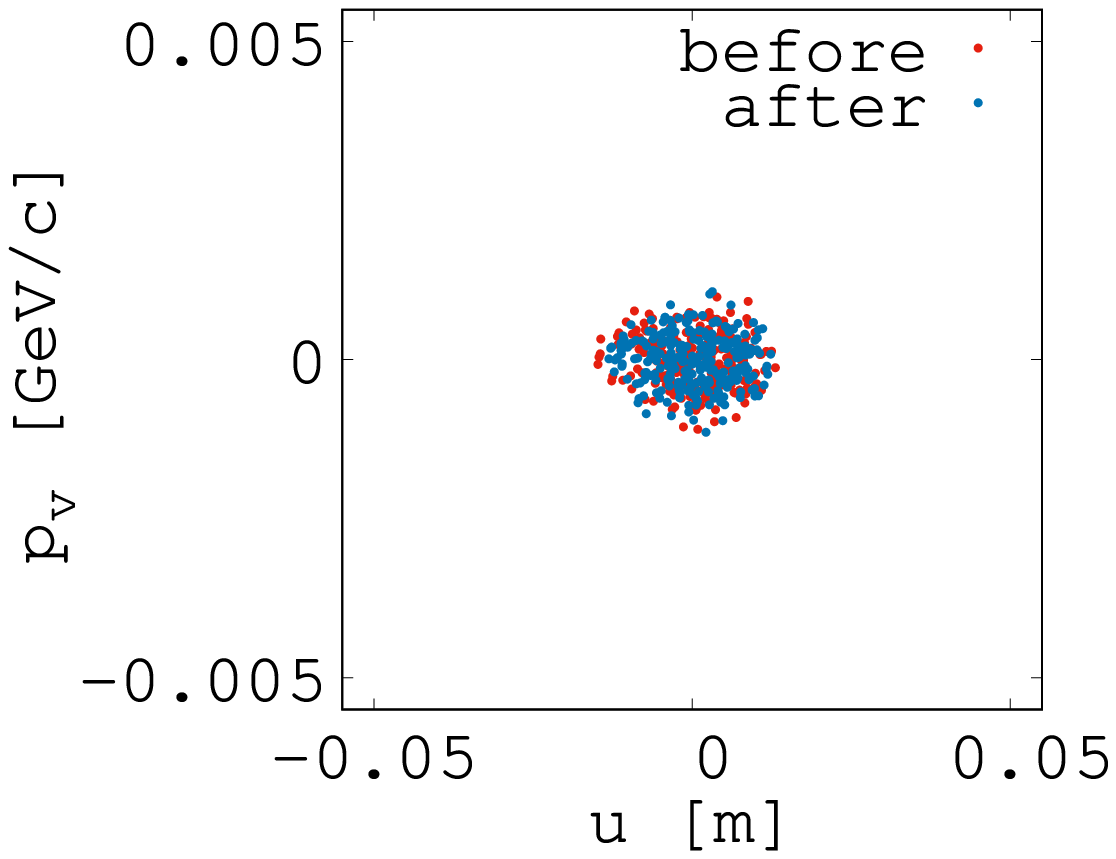}
}
\subfloat[$v$-$p_u$]{%
\includegraphics[width=.45\columnwidth]
{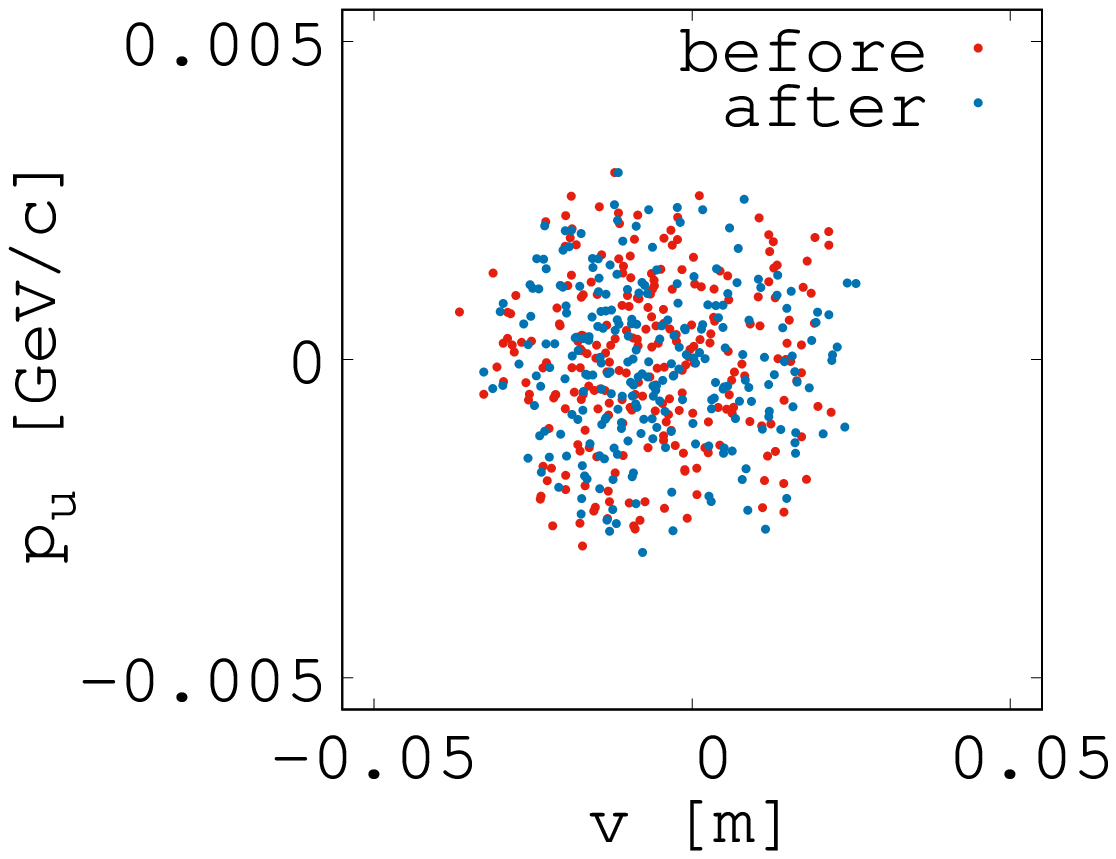}
}
\caption{\label{fig:fig10} Dynamic aperture defined by surviving particles in 
4D space after 1000 turns. 
Red dots show the initial coordinates and blue dots are the final 
coordinates.
Projections onto the decoupled spaces ((a)~$u$-$p_u$ and (b)~$v$-$p_v$)
show that
normalised dynamic aperture is estimated to be about 30\,$\pi$\,mm\,mrad.
Projections of the 4D phase space volume onto other planes are also depicted.}
\end{figure}

In order to explore the accepted phase space more thoroughly, particles were 
tracked having random initial position and momentum in the 4D phase space.
The results of the first step help to define the appropriate volume in 4D 
phase space where the particles should be randomly allocated:
it should be large enough to cover the entire dynamic aperture.
The ability of each particle to survive when tracked for 1000 turns of the ring is checked.
The 4D phase space coordinates of the surviving particles define the 4D phase 
space volume and we identify dynamic aperture through projections onto the 
$u$-$p_u$ and  $v$-$p_v$ linearly decoupled 2D phase spaces~\cite{parzen}.
The coordinates of the particles surviving after tracking are also
plotted.
We assume that if the phase space area before and after the tracking have approximately the same shape and size, we are assured that the dynamic aperture defined above is stable and consistent, as shown in Fig.~\ref{fig:fig10}.
In either plane, the normalised dynamic aperture is about 30\,$\pi$\,mm\,mrad.
Projections of the 4D phase space volume to other planes are also depicted.

\subsection{Physical Aperture}
Knowing the normalised dynamic aperture is about 30$\pi$ mm\,mrad,
the mechanical aperture as well as beam core emittance after multi-turn painting injection are defined in Table~\ref{tab:tab5}.
To convert un-normalised emittance at 3\,MeV to beam size, values
$\beta_{xI,\mathrm{max}}$=1.0\,m and  $\beta_{yII,\mathrm{max}}$=4.6\,m are used according to Fig.~\ref{fig:fig4}.
Contributions to the beam size from other planes are ignored since $\beta_{xII,\mathrm{max}}$ and $\beta_{yI,\mathrm{max}}$ are relatively small.
\begin{table}[h]
\caption{\label{tab:tab5}%
Specifications of beam core emittance and mechanical aperture.}
\begin{ruledtabular}
\begin{tabular}{lccccdr}
 & normalised & un-normalised & $x$ size & $y$ size \\
 & [$\pi$\,mm\,mrad] & [$\pi$\,mm\,mrad] & [mm] & [mm] \\
\colrule
Beam core & 5 -- 10 & 62.5 -- 125 & 8 -- 11 & 17 -- 24\\
Collimator \\ aperture & 10 -- 20 & 125 -- 250 & 11 -- 16 & 24 -- 34\\
Chamber \\ aperture & --  & -- & $\pm$30 & $\pm$350\\
\end{tabular}
\end{ruledtabular}
\end{table}

\section{Summary}
In designing a vFFA, it is necessary to take into account coupling in the
transverse planes, both for calculation of optics and optimisation of accelerator parameters.
Strong coupling arises from the skew quadrupole components in the vFFA magnet
body and also from longitudinal field components in the end fields.
In this paper, we have employed tools to properly analyse these coupled optics.
Assuming an idealised magnetic field model that satisfies Maxwell equations and 
has field strength increasing exponentially in the vertical direction,
a design procedure has been established, with parameters such as the normalised 
field gradient $m$ and the ratio of Bd and Bf field strengths used to tune the 
optics.
A prototype vFFA, accelerating proton beams from 
3\,MeV to 12\,MeV, was used as an example.
The design procedure is applicable to any vFFA since no approximation -- such as small bending angle per cell -- was made.
An optics solution
was found that is robust against small variation of parameters and gives large dynamic aperture.

\begin{acknowledgments}
We wish to acknowledge fruitful discussion with S.~J.~Brooks and the support of members of the ISIS Accelerator and Design Divisions.
We also thank C.~R.~Prior for careful reading of the manuscript.
\end{acknowledgments}

\appendix

\section{Benchmark of Simulation Codes}

The design of the lattice has been performed throughout with SCODE+~\cite{scodep}.
This code was benchmarked against two other codes: FixField (previously named JBT)~\cite{fixfield} and OPAL~\cite{OPAL} comparing eigentunes as shown in
Table~\ref{tab:tab6}.
Eigentunes are constant for the whole range of momentum as expected.

\begin{table}[htb]
\caption{\label{tab:tab6}%
Comparison of results from three different codes.
}
\begin{ruledtabular}
\begin{tabular}{lccdr}
 & $q_u$ & $q_v$ \\
\colrule
SCODE+ & 0.2178 & 0.1342\\
FixField & 0.2198 & 0.1335\\
OPAL & 0.2199 & 0.1340\\
\end{tabular}
\end{ruledtabular}
\end{table}

\section{Stability of Lattices with Slightly Different Geometry}

\begin{figure}[ht]
\centering
\subfloat[Bd=0.16\,m]{
\includegraphics[width=.45\columnwidth]
{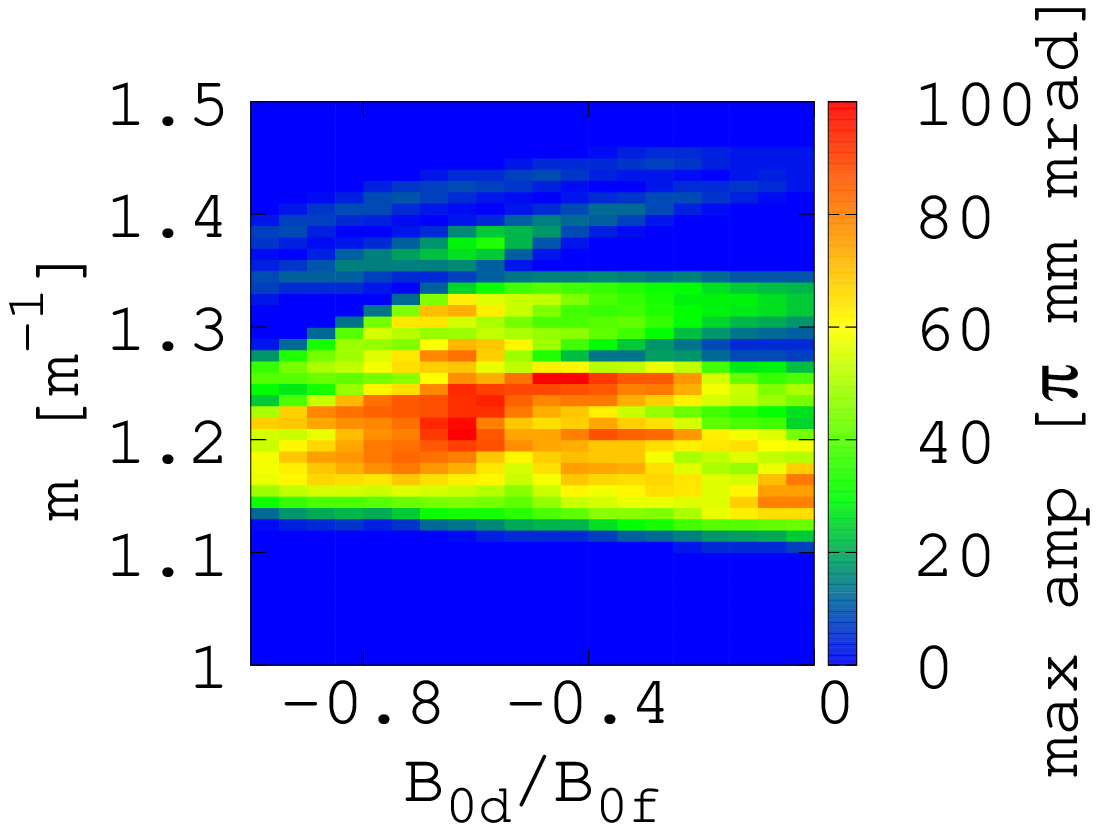}
}
\subfloat[Bd=0.16\,m]{
\includegraphics[width=.45\columnwidth]
{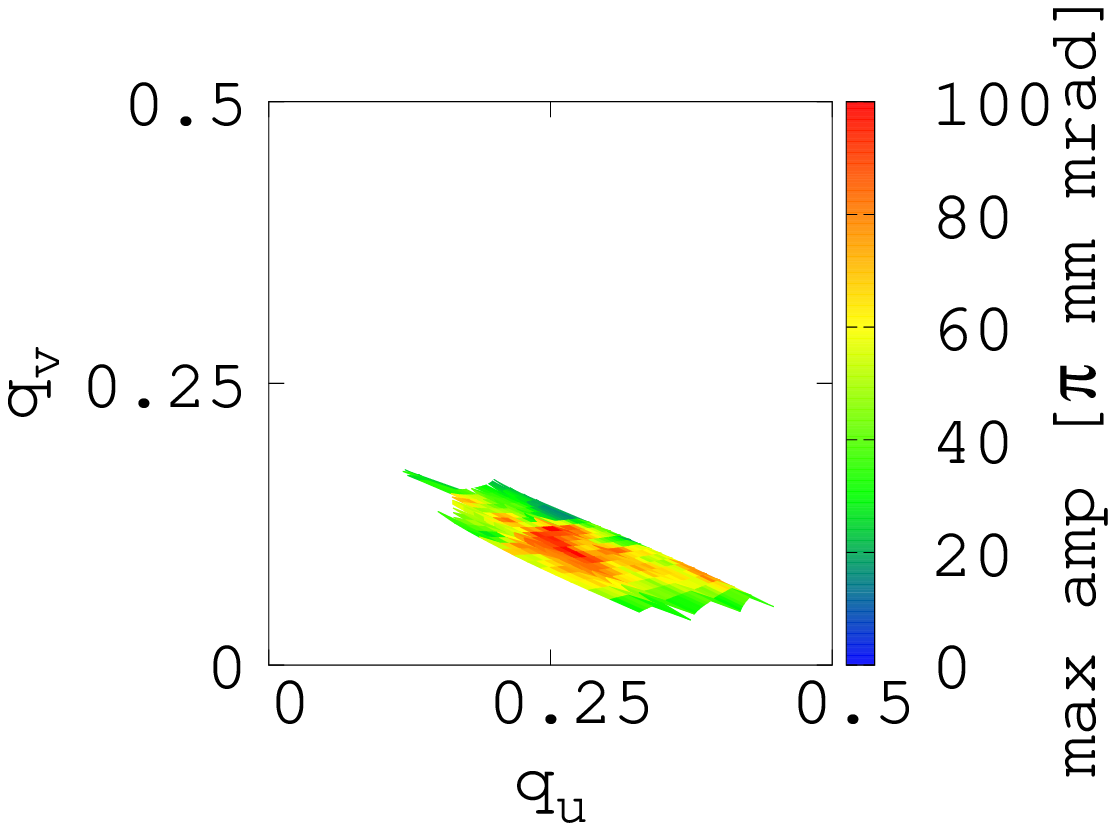}
}
\hspace{0mm}
\subfloat[Bd=0.24\,m]{
\includegraphics[width=.45\columnwidth]
{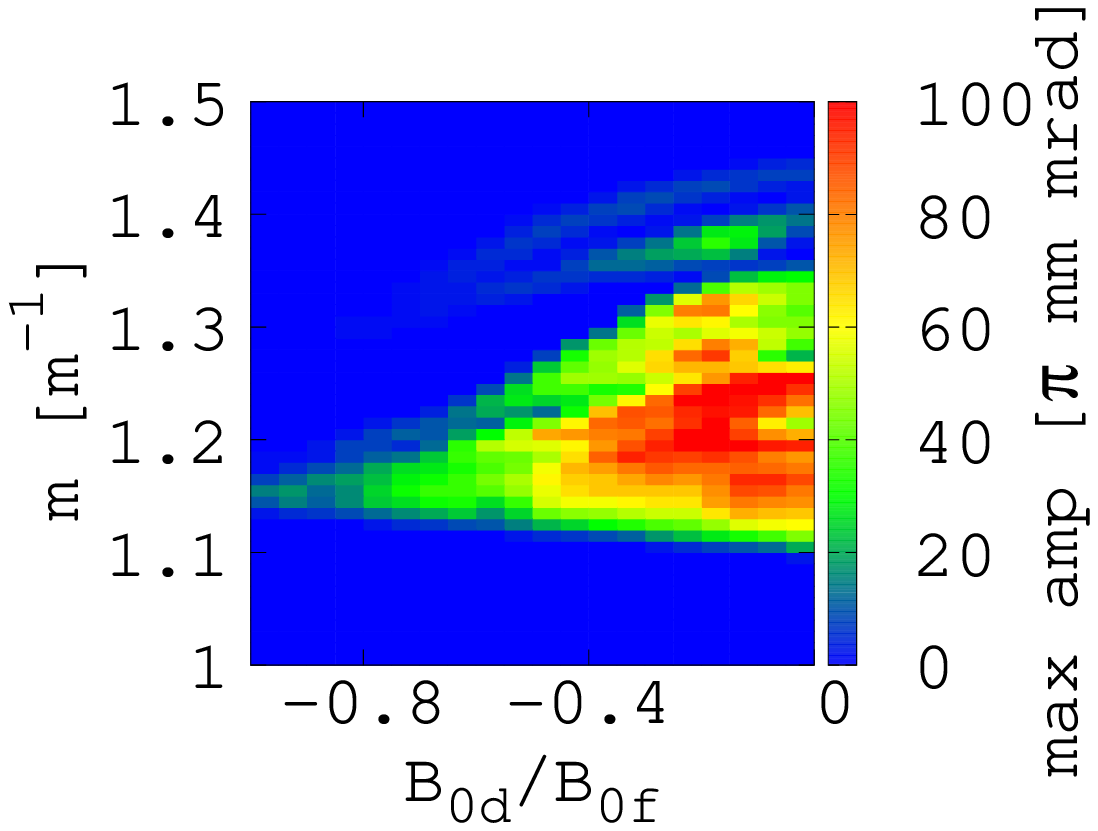}
}
\subfloat[Bd=0.24\,m]{
\includegraphics[width=.45\columnwidth]
{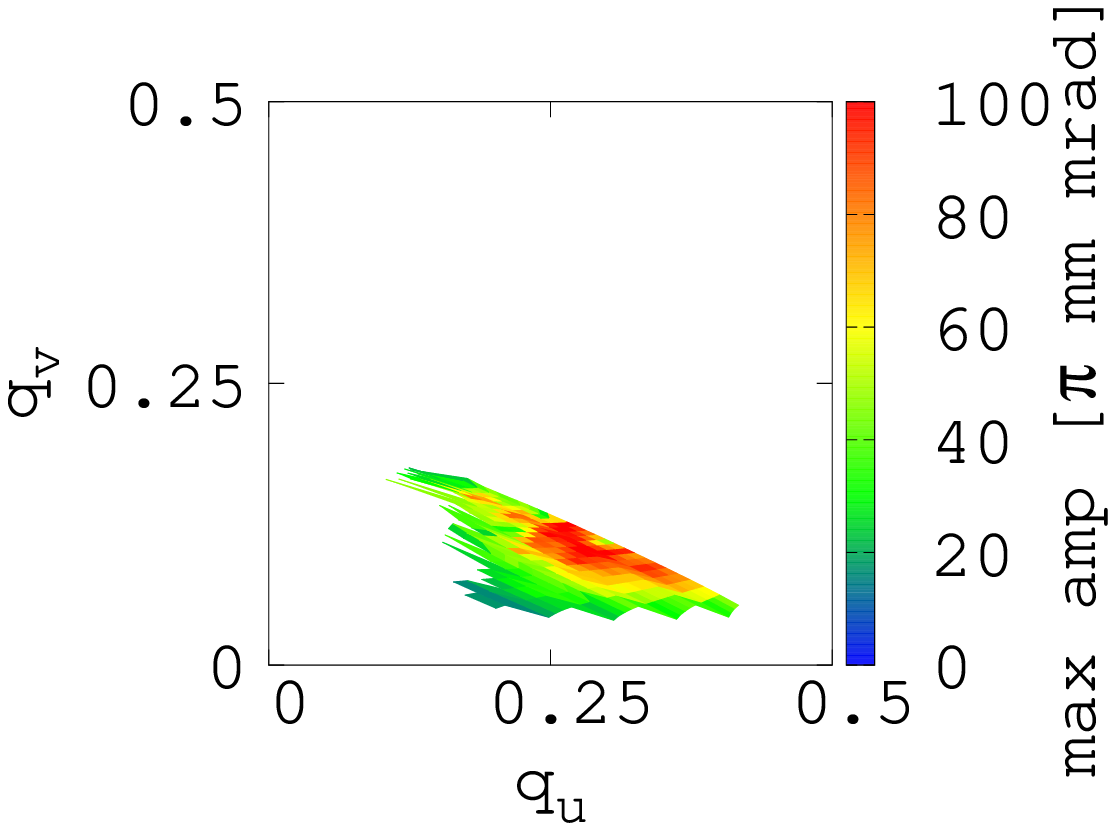}
}
\hspace{0mm}
\subfloat[Bd=0.32\,m]{
\includegraphics[width=.45\columnwidth]
{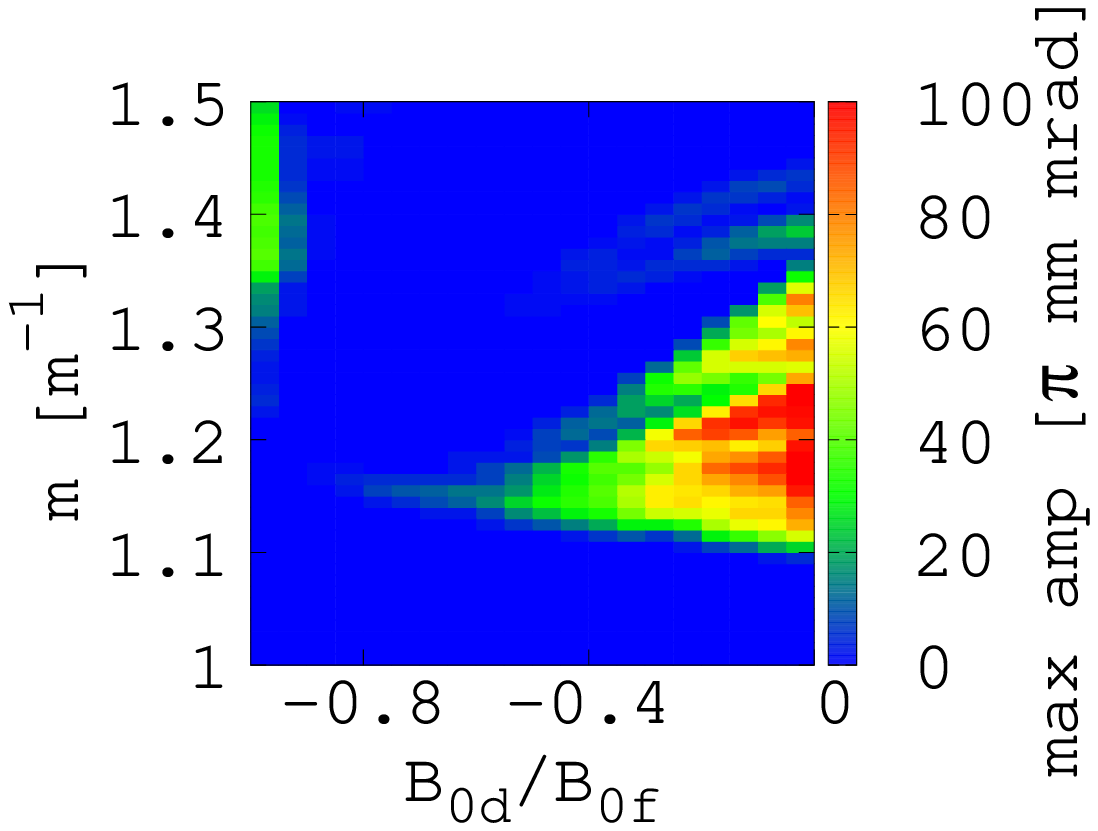}
}
\subfloat[Bd=0.32\,m]{
\includegraphics[width=.45\columnwidth]
{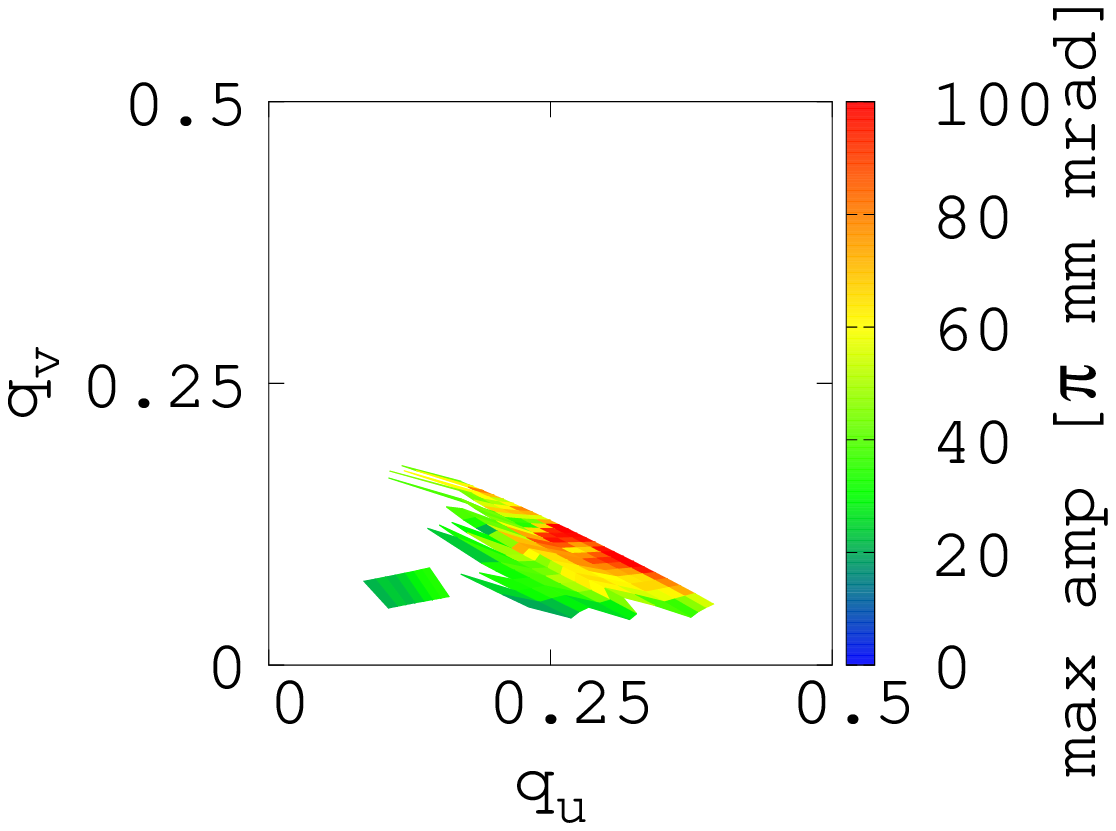}
}
\hspace{0mm}
\subfloat[Bd=0.40\,m]{
\includegraphics[width=.45\columnwidth]
{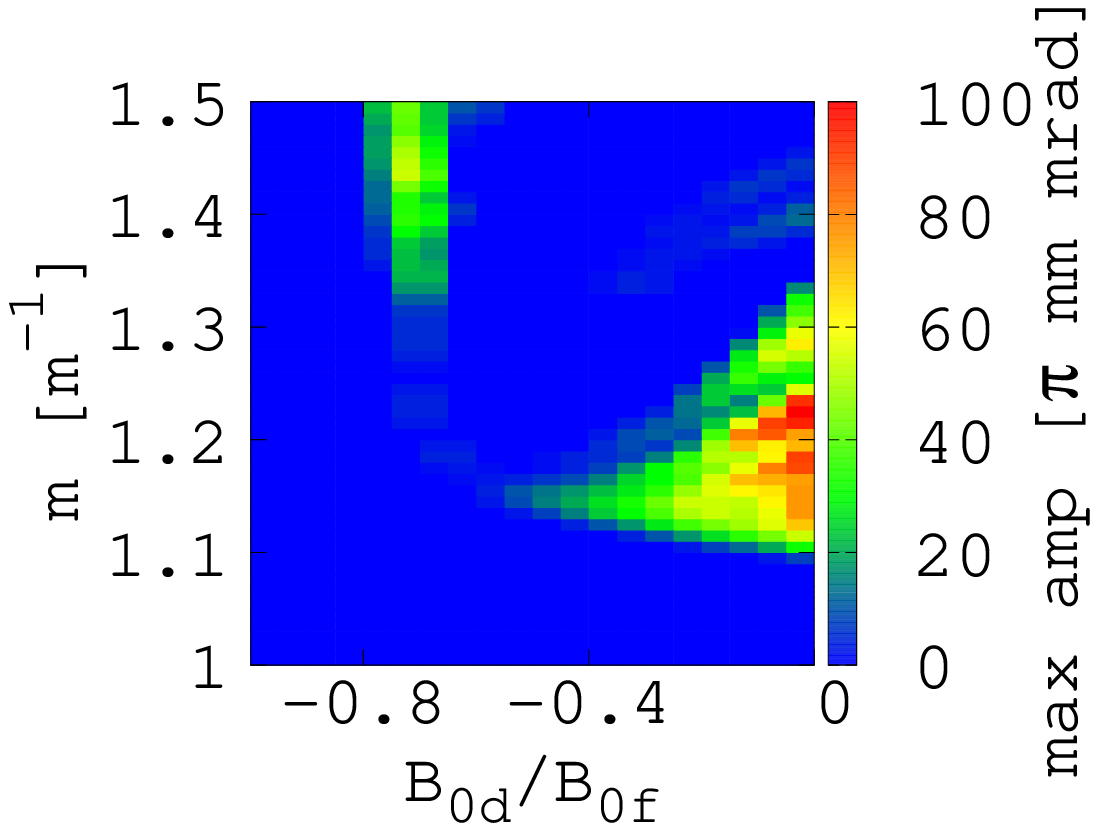}
}
\subfloat[Bd=0.40\,m]{
\includegraphics[width=.45\columnwidth]
{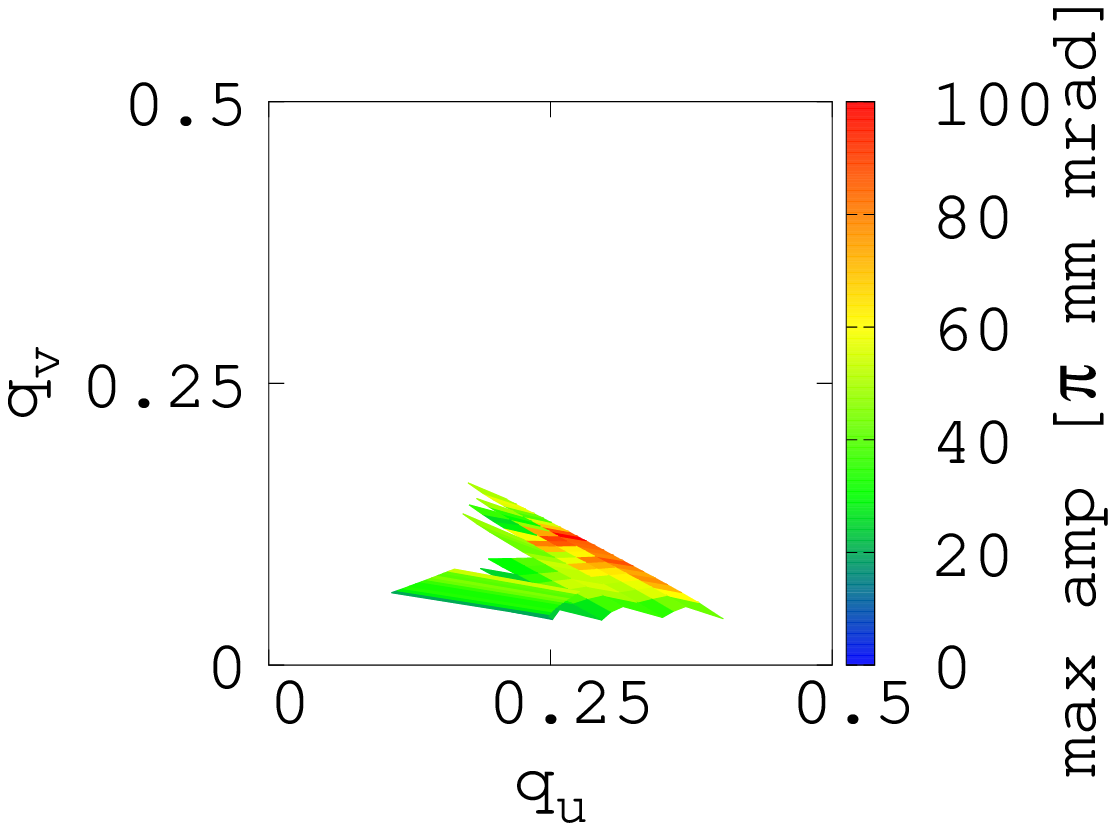}
}
\caption{\label{fig:fig11} Stable area in $m$ and $B_{0d}/B_{0f}$ space and in $q_u$ and $q_v$ space depending on the length of Bd.}
\end{figure}

\begin{figure}[ht]
\centering
\subfloat[Bf=0.24\,m]{
\includegraphics[width=.45\columnwidth]
{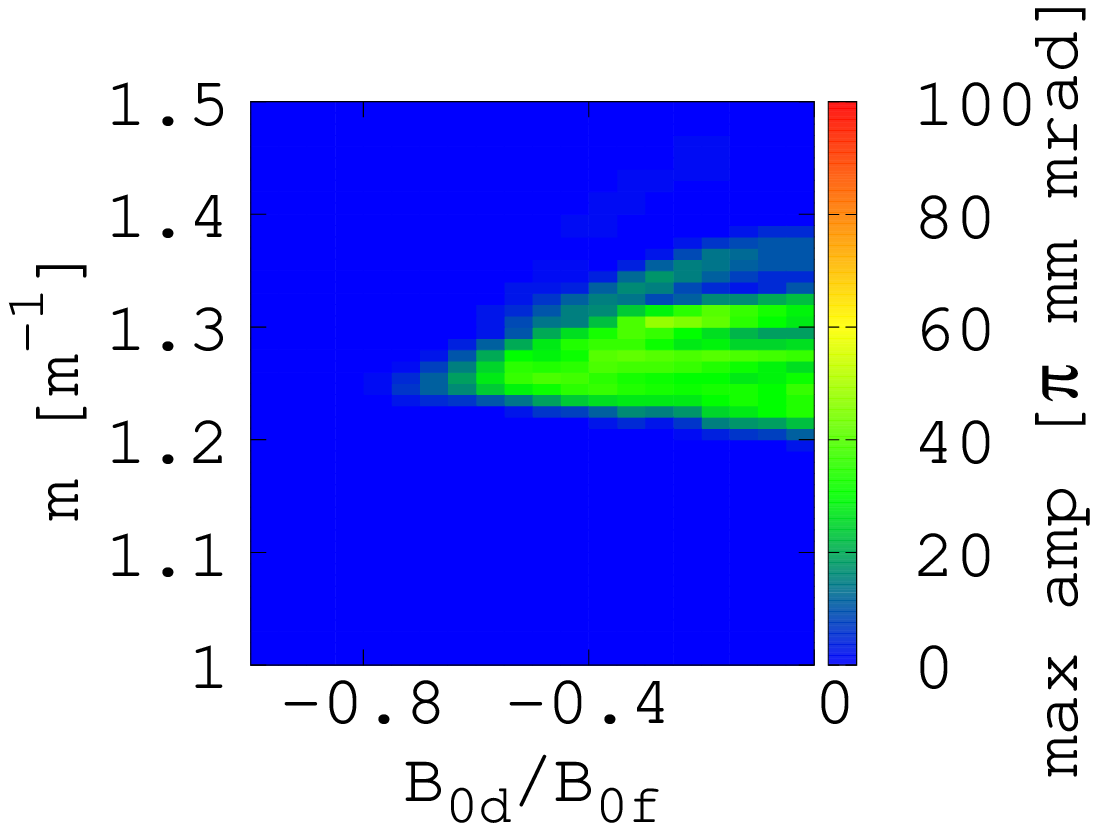}
}
\subfloat[Bf=0.24\,m]{
\includegraphics[width=.45\columnwidth]
{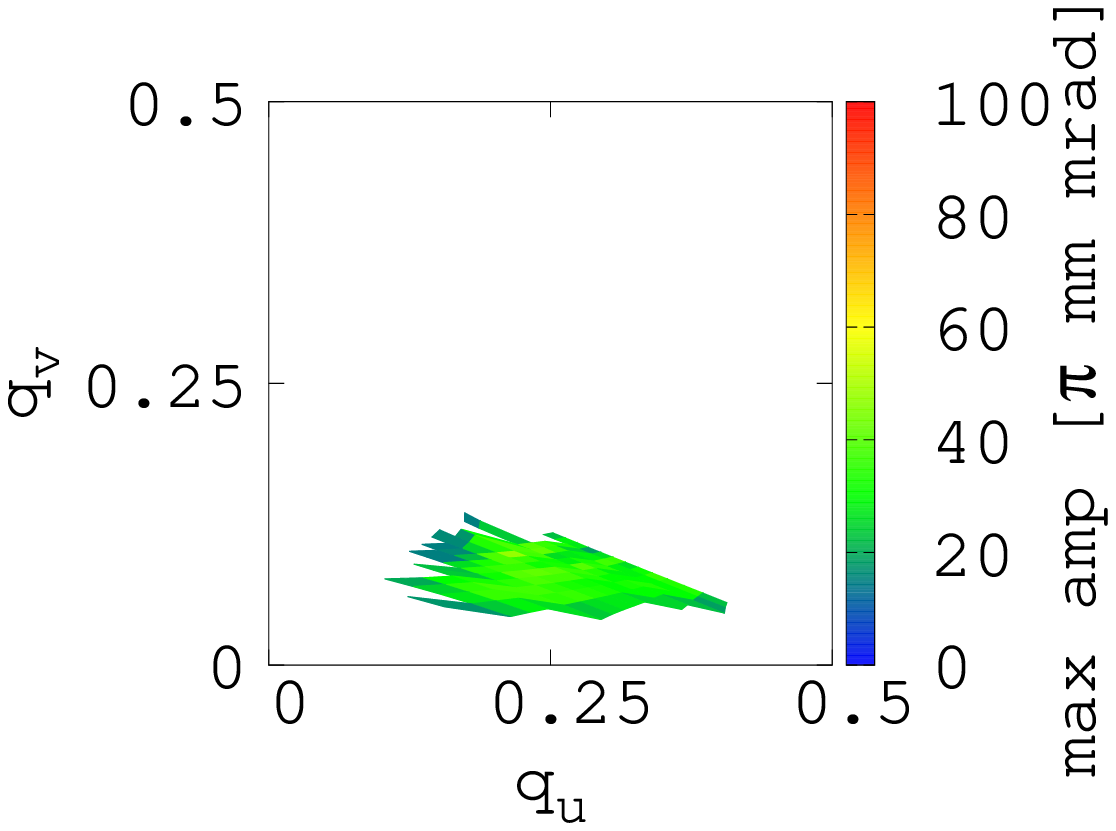}
}
\hspace{0mm}
\subfloat[Bf=0.32\,m]{
\includegraphics[width=.45\columnwidth]
{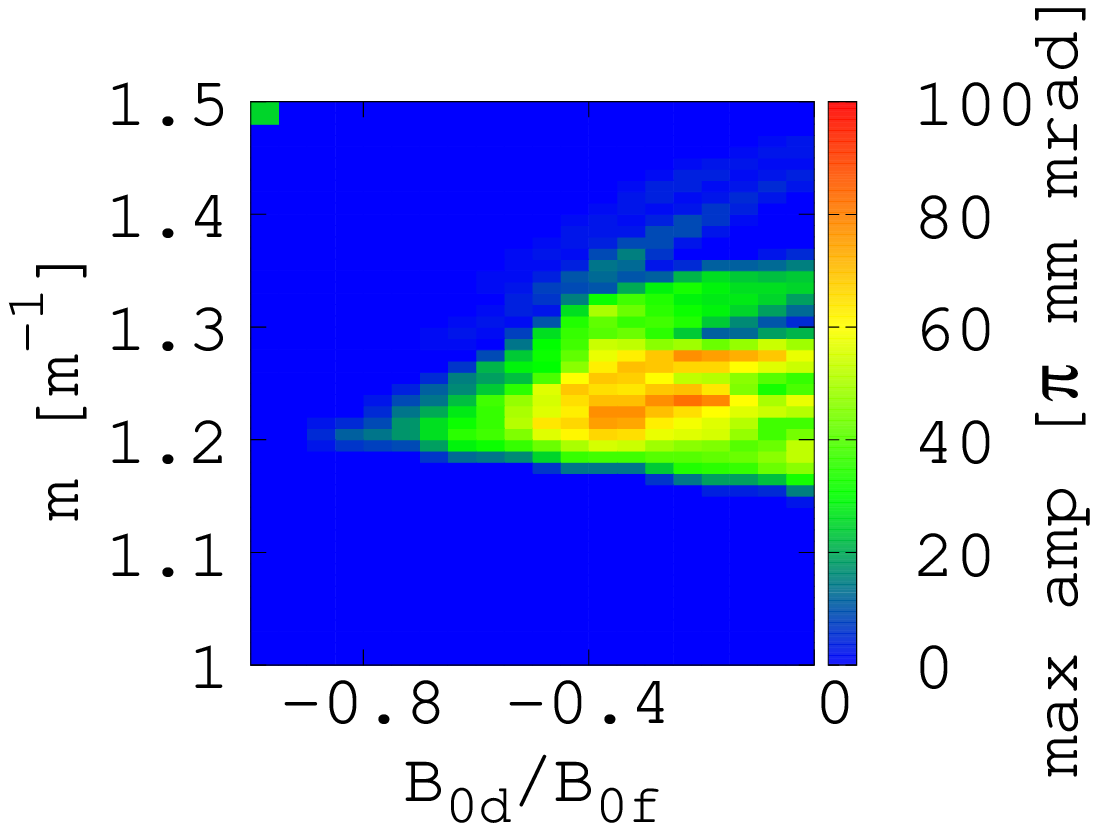}
}
\subfloat[Bf=0.32\,m]{
\includegraphics[width=.45\columnwidth]
{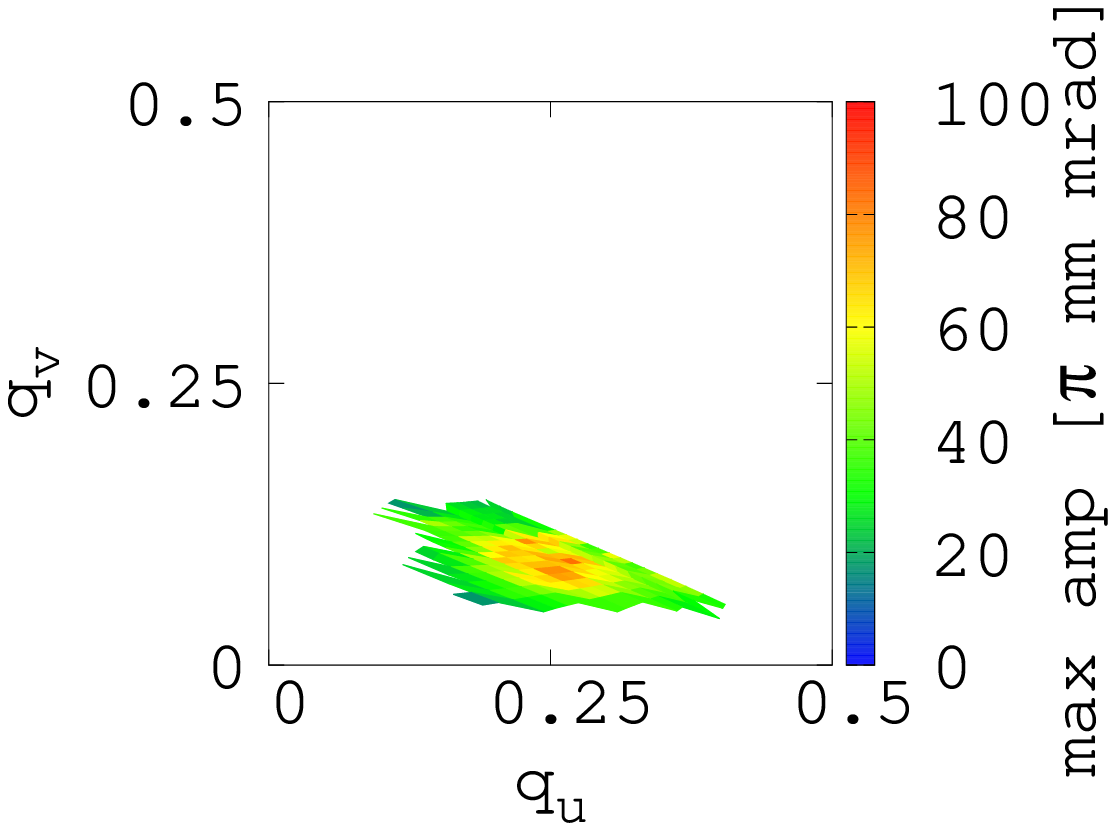}
}
\hspace{0mm}
\subfloat[Bf=0.40\,m]{
\includegraphics[width=.45\columnwidth]
{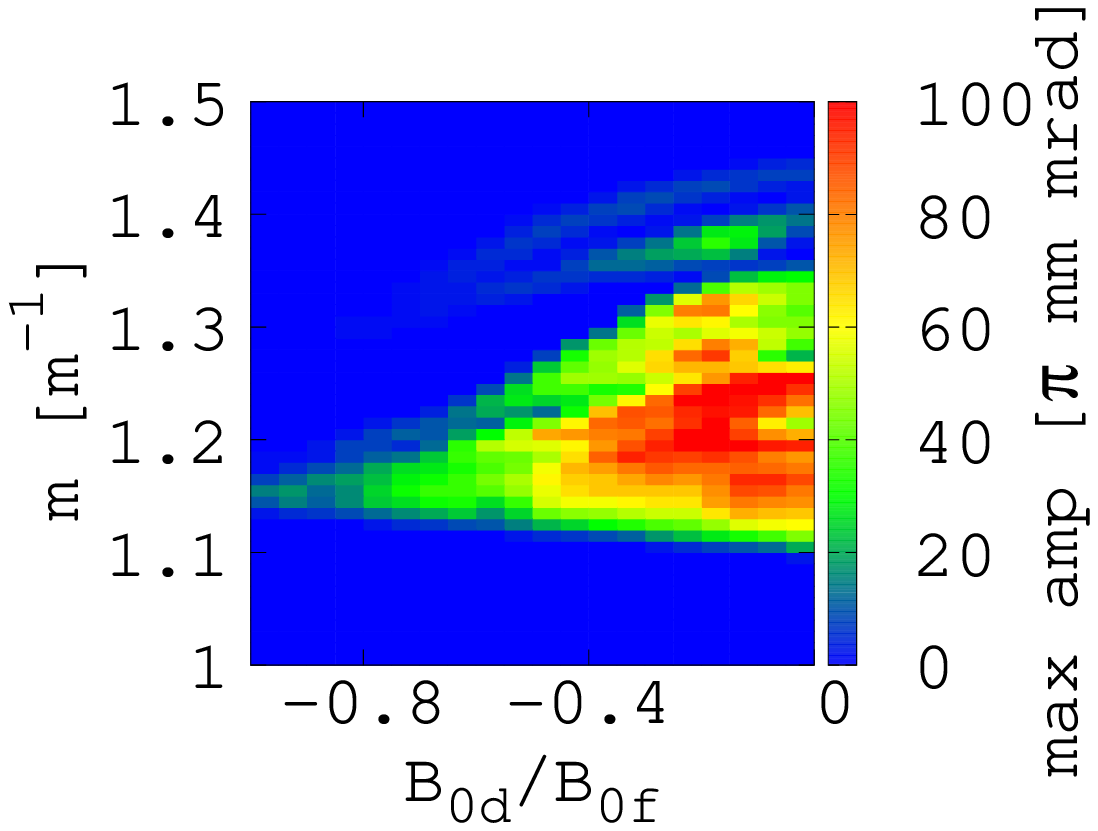}
}
\subfloat[Bf=0.40\,m]{
\includegraphics[width=.45\columnwidth]
{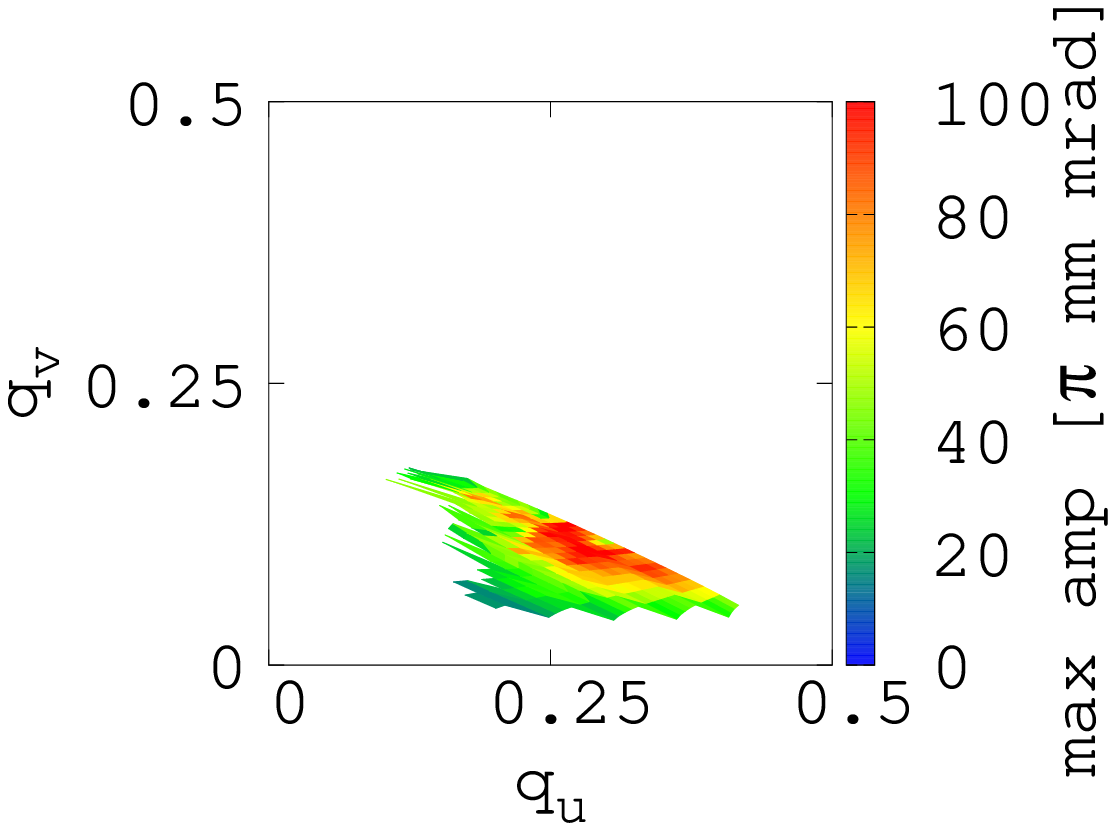}
}
\hspace{0mm}
\subfloat[Bf=0.48\,m]{
\includegraphics[width=.45\columnwidth]
{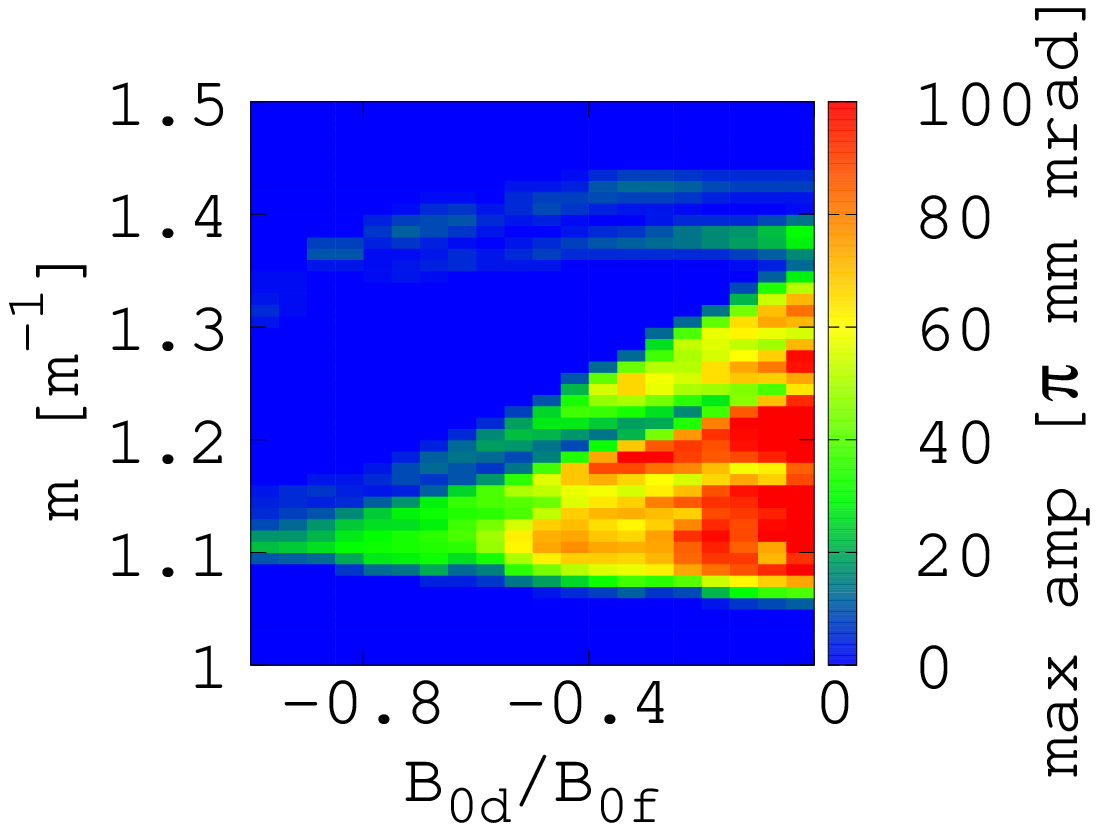}
}
\subfloat[Bf=0.48\,m]{
\includegraphics[width=.45\columnwidth]
{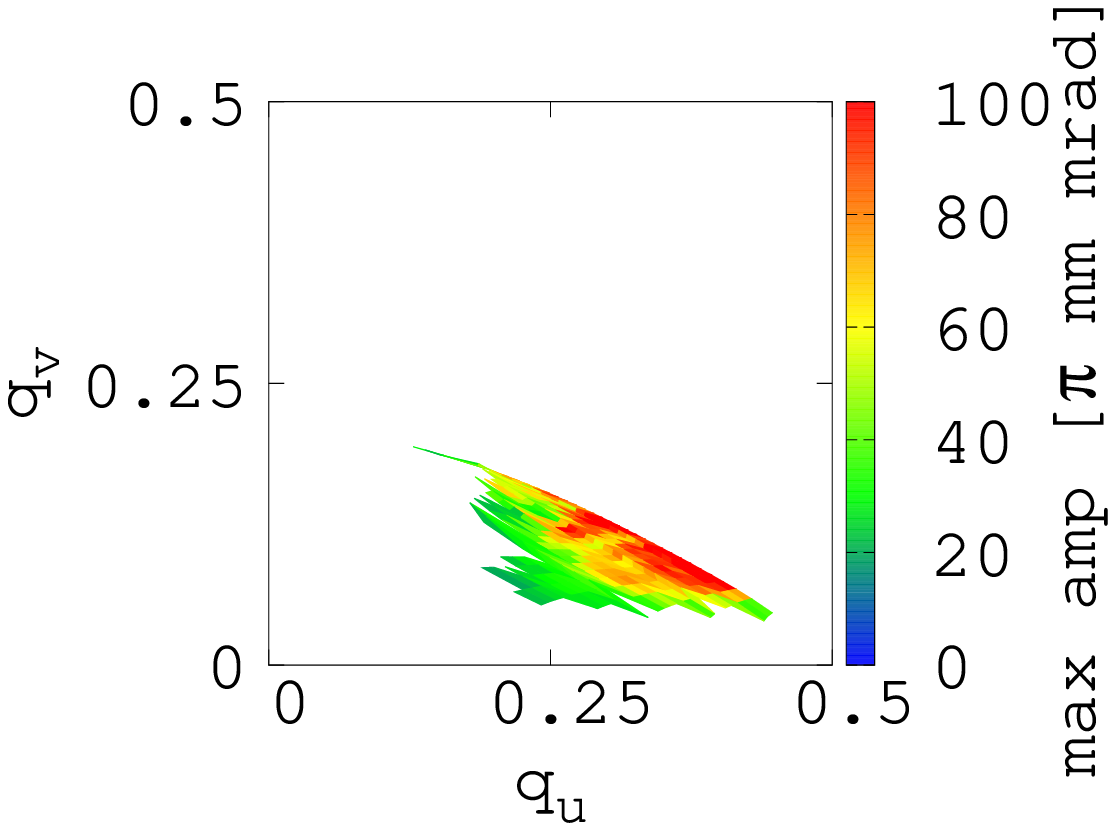}
}
\caption{\label{fig:fig12} Stable area in $m$ and $B_{0d}/B_{0f}$ space and in $q_u$ and $q_v$ space depending on the length of Bf.}
\end{figure}

\begin{figure}[ht]
\centering
\subfloat[L=0.150\,m]{
\includegraphics[width=.45\columnwidth]
{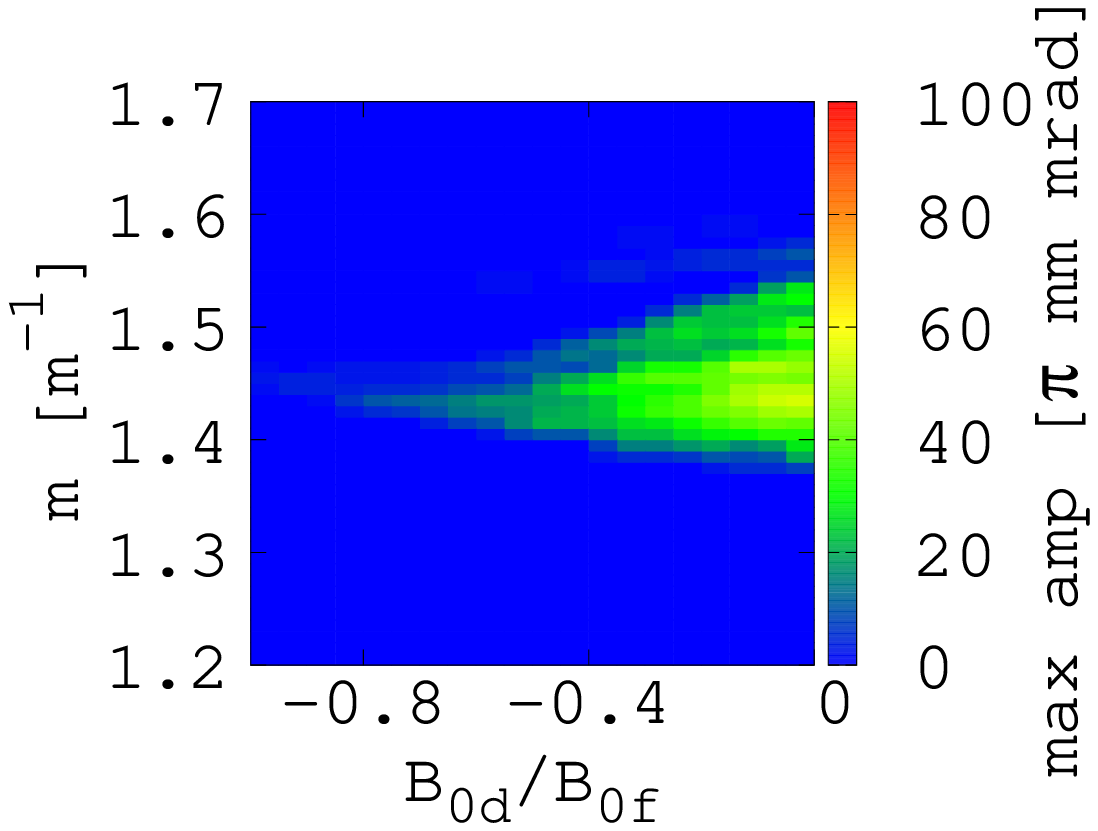}
}
\subfloat[L=0.150\,m]{
\includegraphics[width=.45\columnwidth]
{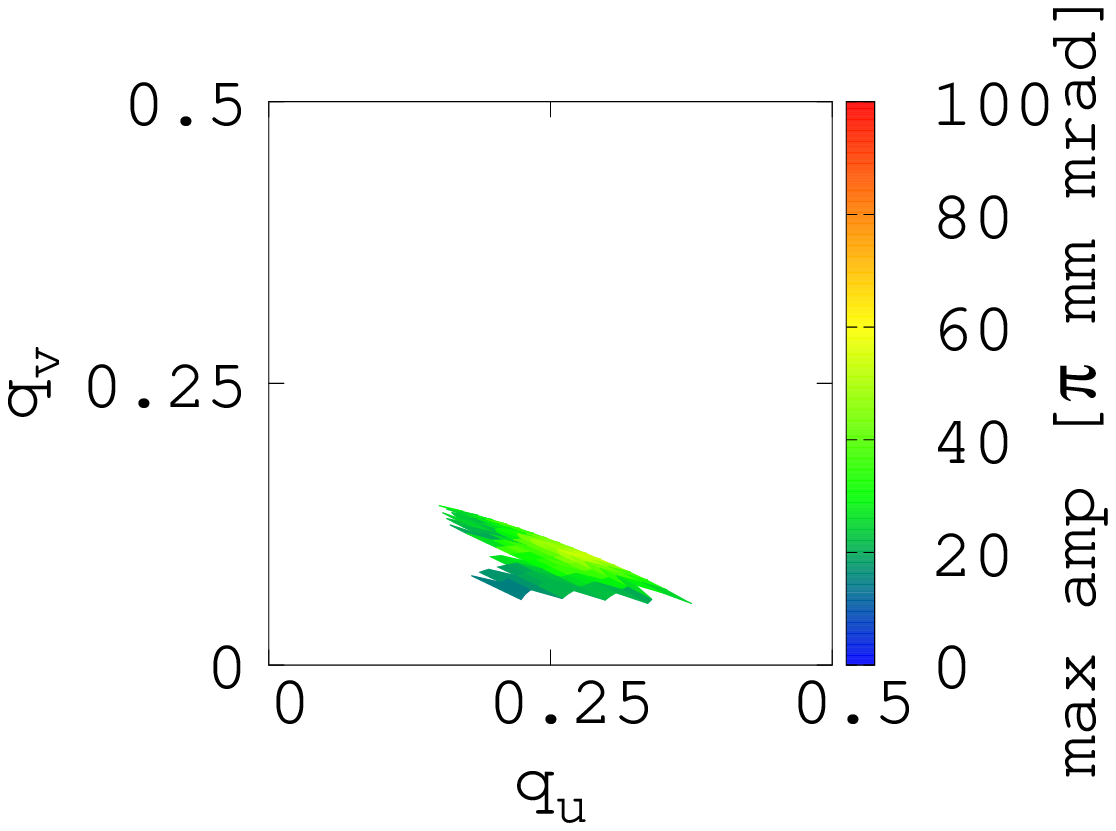}
}
\hspace{0mm}
\subfloat[L=0.175\,m]{
\includegraphics[width=.45\columnwidth]
{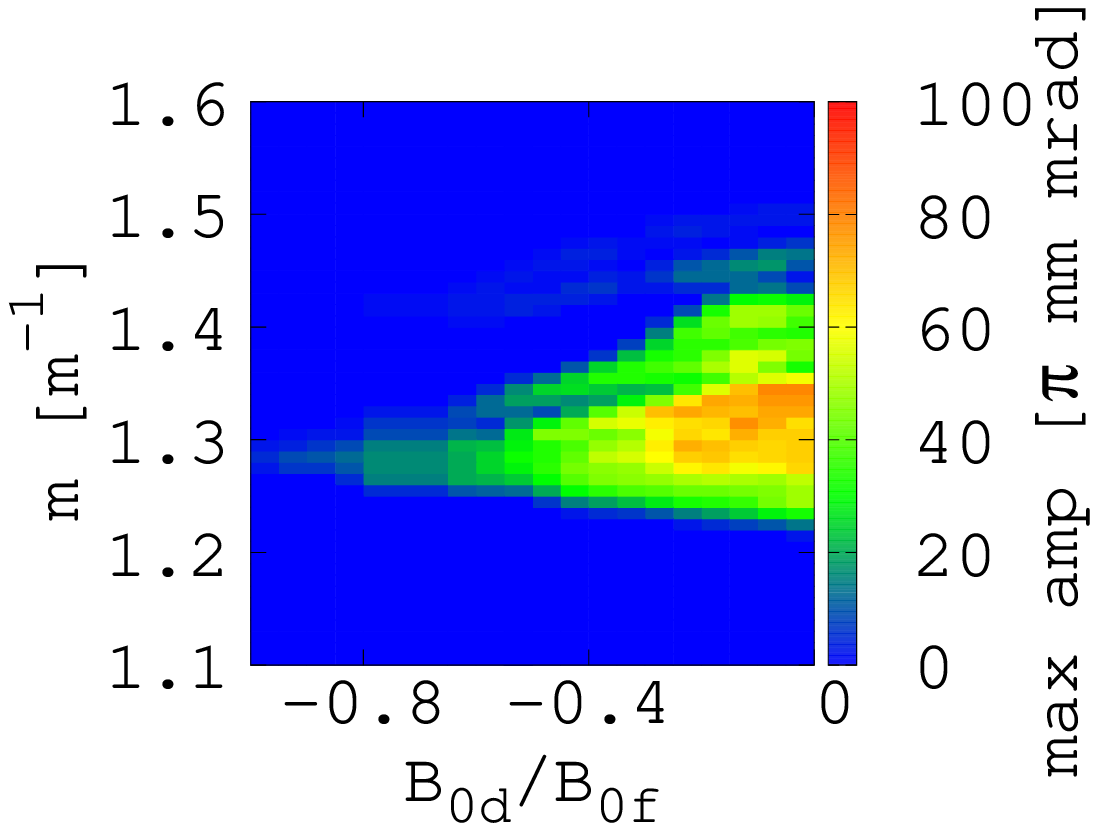}
}
\subfloat[L=0.175\,m]{
\includegraphics[width=.45\columnwidth]
{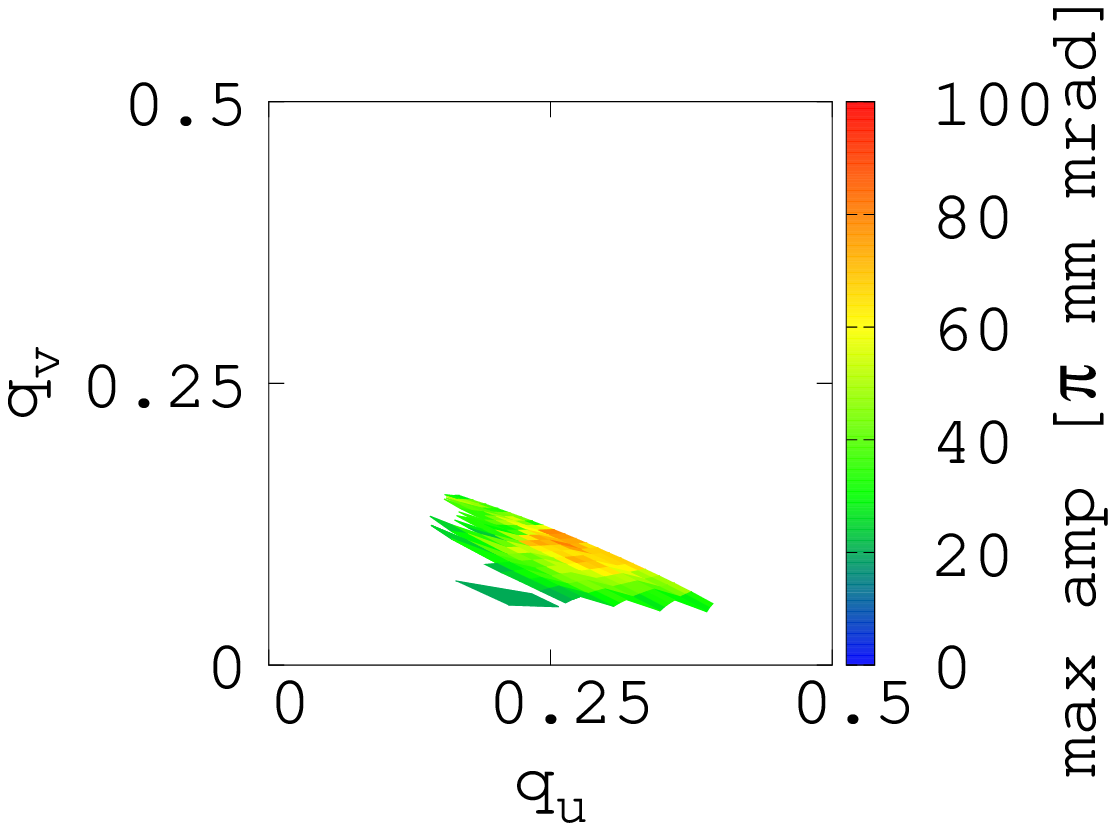}
}
\hspace{0mm}
\subfloat[L=0.200\,m]{
\includegraphics[width=.45\columnwidth]
{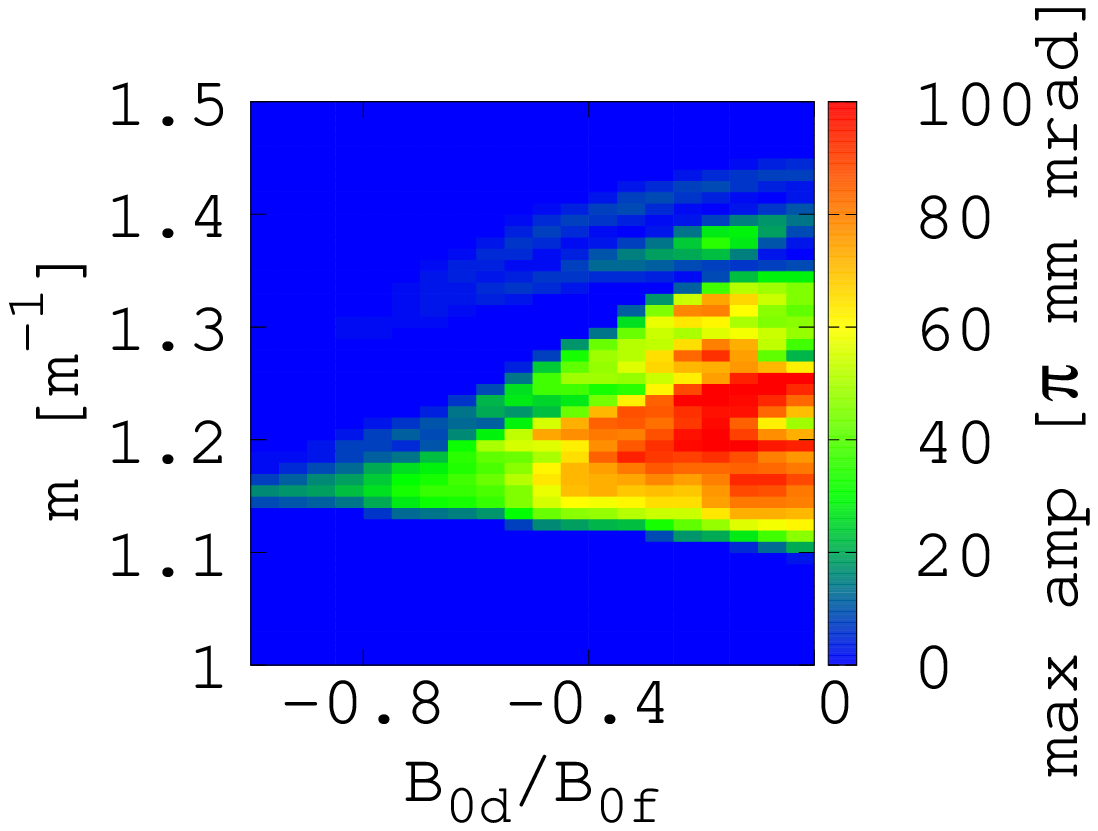}
}
\subfloat[L=0.200\,m]{
\includegraphics[width=.45\columnwidth]
{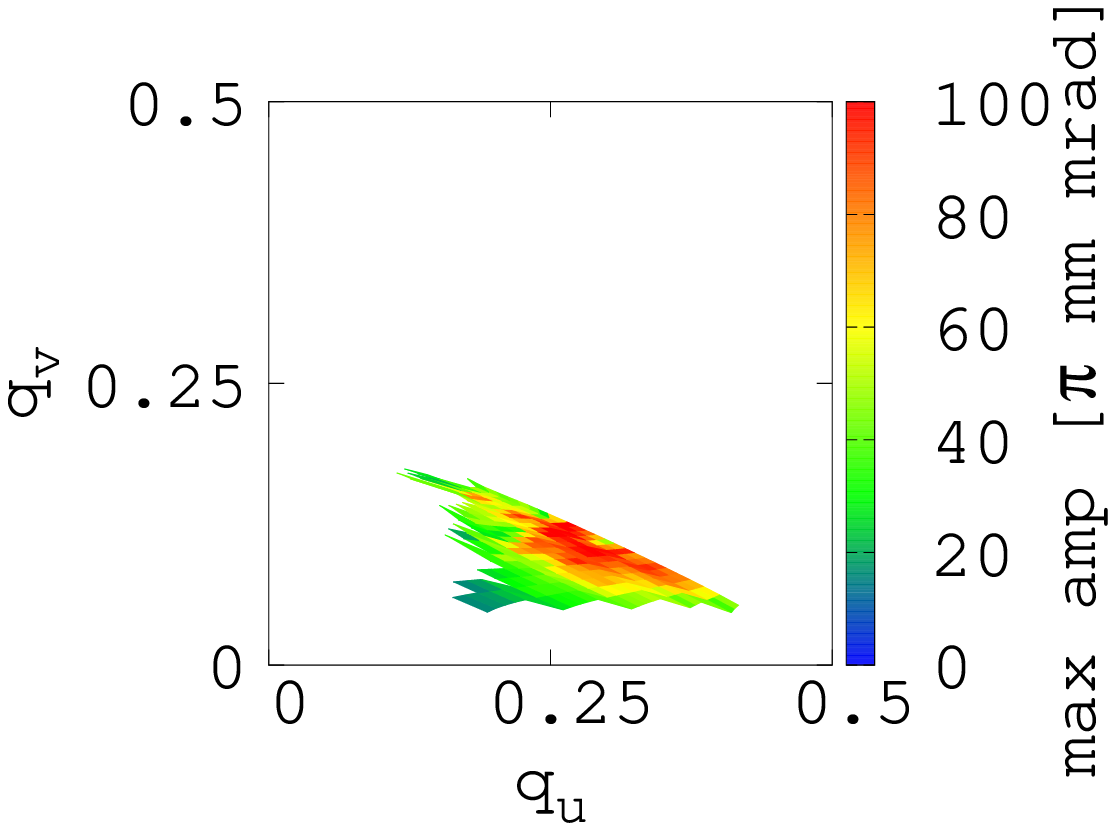}
}
\hspace{0mm}
\subfloat[L=0.225\,m]{
\includegraphics[width=.45\columnwidth]
{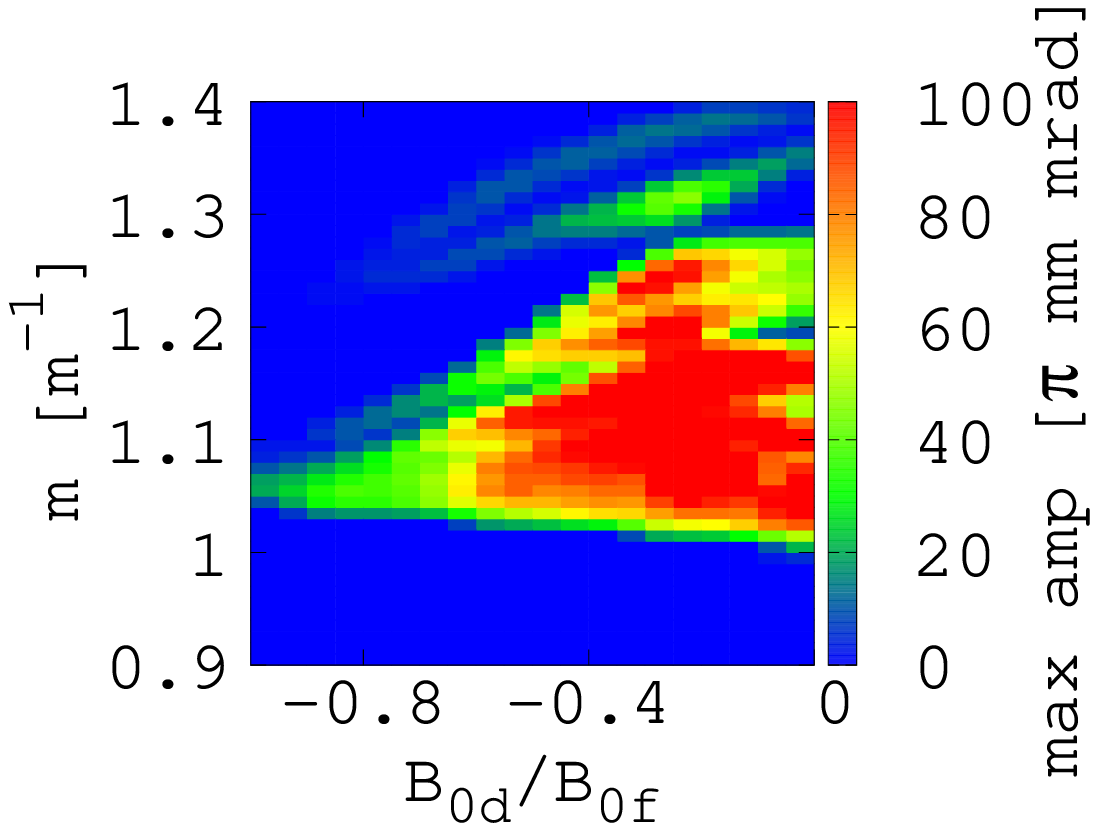}
}
\subfloat[L=0.225\,m]{
\includegraphics[width=.45\columnwidth]
{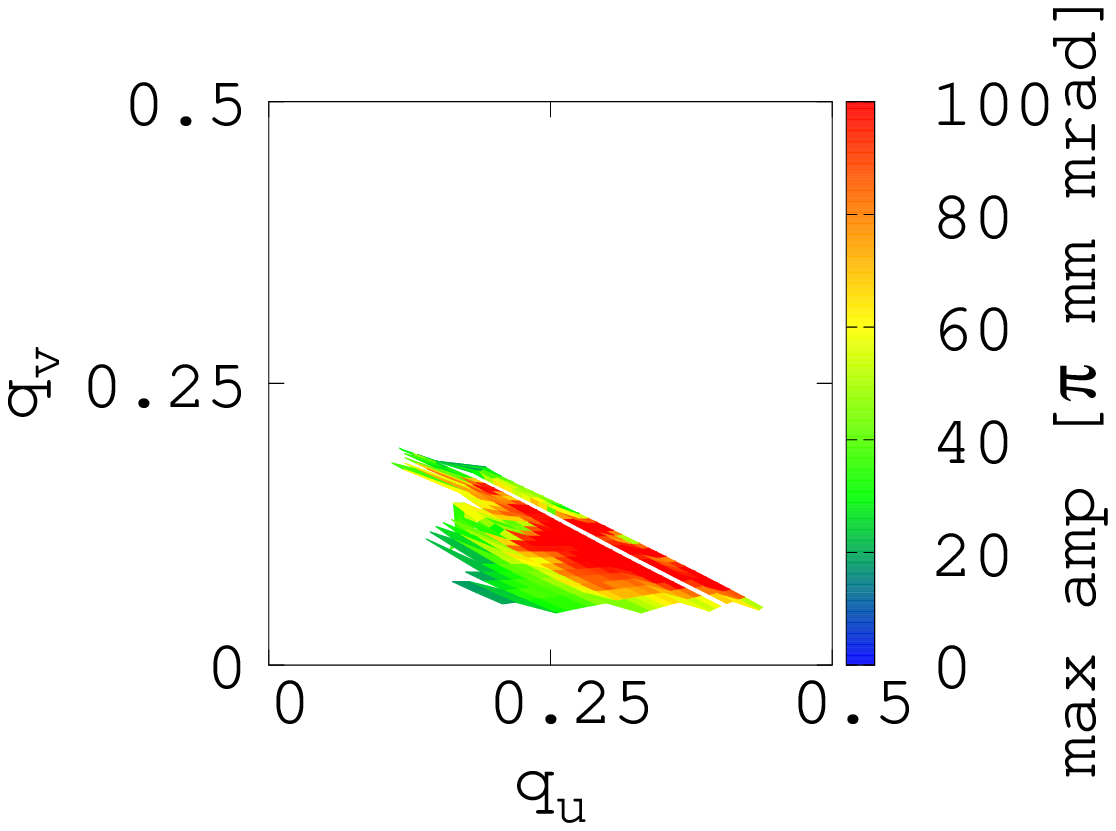}
}
\hspace{0mm}
\subfloat[L=0.250\,m]{
\includegraphics[width=.45\columnwidth]
{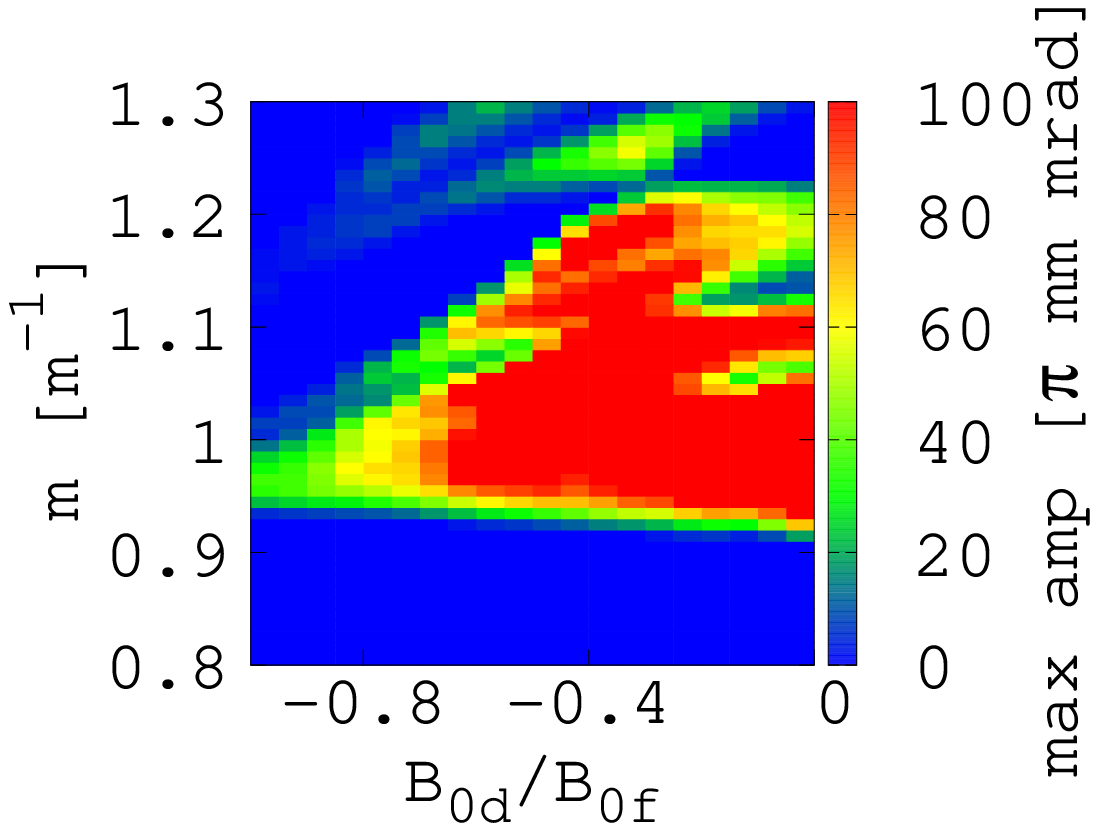}
}
\subfloat[L=0.250\,m]{
\includegraphics[width=.45\columnwidth]
{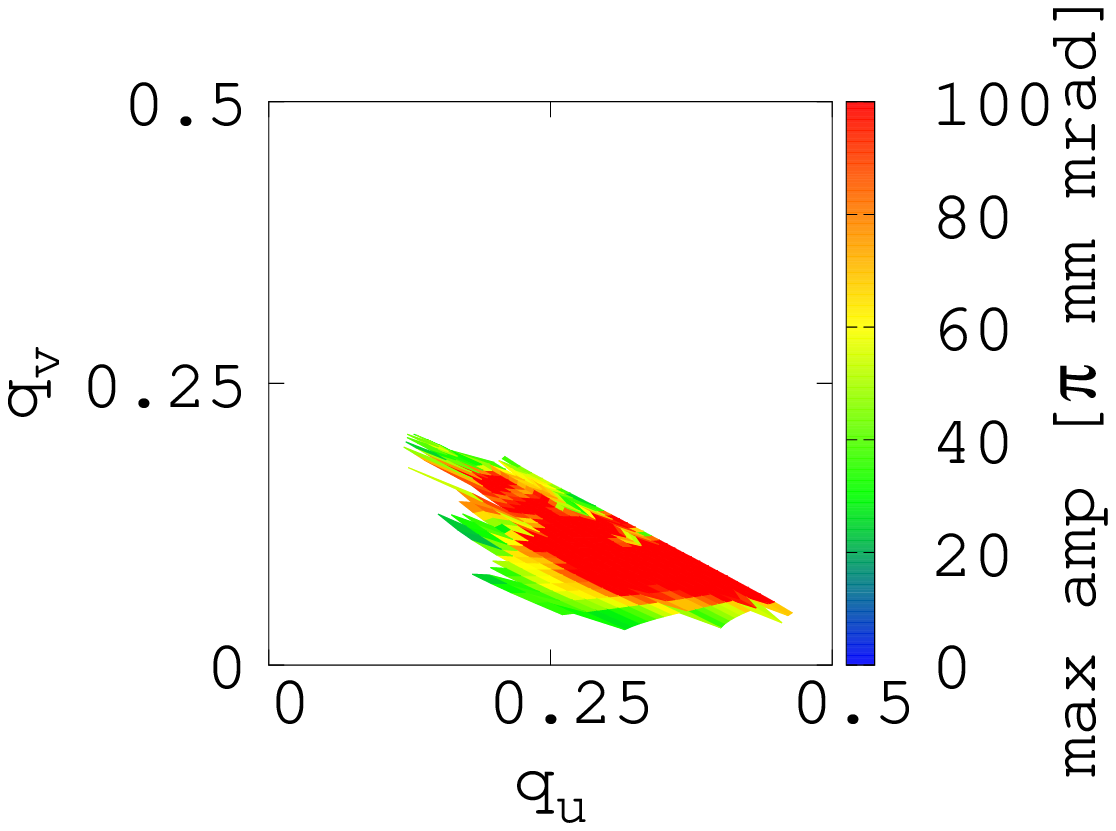}
}
\caption{\label{fig:fig13} Stable area in $m$ and $B_{0d}/B_{0f}$ space and in $q_u$ and $q_v$ space depending on the fringe field extent $L$.}
\end{figure}

The stability of other lattices with slightly different geometry has also been 
investigated.
Figures~\ref{fig:fig11}, \ref{fig:fig12} and \ref{fig:fig13} show the effect of 
varying the lengths of the Bd and Bf magnets and fringe field extent 
respectively. 
There is a non-zero drift space between the Bd and Bf magnets into which the overlapping fringe fields penetrate.

\end{document}